\documentclass[a4paper]{article}

\usepackage{geometry}
\usepackage{color}

\usepackage{upgreek}

\usepackage{graphicx}
\usepackage{array}
\usepackage{multirow}
\usepackage{makecell}

\usepackage{tensor}
\usepackage{mathtools}

\usepackage{amsmath}
\usepackage{mathrsfs}

\usepackage{bm}

\usepackage[noblocks]{authblk}

\usepackage[numbers,sort&compress]{natbib}

\usepackage{CJK}
\usepackage{txfonts}
\usepackage{times}
\usepackage{ragged2e}
\usepackage{indentfirst}

\usepackage{verbatim}

\usepackage{ntheorem}
\newtheorem{theorem}{\textbf{Theorem}}

\newtheorem{corollary}{\textbf{Corollary}}

\renewcommand{\raggedright}{\leftskip=0pt \rightskip=0pt plus 0cm}

\allowdisplaybreaks[4]

\begin{document}

\title{Entanglement swapping for Bell states and Greenberger-Horne-Zeilinger states in qubit systems}

\author{Zhaoxu Ji$^{1,*}$, Peiru Fan$^{1,2,\dag}$, Huanguo Zhang$^{1,\ddagger}$
\\
{\small $^1$Key Laboratory of Aerospace Information Security and Trusted Computing,
Ministry of Education, School of Cyber Science and Engineering,
Wuhan University, Wuhan 430072 China\\
$^2$China Industrial Control Systems Cyber Emergency Response Team, Beijing 100040, China\\
$^{*}$jizhaoxu@whu.edu.cn,$^{\dag}$fanpeiru@whu.edu.cn,$^{\ddagger}$liss@whu.edu.cn}}
\date{}

\maketitle

\begin{abstract}

We introduce a class of two-level multi-particle Greenberger-Horne-Zeilinger (GHZ) states,
and study entanglement swapping between two systems 
for Bell states and the class of GHZ states in qubit systems, respectively.
We give the formulas for the entanglement swapping of Bell states 
and GHZ states in any number of qubit systems.
We further consider entanglement swapping between any number of Bell states and between 
any number of the introduced GHZ states, and propose a series of entanglement swapping schemes in a detailed way.
We illustrate the applications of such schemes in quantum information processing
by proposing quantum protocols for quantum key distribution, quantum secret sharing and quantum private comparison.

\end{abstract}




\section{Introduction}

Quantum entanglement, as one of the most fundamental features of quantum mechanics,
is at the heart of the Einstein-Podolsky-Rosen paradox and the quantum nonlocality
\cite{MacchiavelloC3382004,HorodeckiR8122007}.
Entanglement has been widely considered as one of the most important resources for 
quantum information procession,
such as quantum computation \cite{YangM3662006,GalindoA7422002,WuWQ1672017,WuWQ1442015},
quantum cryptography and quantum communication
\cite{MacchiavelloC3382004,HongL5542020,ZhangH16102019}.
However, the entanglement between two distant sites is extremely fragile
due to the existence of channel noise,
which makes long-distance quantum communication infeasible, let alone the construction of large-scale quantum networks 
\cite{NicolasS8312011}.
Against such a background, quantum repeaters have been proposed, 
which can solve the distance limits by dividing the total communication
distance into shorter channels connected by intermediate nodes
\cite{NicolasS8312011}.

Entanglement swapping is a particularly intriguing and useful method to generate nonlocal correlations
between quantum systems that have never interacted or never had a common past
\cite{ZukowskiM711993,QiangWC389212010},
which makes it a key component of entanglement distribution,
and the key building block for the construction of quantum repeaters
\cite{NicolasS8312011,XuP119172017}.
Entanglement swapping also has important applications in entanglement purification 
\cite{ShiBS6252000,SongW8912004,BoseS6011998},
in various quantum information processing tasks 
\cite{CabelloA6151999,LeeJ7032004,HongLu276562000,
ZhangZ7232005,ZXMan392006,KangMS2492015,ZhouN254462005,LiQ8942014},
and in the creation of multi-particle entangled states \cite{BoseS5721998,HardyL6252000,BoudaJ34202001}.
The study of entanglement swapping was initially focused on discrete variable systems
and then extended to continuous variable systems
\cite{BoseS5721998,HardyL6252000,BoudaJ34202001,KarimipourV652002,BanM37312004,MarshallK1752015}.
In 1993, $\dot{Z}$ukowski et al. proposed the entanglement swapping scheme for two two-level Bell states \cite{ZukowskiM711993}.
Afterwards, Bose et al. studied the entanglement swapping between any number of two-level
multi-particle cat states \cite{BoseS5721998}.
Hardy and Song studied the entanglement swapping chains for d-level bipartite pure states \cite{HardyL6252000}.
Bouda and Buzek presented the entanglement swapping between 
any number of d-level multi-particle cat states \cite{BoudaJ34202001}.
Karimipour et al. investigated the entanglement swapping between a d-level cat state and d-level Bell states \cite{KarimipourV652002}.
Ban considered continuous variable entanglement swapping for bipartite Gaussian states \cite{BanM37312004}.

In this paper, we theoretically study the entanglement swapping of Bell states 
and of Greenberger-Horne-Zeilinger (GHZ) states in qubit systems.
We first consider the entanglement swapping between two Bell states in qubit systems,
and draw the conclusion that two identical Bell states can be generated by entanglement swapping
if the initial two Bell states are the same; otherwise the two Bell states generated are different.
We then introduce a class of multi-particle GHZ states (for simplicity, hereinafter called SGHZ states)
and consider entanglement swapping between them.
Interestingly, we reach a conclusion similar to that of entanglement swapping between two Bell states.
Furthermore, we derive the formulas 
for the entanglement swapping of any number of Bell states and GHZ states in qubit systems.
Then, in a more detailed way, we study the entanglement swapping between any number of qubit systems
for Bell states and SGHZ states to find the schemes that can produce two identical GHZ states.
We show that such schemes can be used to accomplish some quantum information processing tasks
such as quantum key distribution \cite{GalindoA7422002,Brandt4392002},
quantum secret sharing \cite{HilleryM5931999,ZhouP38112007,WangTJ387182008},
and quantum private comparison \cite{YangYG4252009,ZXJi1249112019}.

The rest of this paper is arranged as follows. In Sec. 2, we introduce the SGHZ states
and consider two qubit-system entanglement swapping for Bell states and SGHZ states, respectively.
In Sec. 3, we describe the rules
for the entanglement swapping of Bell states and GHZ states in any number of qubit systems.
We then propose a number of entanglement swapping schemes of any number of Bell states 
and SGHZ states, which can generate two identical GHZ states.
In Sec. 4, we apply entanglement swapping of Bell states and SGHZ states
in quantum key distribution, quantum secret sharing and quantum private comparison.
In Sec. 5, we introduce the main advantages of the entanglement-swapping-based 
quantum protocols over the ones without entanglement swapping,
and then make a comparison between the proposed entanglement swapping schemes and several existing ones.
This paper is summarized in Sec. 6.


\section{Two-state entanglement swapping in qubit systems}

\begin{figure}[t]
\centering
\includegraphics[height=3cm,width=8cm]{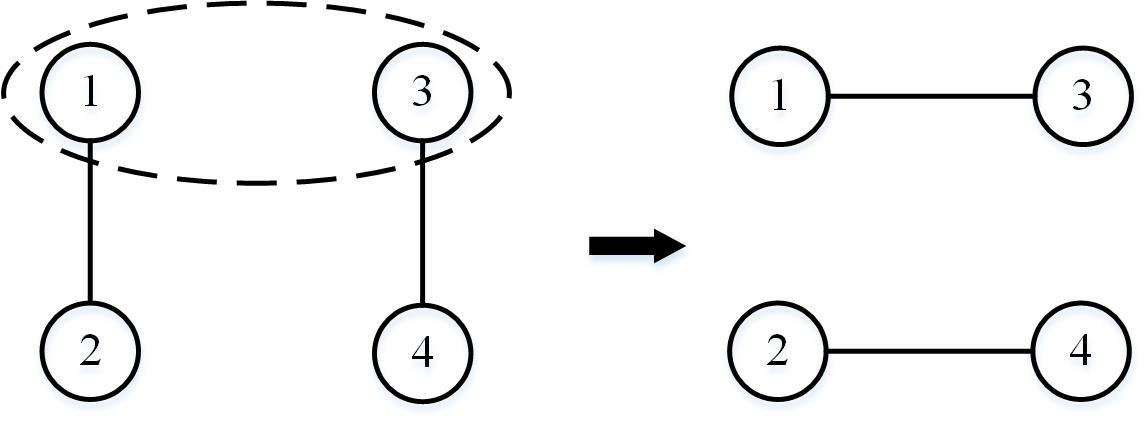}
\caption{
The graphical description of the entanglement swapping between two Bell states. 
The small circle represents a qubit, and the two circles connected by the solid line represents 
a Bell state. The dotted ellipses represents a Bell measurement.
}
\label{ES1}
\end{figure}

Let us start by reviewing the entanglement swapping between two Bell states.
The Bell states can be expressed as \cite{ZukowskiM711993}
\begin{align}
\label{Bell_states}
\left|\phi^{\pm}\right\rangle = \frac 1{\sqrt{2}}
\Big(   \left|00\right\rangle \pm \left|11\right\rangle  \Big),
\quad
\left|\psi^{\pm}\right\rangle = \frac 1{\sqrt{2}} 
\Big(   \left|01\right\rangle {\pm} \left|10\right\rangle  \Big).
\end{align}
One can get
\begin{align}
&\left|00\right\rangle = \frac 1{\sqrt{2}}
\Big(   \left|\phi^{+}\right\rangle + \left|\phi^{-}\right\rangle  \Big), 
\quad
\left|11\right\rangle = \frac 1{\sqrt{2}}
\Big(   \left|\phi^{+}\right\rangle - \left|\phi^{-}\right\rangle  \Big), \notag\\
&\left|01\right\rangle = \frac 1{\sqrt{2}}
\Big(   \left|\psi^{+}\right\rangle + \left|\psi^{-}\right\rangle  \Big), 
\quad
\left|10\right\rangle = \frac 1{\sqrt{2}}
\Big(   \left|\psi^{+}\right\rangle - \left|\psi^{-}\right\rangle  \Big).
\end{align}
We would first like to review the entanglement swapping between two identical Bell states, 
including four cases:
\{$\left|\phi^{+}\right\rangle$, $\left|\phi^{+}\right\rangle$\},
\{$\left|\phi^{-}\right\rangle$, $\left|\phi^{-}\right\rangle$\},
\{$\left|\psi^{+}\right\rangle$, $\left|\psi^{+}\right\rangle$\},
and \{$\left|\psi^{-}\right\rangle$, $\left|\psi^{-}\right\rangle$\}.
For simplicity, let us use $\left|0i\right\rangle \pm \left|1\bar{i}\right\rangle$ to
denote four Bell states, where $i \in \{0,1\}$ and a bar over a bit value indicates its logical negation
(similarly hereinafter).
Suppose that a Bell measurement is performed on the particles (1,3)
(see Fig. \ref{ES1}),
then we can express the entanglement swapping as follows:
\begin{align}
\label{ES_Bell1}
& 	\Big( \left|0i\right\rangle \pm \left|1\bar{i}\right\rangle \Big)_{12} \otimes
	\Big( \left|0i\right\rangle \pm \left|1\bar{i}\right\rangle \Big)_{34}	\notag\\
= & 	\left|0i0i\right\rangle_{1234} \pm \left|0i1\bar{i}\right\rangle_{1234} \pm
	\left|1\bar{i}0i\right\rangle_{1234} + \left|1\bar{i}1\bar{i}\right\rangle_{1234}	 \notag\\
= & 	\left|00ii\right\rangle_{1324} \pm \left|01i\bar{i}\right\rangle_{1324} \pm
	\left|10\bar{i}i\right\rangle_{1324} + \left|11\bar{i}\bar{i}\right\rangle_{1324}	\notag\\
= &	
\begin{cases}	
&	\left|\phi^{+}\right\rangle_{13} \left|\phi^{+}\right\rangle_{24} +
	\left|\phi^{-}\right\rangle_{13} \left|\phi^{-}\right\rangle_{24} \pm
	\left|\psi^{+}\right\rangle_{13} \left|\psi^{+}\right\rangle_{24} \pm
	\left|\psi^{-}\right\rangle_{13} \left|\psi^{-}\right\rangle_{24} \quad \textrm{if} \phantom{i} i = 0;	\\
&	\left|\phi^{+}\right\rangle_{13} \left|\phi^{+}\right\rangle_{24} -
	\left|\phi^{-}\right\rangle_{13} \left|\phi^{-}\right\rangle_{24} \pm
	\left|\psi^{+}\right\rangle_{13} \left|\psi^{+}\right\rangle_{24} \mp
	\left|\psi^{-}\right\rangle_{13} \left|\psi^{-}\right\rangle_{24} \quad \textrm{if} \phantom{i} i = 1,
\end{cases}
\end{align}
where the subscripts (1,2) and (3,4) denote two particles in the two Bell states, respectively.
Note here that we swap the particles 2 and 3 in the second step of the above formula,
and for simplicity, we express the four cases mentioned above by this formula
and ignore the inessential coefficients (similarly hereinafter).

The entanglement swapping between two different Bell states
includes the following 12 different cases:
$ \Big[\Big( \left|0i\right\rangle \pm \left|1\bar{i}\right\rangle \Big)_{12},
\Big( \left|0i\right\rangle \mp \left|1\bar{i}\right\rangle \Big)_{34}\Big]$,
$ \Big[\Big( \left|0i\right\rangle \pm \left|1\bar{i}\right\rangle \Big)_{12},
\Big( \left|0\bar{i}\right\rangle \pm \left|1i\right\rangle \Big)_{34}\Big]$, and
$ \Big[\Big( \left|0i\right\rangle \pm \left|1\bar{i}\right\rangle \Big)_{12},
\Big( \left|0\bar{i}\right\rangle \mp \left|1i\right\rangle \Big)_{34}\Big]$.
We express them by
\begin{align}
\label{ES_Bell2}
& 	\Big( \left|0i\right\rangle \pm \left|1\bar{i}\right\rangle \Big)_{12} \otimes
	\Big( \left|0i\right\rangle \mp \left|1\bar{i}\right\rangle \Big)_{34}	\notag\\
= & 	\left|0i0i\right\rangle_{1234} \mp \left|0i1\bar{i}\right\rangle_{1234} \pm
	\left|1\bar{i}0i\right\rangle_{1234} - \left|1\bar{i}1\bar{i}\right\rangle_{1234}	\notag\\
= & 	\left|00ii\right\rangle_{1324} \mp \left|01i\bar{i}\right\rangle_{1324} \pm
	\left|10\bar{i}i\right\rangle_{1324} - \left|11\bar{i}\bar{i}\right\rangle_{1324}	\notag\\
= &	
\begin{cases}
\phantom{-}	\left|\phi^{+}\right\rangle_{13} \left|\phi^{-}\right\rangle_{24} +
	\left|\phi^{-}\right\rangle_{13} \left|\phi^{+}\right\rangle_{24} \mp
	\left|\psi^{+}\right\rangle_{13} \left|\psi^{-}\right\rangle_{24} \mp
	\left|\psi^{-}\right\rangle_{13} \left|\psi^{+}\right\rangle_{24} \quad \textrm{if} \phantom{i} i = 0;	\\
- \left|\phi^{+}\right\rangle_{13} \left|\phi^{-}\right\rangle_{24} +
	\left|\phi^{-}\right\rangle_{13} \left|\phi^{+}\right\rangle_{24} \pm
	\left|\psi^{+}\right\rangle_{13} \left|\psi^{-}\right\rangle_{24} \mp
	\left|\psi^{-}\right\rangle_{13} \left|\psi^{+}\right\rangle_{24} \quad \textrm{if} \phantom{i} i = 1,
\end{cases}
\end{align}
\begin{align}
\label{ES_Bell3}
& 	\Big( \left|0i\right\rangle \pm \left|1\bar{i}\right\rangle \Big)_{12} \otimes
	\Big( \left|0\bar{i}\right\rangle \pm \left|1i\right\rangle \Big)_{34}	\notag\\
= & 	\left|0i0\bar{i}\right\rangle_{1234} \pm \left|0i1i\right\rangle_{1234} \pm
	\left|1\bar{i}0\bar{i}\right\rangle_{1234} + \left|1\bar{i}1i\right\rangle_{1234}	\notag\\
= & 	\left|00i\bar{i}\right\rangle_{1324} \pm \left|01ii\right\rangle_{1324} \pm
	\left|10\bar{i}\bar{i}\right\rangle_{1324} + \left|11\bar{i}i\right\rangle_{1324}	\notag\\
= &	
\begin{cases}
&	\left|\phi^{+}\right\rangle_{13} \left|\psi^{+}\right\rangle_{24} +
	\left|\phi^{-}\right\rangle_{13} \left|\psi^{-}\right\rangle_{24} \pm
	\left|\psi^{+}\right\rangle_{13} \left|\phi^{+}\right\rangle_{24} \pm
	\left|\psi^{-}\right\rangle_{13} \left|\phi^{-}\right\rangle_{24} \quad \textrm{if} \phantom{i} i = 0;	\\
&	\left|\phi^{+}\right\rangle_{13} \left|\psi^{+}\right\rangle_{24} -
	\left|\phi^{-}\right\rangle_{13} \left|\psi^{-}\right\rangle_{24} \pm
	\left|\psi^{+}\right\rangle_{13} \left|\phi^{+}\right\rangle_{24} \mp
	\left|\psi^{-}\right\rangle_{13} \left|\phi^{-}\right\rangle_{24} \quad \textrm{if} \phantom{i} i = 1,
\end{cases}
\end{align}
and
\begin{align}
\label{ES_Bell4}
& 	\Big( \left|0i\right\rangle \pm \left|1\bar{i}\right\rangle \Big)_{12} \otimes
	\Big( \left|0\bar{i}\right\rangle \mp \left|1i\right\rangle \Big)_{34}	\notag\\
= & 	\left|0i0\bar{i}\right\rangle_{1234} \mp \left|0i1i\right\rangle_{1234} \pm
	\left|1\bar{i}0\bar{i}\right\rangle_{1234} - \left|1\bar{i}1i\right\rangle_{1234}	\notag\\
= & 	\left|00i\bar{i}\right\rangle_{1324} \mp \left|01ii\right\rangle_{1324} \pm
	\left|10\bar{i}\bar{i}\right\rangle_{1324} - \left|11\bar{i}i\right\rangle_{1324}	\notag\\
= &	
\begin{cases}
\phantom{-}	\left|\phi^{+}\right\rangle_{13} \left|\psi^{-}\right\rangle_{24} +
	\left|\phi^{-}\right\rangle_{13} \left|\psi^{+}\right\rangle_{24} \mp
	\left|\psi^{+}\right\rangle_{13} \left|\phi^{-}\right\rangle_{24} \mp
	\left|\psi^{-}\right\rangle_{13} \left|\phi^{+}\right\rangle_{24} \quad \textrm{if} \phantom{i} i = 0;	\\
-	\left|\phi^{+}\right\rangle_{13} \left|\psi^{-}\right\rangle_{24} +
	\left|\phi^{-}\right\rangle_{13} \left|\psi^{+}\right\rangle_{24} \pm
	\left|\psi^{+}\right\rangle_{13} \left|\phi^{-}\right\rangle_{24} \mp
	\left|\psi^{-}\right\rangle_{13} \left|\phi^{+}\right\rangle_{24} \quad \textrm{if} \phantom{i} i = 1.
\end{cases}
\end{align}
From Eqs. (\ref{ES_Bell1}) to (\ref{ES_Bell4}), 
we find that if the initial two Bell states are the same, 
then the two Bell states obtained by entanglement swapping are the same
[i.e., the Bell state obtained by performing the measurement on the particles (1,3)
is the same as the one that the particles (2,4) collapse into], otherwise they are different.

In what follows, we will introduce a class of multi-particle GHZ states 
and consider entanglement swapping.
Let us first introduce the full set of canonical orthonormal $m$-qubit 
($m \in \rm N_+$ and $m \ge 3$) GHZ states \cite{ParasharP8332011,JiZX19112019}:
\begin{align}
\label{GHZ1}
\left|\mathcal{G}_{d}^{\pm}\right\rangle = \frac 1{\sqrt{2}}
\Big( \left|\mathbb{B}\big(d\big)\right\rangle \pm 
			\left|\mathbb{B}\big(2^{m}-d-1 \big)\right\rangle \Big),
\end{align}
where $d = 0,1,\ldots,2^{m-1}-1$, and $\mathbb{B}(d) = 0b_2b_3\cdots b_m$ is the
binary representation of $d$ in an $m$-bit string, thus
$d = \sum_{k=2}^{m} b_k\cdot2^{m-k}$.
These states are complete and orthonormal,
\begin{align}
\begin{cases}
\phantom{i} \langle \mathcal{G}_{d}^{\pm}|\mathcal{G}_{d^{\prime}}^{\pm} \rangle = \delta_{d,d^{\prime}}, \\
\phantom{i} \left|\mathbb{B}\big(d\big)\right\rangle = \frac 1{\sqrt{2}}
\Big( \left|\mathcal{G}_{d}^{+}\right\rangle + \left|\mathcal{G}_{d}^{-}\right\rangle \Big), \\
\phantom{i} \left|\mathbb{B}\big(2^{m}-d-1 \big)\right\rangle = \frac 1{\sqrt{2}}
\Big( \left|\mathcal{G}_{d}^{+}\right\rangle - \left|\mathcal{G}_{d}^{-}\right\rangle \Big),
\end{cases}
\end{align}
and they can be written in a more concise form:
\begin{align}
\label{GHZ2}
\left|\mathcal{G}_{d}^{\pm}\right\rangle = \frac 1{\sqrt{2}}
\Big( \left|0\right\rangle \bigotimes_{k=2}^{m} \left|b_k\right\rangle \pm
\left|1\right\rangle \bigotimes_{k=2}^{m}  \left|\bar{b}_k\right\rangle \Big).
\end{align}
Let us now introduce the class of GHZ states
(hereinafter called SGHZ states for simplicity),
\begin{align}
\label{2n-GHZ}
\left|\mathcal{G}_{e}^{\pm}\right\rangle = \frac 1{\sqrt{2}}
\Big( \left|0\right\rangle \bigotimes_{k=2}^{2n} \left|i_k\right\rangle \pm
\left|1\right\rangle \bigotimes_{k=2}^{2n}  \left|\bar{i}_k\right\rangle \Big), 
\end{align}
each of which contains $2n$ qubits,
where $n \in \rm N_+$ and $n \ge 2$,
$e = \sum_{k=2}^{2n} i_k\cdot2^{2n-k}$,
and $ 0i_2i_3\cdots i_n = i_{n+1}i_{n+2}\cdots i_{2n} $ or 
$ 0i_2i_3\cdots i_n = \bar{i}_{n+1}\bar{i}_{n+2}\cdots\bar{i}_{2n} $
(this is the characteristic of SGHZ states).
One can get
\begin{align}
\left|0\right\rangle \bigotimes_{k=2}^{2n} \left|i_k\right\rangle = \frac 1{\sqrt{2}}
\Big(   \left|\mathcal{G}_{e}^{+}\right\rangle + \left|\mathcal{G}_{e}^{-}\right\rangle  \Big),
\quad
\left|1\right\rangle \bigotimes_{k=2}^{2n}  \left|\bar{i}_k\right\rangle =
\frac 1{\sqrt{2}}
\Big(   \left|\mathcal{G}_{e}^{+}\right\rangle - \left|\mathcal{G}_{e}^{-}\right\rangle  \Big).
\end{align}

\begin{figure}[t]
\centering
\includegraphics[height=3cm,width=11cm]{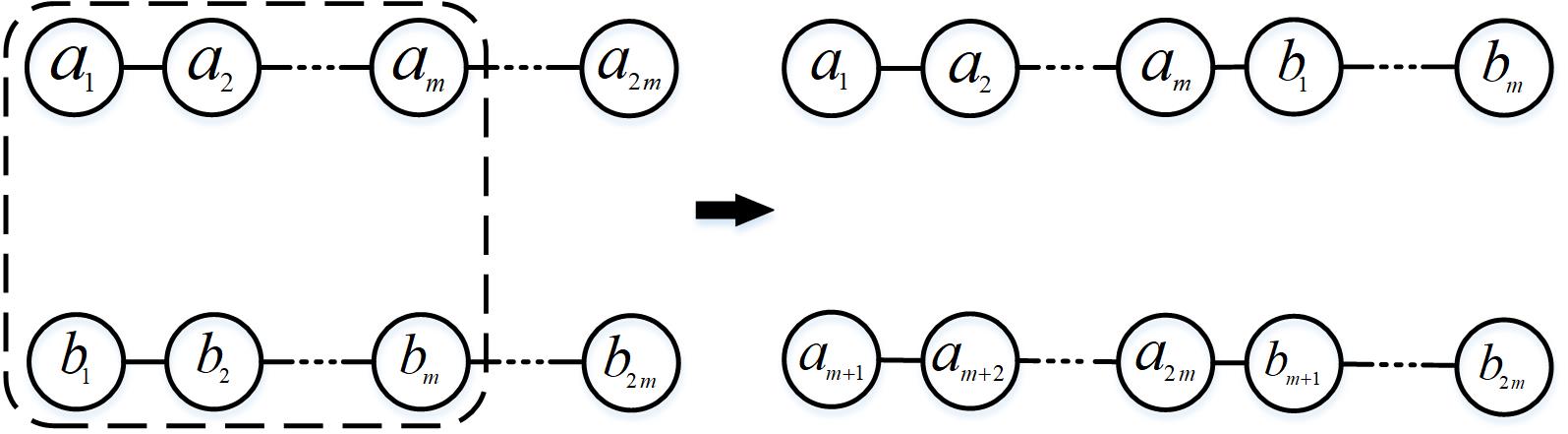}
\caption{
The graphical description of the entanglement swapping between two SGHZ states,
each composed of $2m$ particles.
The marks $a_1,a_2,\ldots,a_{2m}$ and $b_1,b_2,\ldots,b_{2m}$ 
represents the particles in the two SGHZ states, respectively.
The dotted box represents the GHZ measurement which is performed on
the first $m$ particles in each state.
}
\label{ES2}
\end{figure}

We now consider the entanglement swapping between two SGHZ states
containing the same number of particles.
Let us first derive formulas for the entanglement swapping between two identical SGHZ states,
that is, \{$\left|\mathcal{G}_{e}^{+}\right\rangle_{12}$,
$\left|\mathcal{G}_{e}^{+}\right\rangle_{34}$\}
and \{$\left|\mathcal{G}_{e}^{-}\right\rangle_{12}, \left|\mathcal{G}_{e}^{-}\right\rangle_{34}$\},
where the subscripts 1 and 3 denote respectively the first $n$ particles
in $\left|G_{e}^{\pm}\right\rangle_{12}$ and $\left|G_{e}^{\pm}\right\rangle_{34}$,
and 2 and 4 the last $n$ particles.
Suppose that a GHZ measurement is performed on the particles 1 and 3
(see Fig. \ref{ES2}, similarly hereinafter), then we arrive at
\begin{align}
\label{2nGHZ1}
&\left|\mathcal{G}_{e}^{\pm}\right\rangle_{12} \otimes 
	\left|\mathcal{G}_{e}^{\pm}\right\rangle_{34}	\notag\\
= &\quad	\Big( \left|0\right\rangle \bigotimes_{k=2}^{2n} \left|i_k\right\rangle \pm
\left|1\right\rangle \bigotimes_{k=2}^{2n}  \left|\bar{i}_k\right\rangle \Big)_{12} \bigotimes
\Big( \left|0\right\rangle \bigotimes_{k=2}^{2n} \left|i_k\right\rangle \pm
\left|1\right\rangle \bigotimes_{k=2}^{2n}  \left|\bar{i}_k\right\rangle \Big)_{34}	\notag\\
= &\quad \Big( \left|0\right\rangle \bigotimes_{k=2}^{2n} \left|i_k\right\rangle \Big)_{12} \bigotimes
	\Big( \left|0\right\rangle \bigotimes_{k=2}^{2n} \left|i_k\right\rangle \Big)_{34}	\pm
	\Big( \left|0\right\rangle \bigotimes_{k=2}^{2n} \left|i_k\right\rangle \Big)_{12} \bigotimes
	\Big( \left|1\right\rangle \bigotimes_{k=2}^{2n}  \left|\bar{i}_k\right\rangle \Big)_{34}
	\notag\\
&\pm 
	\Big( \left|1\right\rangle \bigotimes_{k=2}^{2n}  \left|\bar{i}_k\right\rangle \Big)_{12}	\bigotimes
	\Big( \left|0\right\rangle \bigotimes_{k=2}^{2n} \left|i_k\right\rangle \Big)_{34} +
	 \Big( \left|1\right\rangle \bigotimes_{k=2}^{2n}  \left|\bar{i}_k\right\rangle \Big)_{12}	\bigotimes
	 \Big( \left|1\right\rangle \bigotimes_{k=2}^{2n}  \left|\bar{i}_k\right\rangle \Big)_{34}
	 \notag\\
= &\quad
	\Big( \left|0\right\rangle \bigotimes_{k=2}^{n} \left|i_k\right\rangle
	\left|0\right\rangle \bigotimes_{k=2}^{n} \left|i_k\right\rangle	\Big)_{13}	\bigotimes
	\Big( \bigotimes_{k=n+1}^{2n} \left|i_k\right\rangle
	\bigotimes_{k=n+1}^{2n} \left|i_k\right\rangle \Big)_{24}
	\pm	
	\Big( \left|0\right\rangle \bigotimes_{k=2}^{n} \left|i_k\right\rangle
	\left|1\right\rangle \bigotimes_{k=2}^{n} \left|\bar{i}_k\right\rangle	\Big)_{13}	\bigotimes
	\Big( \bigotimes_{k=n+1}^{2n} \left|i_k\right\rangle
	\bigotimes_{k=n+1}^{2n} \left|\bar{i}_k\right\rangle \Big)_{24}
	\notag\\
&\pm 
	\Big( \left|1\right\rangle \bigotimes_{k=2}^{n} \left|\bar{i}_k\right\rangle
	\left|0\right\rangle \bigotimes_{k=2}^{n} \left|i_k\right\rangle	\Big)_{13}	\bigotimes
	\Big( \bigotimes_{k=n+1}^{2n} \left|\bar{i}_k\right\rangle
	\bigotimes_{k=n+1}^{2n} \left|i_k\right\rangle \Big)_{24}
	+
	\Big( \left|1\right\rangle \bigotimes_{k=2}^{n} \left|\bar{i}_k\right\rangle
	\left|1\right\rangle \bigotimes_{k=2}^{n} \left|\bar{i}_k\right\rangle	\Big)_{13}	\bigotimes
	\Big( \bigotimes_{k=n+1}^{2n} \left|\bar{i}_k\right\rangle
	\bigotimes_{k=n+1}^{2n} \left|\bar{i}_k\right\rangle \Big)_{24}
	\notag\\
= &	
\begin{cases}
		\phantom{\pm}
	\Big( \left|\mathcal{G}_e^+\right\rangle + \left|\mathcal{G}_e^-\right\rangle \Big)_{13}
	\Big( \left|\mathcal{G}_e^+\right\rangle + \left|\mathcal{G}_e^-\right\rangle \Big)_{24}	\pm
	\Big( \left|\mathcal{G}_g^+\right\rangle + \left|\mathcal{G}_g^-\right\rangle \Big)_{13}
	\Big( \left|\mathcal{G}_g^+\right\rangle + \left|\mathcal{G}_g^-\right\rangle \Big)_{24}	\\
	\pm	\Big( \left|\mathcal{G}_g^+\right\rangle - \left|\mathcal{G}_g^-\right\rangle \Big) _{13}
	\Big( \left|\mathcal{G}_g^+\right\rangle - \left|\mathcal{G}_g^-\right\rangle \Big)_{24}	+
	\Big( \left|\mathcal{G}_e^+\right\rangle - \left|\mathcal{G}_e^-\right\rangle \Big) _{13}
	\Big( \left|\mathcal{G}_e^+\right\rangle - \left|\mathcal{G}_e^-\right\rangle \Big)_{24}
	& \textrm{if} \phantom{i} 0i_2i_3 \cdots i_n = i_{n+1}i_{n+2}\cdots i_{2n};	\\
		\\
		\phantom{\pm}
	\Big( \left|\mathcal{G}_e^+\right\rangle + \left|\mathcal{G}_e^-\right\rangle \Big)_{13}
	\Big( \left|\mathcal{G}_e^+\right\rangle - \left|\mathcal{G}_e^-\right\rangle \Big)_{24}	\pm
	\Big( \left|\mathcal{G}_g^+\right\rangle + \left|\mathcal{G}_g^-\right\rangle \Big)_{13}
	\Big( \left|\mathcal{G}_g^+\right\rangle - \left|\mathcal{G}_g^-\right\rangle \Big)_{24}	\\
	\pm	\Big( \left|\mathcal{G}_g^+\right\rangle - \left|\mathcal{G}_g^-\right\rangle \Big) _{13}
	\Big( \left|\mathcal{G}_g^+\right\rangle + \left|\mathcal{G}_g^-\right\rangle \Big)_{24}	+
	\Big( \left|\mathcal{G}_e^+\right\rangle - \left|\mathcal{G}_e^-\right\rangle \Big) _{13}
	\Big( \left|\mathcal{G}_e^+\right\rangle + \left|\mathcal{G}_e^-\right\rangle \Big)_{24}
	& \textrm{if} \phantom{i} 0i_2i_3 \cdots i_n = \bar{i}_{n+1}\bar{i}_{n+2}\cdots\bar{i}_{2n};	\\
\end{cases}\notag\\
= &	
\begin{cases}
		\phantom{\pm}
		\left|\mathcal{G}_e^+\right\rangle_{13} \left|\mathcal{G}_e^+\right\rangle_{24} +
		\left|\mathcal{G}_e^-\right\rangle_{13} \left|\mathcal{G}_e^-\right\rangle_{24} \pm
		\left|\mathcal{G}_g^+\right\rangle_{13} \left|\mathcal{G}_g^+\right\rangle_{24} \mp
		\left|\mathcal{G}_g^-\right\rangle_{13} \left|\mathcal{G}_g^-\right\rangle_{24};\\
		\\
		\phantom{\pm}
		\left|\mathcal{G}_e^+\right\rangle_{13} \left|\mathcal{G}_e^+\right\rangle_{24} -
		\left|\mathcal{G}_e^-\right\rangle_{13} \left|\mathcal{G}_e^-\right\rangle_{24} \pm
		\left|\mathcal{G}_g^+\right\rangle_{13} \left|\mathcal{G}_g^+\right\rangle_{24} \mp
		\left|\mathcal{G}_g^-\right\rangle_{13} \left|\mathcal{G}_g^-\right\rangle_{24},
\end{cases}
\end{align}
where
\begin{align}
\left|\mathcal{G}_{g}^{\pm}\right\rangle = \frac 1{\sqrt{2}} 
\Big(
\left|0\right\rangle \bigotimes_{k=2}^{n} \left|i_k\right\rangle 
\bigotimes_{k=n+1}^{2n} \left|\bar{i}_k\right\rangle
\pm 
\left|1\right\rangle \bigotimes_{k=2}^{n} \left|\bar{i}_k\right\rangle 
\bigotimes_{k=n+1}^{2n} \left|i_k\right\rangle	
\Big).
\end{align}
Let us then derive formulas for the entanglement swapping between two different SGHZ states,
that is, \{$\left|\mathcal{G}_{e}^{\pm}\right\rangle, 
\left|\mathcal{G}_{e}^{\mp}\right\rangle$\},
\{$\left|\mathcal{G}_{e}^{\pm}\right\rangle,
\left|\mathcal{G}_{e^{\prime}}^{\pm}\right\rangle$\},
and \{$\left|\mathcal{G}_{e}^{\pm}\right\rangle,
\left|\mathcal{G}_{e^{\prime}}^{\mp}\right\rangle$\},
where $e \ne e^{\prime}$ and
\begin{align}
\left|\mathcal{G}_{e^{\prime}}^{\pm}\right\rangle = \frac 1{\sqrt{2}} 
\Big(
\left|0\right\rangle \bigotimes_{k=2}^{2n} \left|j_k\right\rangle 
\pm 
\left|1\right\rangle \bigotimes_{k=2}^{2n} \left|\bar{j}_k\right\rangle
\Big).
\end{align}
We can get
\begin{align}
\label{2nGHZ2}
&\left|\mathcal{G}_{e}^{\pm}\right\rangle_{12} \otimes
	\left|\mathcal{G}_{e}^{\mp}\right\rangle_{34}	\notag\\
= &\quad	\Big( \left|0\right\rangle \bigotimes_{k=2}^{2n} \left|i_k\right\rangle \pm
\left|1\right\rangle \bigotimes_{k=2}^{2n}  \left|\bar{i}_k\right\rangle \Big)_{12} \bigotimes
\Big( \left|0\right\rangle \bigotimes_{k=2}^{2n} \left|i_k\right\rangle \mp
\left|1\right\rangle \bigotimes_{k=2}^{2n}  \left|\bar{i}_k\right\rangle \Big)_{34}	\notag\\
= &\quad \Big( \left|0\right\rangle \bigotimes_{k=2}^{2n} \left|i_k\right\rangle \Big)_{12} \bigotimes
	\Big( \left|0\right\rangle \bigotimes_{k=2}^{2n} \left|i_k\right\rangle \Big)_{34}	\mp
	\Big( \left|0\right\rangle \bigotimes_{k=2}^{2n} \left|i_k\right\rangle \Big)_{12} \bigotimes
	\Big( \left|1\right\rangle \bigotimes_{k=2}^{2n}  \left|\bar{i}_k\right\rangle \Big)_{34}
	\notag\\
&\pm 
	\Big( \left|1\right\rangle \bigotimes_{k=2}^{2n}  \left|\bar{i}_k\right\rangle \Big)_{12}	\bigotimes
	\Big( \left|0\right\rangle \bigotimes_{k=2}^{2n} \left|i_k\right\rangle \Big)_{34} -
	 \Big( \left|1\right\rangle \bigotimes_{k=2}^{2n}  \left|\bar{i}_k\right\rangle \Big)_{12}	\bigotimes
	 \Big( \left|1\right\rangle \bigotimes_{k=2}^{2n}  \left|\bar{i}_k\right\rangle \Big)_{34}
	 \notag\\
= &\quad
	\Big( \left|0\right\rangle \bigotimes_{k=2}^{n} \left|i_k\right\rangle
	\left|0\right\rangle \bigotimes_{k=2}^{n} \left|i_k\right\rangle	\Big)_{13}	\bigotimes
	\Big( \bigotimes_{k=n+1}^{2n} \left|i_k\right\rangle
	\bigotimes_{k=n+1}^{2n} \left|i_k\right\rangle \Big)_{24}
	\mp	
	\Big( \left|0\right\rangle \bigotimes_{k=2}^{n} \left|i_k\right\rangle
	\left|1\right\rangle \bigotimes_{k=2}^{n} \left|\bar{i}_k\right\rangle	\Big)_{13}	\bigotimes
	\Big( \bigotimes_{k=n+1}^{2n} \left|i_k\right\rangle
	\bigotimes_{k=n+1}^{2n} \left|\bar{i}_k\right\rangle \Big)_{24}
	\notag\\
&\pm 
	\Big( \left|1\right\rangle \bigotimes_{k=2}^{n} \left|\bar{i}_k\right\rangle
	\left|0\right\rangle \bigotimes_{k=2}^{n} \left|i_k\right\rangle	\Big)_{13}	\bigotimes
	\Big( \bigotimes_{k=n+1}^{2n} \left|\bar{i}_k\right\rangle
	\bigotimes_{k=n+1}^{2n} \left|i_k\right\rangle \Big)_{24}
	-
	\Big( \left|1\right\rangle \bigotimes_{k=2}^{n} \left|\bar{i}_k\right\rangle
	\left|1\right\rangle \bigotimes_{k=2}^{n} \left|\bar{i}_k\right\rangle	\Big)_{13}	\bigotimes
	\Big( \bigotimes_{k=n+1}^{2n} \left|\bar{i}_k\right\rangle
	\bigotimes_{k=n+1}^{2n} \left|\bar{i}_k\right\rangle \Big)_{24}
	\notag\\
= &	
\begin{cases}
		\phantom{\pm}
		\left|\mathcal{G}_e^+\right\rangle_{13} \left|\mathcal{G}_e^-\right\rangle_{24} +
		\left|\mathcal{G}_e^-\right\rangle_{13} \left|\mathcal{G}_e^+\right\rangle_{24} \mp
		\left|\mathcal{G}_g^+\right\rangle_{13} \left|\mathcal{G}_g^-\right\rangle_{24} \mp
		\left|\mathcal{G}_g^-\right\rangle_{13} \left|\mathcal{G}_g^+\right\rangle_{24}
	&\textrm{if} \phantom{i} 0i_2i_3 \cdots i_n = i_{n+1}i_{n+2}\cdots i_{2n};
		\\
- 		\left|\mathcal{G}_e^+\right\rangle_{13} \left|\mathcal{G}_e^-\right\rangle_{24} +
		\left|\mathcal{G}_e^-\right\rangle_{13} \left|\mathcal{G}_e^+\right\rangle_{24} \pm
		\left|\mathcal{G}_g^+\right\rangle_{13} \left|\mathcal{G}_g^-\right\rangle_{24} \mp
		\left|\mathcal{G}_g^-\right\rangle_{13} \left|\mathcal{G}_g^+\right\rangle_{24}
	&\textrm{if} \phantom{i} 0i_2i_3 \cdots i_n = \bar{i}_{n+1}\bar{i}_{n+2}\cdots\bar{i}_{2n},
\end{cases}
\end{align}
\begin{align}
\label{2nGHZ3}
&\left|\mathcal{G}_{e}^{\pm}\right\rangle_{12} \otimes 
	\left|\mathcal{G}_{e^{\prime}}^{\pm}\right\rangle_{34}	\notag\\
= &\quad	\Big( \left|0\right\rangle \bigotimes_{k=2}^{2n} \left|i_k\right\rangle \pm
\left|1\right\rangle \bigotimes_{k=2}^{2n}  \left|\bar{i}_k\right\rangle \Big)_{12} \bigotimes
\Big( \left|0\right\rangle \bigotimes_{k=2}^{2n} \left|j_k\right\rangle \mp
\left|1\right\rangle \bigotimes_{k=2}^{2n}  \left|\bar{j}_k\right\rangle \Big)_{34}	\notag\\
= &\quad \Big( \left|0\right\rangle \bigotimes_{k=2}^{2n} \left|i_k\right\rangle \Big)_{12} \bigotimes
	\Big( \left|0\right\rangle \bigotimes_{k=2}^{2n} \left|j_k\right\rangle \Big)_{34}	
	\pm
	\Big( \left|0\right\rangle \bigotimes_{k=2}^{2n} \left|i_k\right\rangle \Big)_{12} \bigotimes
	\Big( \left|1\right\rangle \bigotimes_{k=2}^{2n}  \left|\bar{j}_k\right\rangle \Big)_{34}
	\notag\\
&	\pm 
	\Big( \left|1\right\rangle \bigotimes_{k=2}^{2n}  \left|\bar{i}_k\right\rangle \Big)_{12}	\bigotimes
	\Big( \left|0\right\rangle \bigotimes_{k=2}^{2n} \left|j_k\right\rangle \Big)_{34} 
	+
	 \Big( \left|1\right\rangle \bigotimes_{k=2}^{2n}  \left|\bar{i}_k\right\rangle \Big)_{12}	\bigotimes
	 \Big( \left|1\right\rangle \bigotimes_{k=2}^{2n}  \left|\bar{j}_k\right\rangle \Big)_{34}
	 \notag\\
= &\quad
	\Big( \left|0\right\rangle \bigotimes_{k=2}^{n} \left|i_k\right\rangle
	\left|0\right\rangle \bigotimes_{k=2}^{n} \left|j_k\right\rangle	\Big)_{13}	\bigotimes
	\Big( \bigotimes_{k=n+1}^{2n} \left|i_k\right\rangle
	\bigotimes_{k=n+1}^{2n} \left|j_k\right\rangle \Big)_{24}
	\pm	
	\Big( \left|0\right\rangle \bigotimes_{k=2}^{n} \left|i_k\right\rangle
	\left|1\right\rangle \bigotimes_{k=2}^{n} \left|\bar{j}_k\right\rangle	\Big)_{13}	\bigotimes
	\Big( \bigotimes_{k=n+1}^{2n} \left|i_k\right\rangle
	\bigotimes_{k=n+1}^{2n} \left|\bar{j}_k\right\rangle \Big)_{24}
	\notag\\
&	\pm 
	\Big( \left|1\right\rangle \bigotimes_{k=2}^{n} \left|\bar{i}_k\right\rangle
	\left|0\right\rangle \bigotimes_{k=2}^{n} \left|j_k\right\rangle	\Big)_{13}	\bigotimes
	\Big( \bigotimes_{k=n+1}^{2n} \left|\bar{i}_k\right\rangle
	\bigotimes_{k=n+1}^{2n} \left|j_k\right\rangle \Big)_{24}
	+
	\Big( \left|1\right\rangle \bigotimes_{k=2}^{n} \left|\bar{i}_k\right\rangle
	\left|1\right\rangle \bigotimes_{k=2}^{n} \left|\bar{j}_k\right\rangle	\Big)_{13}	\bigotimes
	\Big( \bigotimes_{k=n+1}^{2n} \left|\bar{i}_k\right\rangle
	\bigotimes_{k=n+1}^{2n} \left|\bar{j}_k\right\rangle \Big)_{24}
	\notag\\
= &	
\begin{cases}
	\phantom{\pm}
	\Big( \left|\mathcal{G}_p^+\right\rangle + \left|\mathcal{G}_p^-\right\rangle \Big)_{13}
	\Big( \left|\mathcal{G}_p^+\right\rangle + \left|\mathcal{G}_p^-\right\rangle \Big)_{24}	
	\pm
	\Big( \left|\mathcal{G}_q^+\right\rangle + \left|\mathcal{G}_q^-\right\rangle \Big)_{13}
	\Big( \left|\mathcal{G}_q^+\right\rangle + \left|\mathcal{G}_q^-\right\rangle \Big)_{24}	\\
	\pm	
	\Big( \left|\mathcal{G}_q^+\right\rangle - \left|\mathcal{G}_q^-\right\rangle \Big)_{13}
	\Big( \left|\mathcal{G}_q^+\right\rangle - \left|\mathcal{G}_q^-\right\rangle \Big)_{24}	
	+
	\Big( \left|\mathcal{G}_p^+\right\rangle - \left|\mathcal{G}_p^-\right\rangle \Big)_{13}
	\Big( \left|\mathcal{G}_p^+\right\rangle - \left|\mathcal{G}_p^-\right\rangle \Big)_{24}
	& \textrm{if} \phantom{i} 0i_2i_3 \cdots i_n = i_{n+1}i_{n+2}\cdots i_{2n}	\\
	&	\textrm{and} \phantom{i} 0j_2j_3 \cdots j_n = j_{n+1}j_{n+2}\cdots j_{2n};	\\
	\phantom{\pm}
	\Big( \left|\mathcal{G}_p^+\right\rangle + \left|\mathcal{G}_p^-\right\rangle \Big)_{13}
	\Big( \left|\mathcal{G}_q^+\right\rangle + \left|\mathcal{G}_q^-\right\rangle \Big)_{24}	
	\pm
	\Big( \left|\mathcal{G}_q^+\right\rangle + \left|\mathcal{G}_q^-\right\rangle \Big)_{13}
	\Big( \left|\mathcal{G}_p^+\right\rangle + \left|\mathcal{G}_p^-\right\rangle \Big)_{24}	\\
	\pm	
	\Big( \left|\mathcal{G}_q^+\right\rangle - \left|\mathcal{G}_q^-\right\rangle \Big)_{13}
	\Big( \left|\mathcal{G}_p^+\right\rangle - \left|\mathcal{G}_p^-\right\rangle \Big)_{24}	
	+
	\Big( \left|\mathcal{G}_p^+\right\rangle - \left|\mathcal{G}_p^-\right\rangle \Big)_{13}
	\Big( \left|\mathcal{G}_p^+\right\rangle - \left|\mathcal{G}_p^-\right\rangle \Big)_{24}
	& \textrm{if} \phantom{i} 0i_2i_3 \cdots i_n = i_{n+1}i_{n+2}\cdots i_{2n}	\\
	&	\textrm{and} \phantom{i} 0j_2j_3 \cdots j_n = \bar{j}_{n+1}\bar{j}_{n+2}\cdots \bar{j}_{2n};	\\
	\phantom{\pm}
	\Big( \left|\mathcal{G}_p^+\right\rangle + \left|\mathcal{G}_p^-\right\rangle \Big)_{13}
	\Big( \left|\mathcal{G}_q^+\right\rangle - \left|\mathcal{G}_q^-\right\rangle \Big)_{24}	
	\pm
	\Big( \left|\mathcal{G}_q^+\right\rangle + \left|\mathcal{G}_q^-\right\rangle \Big)_{13}
	\Big( \left|\mathcal{G}_p^+\right\rangle - \left|\mathcal{G}_p^-\right\rangle \Big)_{24}	\\
	\pm	
	\Big( \left|\mathcal{G}_q^+\right\rangle - \left|\mathcal{G}_q^-\right\rangle \Big) _{13}
	\Big( \left|\mathcal{G}_p^+\right\rangle + \left|\mathcal{G}_p^-\right\rangle \Big)_{24}	
	+
	\Big( \left|\mathcal{G}_p^+\right\rangle - \left|\mathcal{G}_p^-\right\rangle \Big) _{13}
	\Big( \left|\mathcal{G}_q^+\right\rangle + \left|\mathcal{G}_q^-\right\rangle \Big)_{24}
	& \textrm{if} \phantom{i} 0i_2i_3 \cdots i_n = \bar{i}_{n+1}\bar{i}_{n+2}\cdots \bar{i}_{2n}	\\
	&	\textrm{and} \phantom{i} 0j_2j_3 \cdots j_n = j_{n+1}j_{n+2}\cdots j_{2n};	\\
	\phantom{\pm}
	\Big( \left|\mathcal{G}_p^+\right\rangle + \left|\mathcal{G}_p^-\right\rangle \Big)_{13}
	\Big( \left|\mathcal{G}_p^+\right\rangle - \left|\mathcal{G}_p^-\right\rangle \Big)_{24}	
	\pm
	\Big( \left|\mathcal{G}_q^+\right\rangle + \left|\mathcal{G}_q^-\right\rangle \Big)_{13}
	\Big( \left|\mathcal{G}_q^+\right\rangle - \left|\mathcal{G}_q^-\right\rangle \Big)_{24}	\\
	\pm	
	\Big( \left|\mathcal{G}_q^+\right\rangle - \left|\mathcal{G}_q^-\right\rangle \Big) _{13}
	\Big( \left|\mathcal{G}_q^+\right\rangle + \left|\mathcal{G}_q^-\right\rangle \Big)_{24}	
	+
	\Big( \left|\mathcal{G}_p^+\right\rangle - \left|\mathcal{G}_p^-\right\rangle \Big) _{13}
	\Big( \left|\mathcal{G}_p^+\right\rangle + \left|\mathcal{G}_p^-\right\rangle \Big)_{24}
	& \textrm{if} \phantom{i} 0i_2i_3 \cdots i_n =\bar{i}_{n+1}\bar{i}_{n+2}\cdots \bar{i}_{2n}	\\
	&	\textrm{and} \phantom{i} 0j_2j_3 \cdots j_n = \bar{j}_{n+1}\bar{j}_{n+2}\cdots \bar{j}_{2n}
\end{cases}\notag\\
= &	
\begin{cases}
		\phantom{\pm}
		\left|\mathcal{G}_p^+\right\rangle_{13} \left|\mathcal{G}_p^+\right\rangle_{24} +
		\left|\mathcal{G}_p^-\right\rangle_{13} \left|\mathcal{G}_p^-\right\rangle_{24} \pm
		\left|\mathcal{G}_q^+\right\rangle_{13} \left|\mathcal{G}_q^+\right\rangle_{24} \pm
		\left|\mathcal{G}_q^-\right\rangle_{13} \left|\mathcal{G}_q^-\right\rangle_{24};
		\\
		\\
		\phantom{\pm}
		\left|\mathcal{G}_p^+\right\rangle_{13} \left|\mathcal{G}_q^+\right\rangle_{24} +
		\left|\mathcal{G}_p^-\right\rangle_{13} \left|\mathcal{G}_q^-\right\rangle_{24} \pm
		\left|\mathcal{G}_q^+\right\rangle_{13} \left|\mathcal{G}_p^+\right\rangle_{24} \pm
		\left|\mathcal{G}_q^-\right\rangle_{13} \left|\mathcal{G}_p^-\right\rangle_{24};
		\\
		\\
		\phantom{\pm}
		\left|\mathcal{G}_p^+\right\rangle_{13} \left|\mathcal{G}_q^+\right\rangle_{24} -
		\left|\mathcal{G}_p^-\right\rangle_{13} \left|\mathcal{G}_q^-\right\rangle_{24} \pm
		\left|\mathcal{G}_q^+\right\rangle_{13} \left|\mathcal{G}_p^+\right\rangle_{24} \mp
		\left|\mathcal{G}_q^-\right\rangle_{13} \left|\mathcal{G}_p^-\right\rangle_{24};
		\\
		\\
		\phantom{\pm}
		\left|\mathcal{G}_p^+\right\rangle_{13} \left|\mathcal{G}_p^+\right\rangle_{24} -
		\left|\mathcal{G}_p^-\right\rangle_{13} \left|\mathcal{G}_p^-\right\rangle_{24} \pm
		\left|\mathcal{G}_q^+\right\rangle_{13} \left|\mathcal{G}_q^+\right\rangle_{24} \mp
		\left|\mathcal{G}_q^-\right\rangle_{13} \left|\mathcal{G}_q^-\right\rangle_{24},
\end{cases}
\end{align}
and
\begin{align}
\label{2nGHZ4}
&\left|\mathcal{G}_{e}^{\pm}\right\rangle_{12} \otimes 
	\left|\mathcal{G}_{e^{\prime}}^{\mp}\right\rangle_{34}	\notag\\
= &\quad	\Big( \left|0\right\rangle \bigotimes_{k=2}^{2n} \left|i_k\right\rangle \pm
\left|1\right\rangle \bigotimes_{k=2}^{2n}  \left|\bar{i}_k\right\rangle \Big)_{12} \bigotimes
\Big( \left|0\right\rangle \bigotimes_{k=2}^{2n} \left|j_k\right\rangle \mp
\left|1\right\rangle \bigotimes_{k=2}^{2n}  \left|\bar{j}_k\right\rangle \Big)_{34}	\notag\\
= &\quad \Big( \left|0\right\rangle \bigotimes_{k=2}^{2n} \left|i_k\right\rangle \Big)_{12} \bigotimes
	\Big( \left|0\right\rangle \bigotimes_{k=2}^{2n} \left|j_k\right\rangle \Big)_{34}	
	\mp
	\Big( \left|0\right\rangle \bigotimes_{k=2}^{2n} \left|i_k\right\rangle \Big)_{12} \bigotimes
	\Big( \left|1\right\rangle \bigotimes_{k=2}^{2n}  \left|\bar{j}_k\right\rangle \Big)_{34}
	\notag\\
&	\pm 
	\Big( \left|1\right\rangle \bigotimes_{k=2}^{2n}  \left|\bar{i}_k\right\rangle \Big)_{12}	\bigotimes
	\Big( \left|0\right\rangle \bigotimes_{k=2}^{2n} \left|j_k\right\rangle \Big)_{34} 
	-
	 \Big( \left|1\right\rangle \bigotimes_{k=2}^{2n}  \left|\bar{i}_k\right\rangle \Big)_{12}	\bigotimes
	 \Big( \left|1\right\rangle \bigotimes_{k=2}^{2n}  \left|\bar{j}_k\right\rangle \Big)_{34}
	 \notag\\
= &\quad
	\Big( \left|0\right\rangle \bigotimes_{k=2}^{n} \left|i_k\right\rangle
	\left|0\right\rangle \bigotimes_{k=2}^{n} \left|j_k\right\rangle	\Big)_{13}	\bigotimes
	\Big( \bigotimes_{k=n+1}^{2n} \left|i_k\right\rangle
	\bigotimes_{k=n+1}^{2n} \left|j_k\right\rangle \Big)_{24}
	\mp	
	\Big( \left|0\right\rangle \bigotimes_{k=2}^{n} \left|i_k\right\rangle
	\left|1\right\rangle \bigotimes_{k=2}^{n} \left|\bar{j}_k\right\rangle	\Big)_{13}	\bigotimes
	\Big( \bigotimes_{k=n+1}^{2n} \left|i_k\right\rangle
	\bigotimes_{k=n+1}^{2n} \left|\bar{j}_k\right\rangle \Big)_{24}
	\notag\\
&	\pm 
	\Big( \left|1\right\rangle \bigotimes_{k=2}^{n} \left|\bar{i}_k\right\rangle
	\left|0\right\rangle \bigotimes_{k=2}^{n} \left|j_k\right\rangle	\Big)_{13}	\bigotimes
	\Big( \bigotimes_{k=n+1}^{2n} \left|\bar{i}_k\right\rangle
	\bigotimes_{k=n+1}^{2n} \left|j_k\right\rangle \Big)_{24}
	-
	\Big( \left|1\right\rangle \bigotimes_{k=2}^{n} \left|\bar{i}_k\right\rangle
	\left|1\right\rangle \bigotimes_{k=2}^{n} \left|\bar{j}_k\right\rangle	\Big)_{13}	\bigotimes
	\Big( \bigotimes_{k=n+1}^{2n} \left|\bar{i}_k\right\rangle
	\bigotimes_{k=n+1}^{2n} \left|\bar{j}_k\right\rangle \Big)_{24}
	\notag\\
= &	
\begin{cases}
	\phantom{\pm}
	\Big( \left|\mathcal{G}_p^+\right\rangle + \left|\mathcal{G}_p^-\right\rangle \Big)_{13}
	\Big( \left|\mathcal{G}_p^+\right\rangle + \left|\mathcal{G}_p^-\right\rangle \Big)_{24}	
	\mp
	\Big( \left|\mathcal{G}_q^+\right\rangle + \left|\mathcal{G}_q^-\right\rangle \Big)_{13}
	\Big( \left|\mathcal{G}_q^+\right\rangle + \left|\mathcal{G}_q^-\right\rangle \Big)_{24}	\\
	\pm	
	\Big( \left|\mathcal{G}_q^+\right\rangle - \left|\mathcal{G}_q^-\right\rangle \Big)_{13}
	\Big( \left|\mathcal{G}_q^+\right\rangle - \left|\mathcal{G}_q^-\right\rangle \Big)_{24}	
	-
	\Big( \left|\mathcal{G}_p^+\right\rangle - \left|\mathcal{G}_p^-\right\rangle \Big)_{13}
	\Big( \left|\mathcal{G}_p^+\right\rangle - \left|\mathcal{G}_p^-\right\rangle \Big)_{24}
	& \textrm{if} \phantom{i} 0i_2i_3 \cdots i_n = i_{n+1}i_{n+2}\cdots i_{2n}	\\
	&	\textrm{and} \phantom{i} 0j_2j_3 \cdots j_n = j_{n+1}j_{n+2}\cdots j_{2n};	\\
	\phantom{\pm}
	\Big( \left|\mathcal{G}_p^+\right\rangle + \left|\mathcal{G}_p^-\right\rangle \Big)_{13}
	\Big( \left|\mathcal{G}_q^+\right\rangle + \left|\mathcal{G}_q^-\right\rangle \Big)_{24}	
	\mp
	\Big( \left|\mathcal{G}_q^+\right\rangle + \left|\mathcal{G}_q^-\right\rangle \Big)_{13}
	\Big( \left|\mathcal{G}_p^+\right\rangle + \left|\mathcal{G}_p^-\right\rangle \Big)_{24}	\\
	\pm	
	\Big( \left|\mathcal{G}_q^+\right\rangle - \left|\mathcal{G}_q^-\right\rangle \Big)_{13}
	\Big( \left|\mathcal{G}_p^+\right\rangle - \left|\mathcal{G}_p^-\right\rangle \Big)_{24}	
	-
	\Big( \left|\mathcal{G}_p^+\right\rangle - \left|\mathcal{G}_p^-\right\rangle \Big)_{13}
	\Big( \left|\mathcal{G}_p^+\right\rangle - \left|\mathcal{G}_p^-\right\rangle \Big)_{24}
	& \textrm{if} \phantom{i} 0i_2i_3 \cdots i_n = i_{n+1}i_{n+2}\cdots i_{2n}	\\
	&	\textrm{and} \phantom{i} 0j_2j_3 \cdots j_n = \bar{j}_{n+1}\bar{j}_{n+2}\cdots \bar{j}_{2n};	\\
	\phantom{\pm}
	\Big( \left|\mathcal{G}_p^+\right\rangle + \left|\mathcal{G}_p^-\right\rangle \Big)_{13}
	\Big( \left|\mathcal{G}_q^+\right\rangle - \left|\mathcal{G}_q^-\right\rangle \Big)_{24}	
	\mp
	\Big( \left|\mathcal{G}_q^+\right\rangle + \left|\mathcal{G}_q^-\right\rangle \Big)_{13}
	\Big( \left|\mathcal{G}_p^+\right\rangle - \left|\mathcal{G}_p^-\right\rangle \Big)_{24}	\\
	\pm	
	\Big( \left|\mathcal{G}_q^+\right\rangle - \left|\mathcal{G}_q^-\right\rangle \Big) _{13}
	\Big( \left|\mathcal{G}_p^+\right\rangle + \left|\mathcal{G}_p^-\right\rangle \Big)_{24}	
	-
	\Big( \left|\mathcal{G}_p^+\right\rangle - \left|\mathcal{G}_p^-\right\rangle \Big) _{13}
	\Big( \left|\mathcal{G}_q^+\right\rangle + \left|\mathcal{G}_q^-\right\rangle \Big)_{24}
	& \textrm{if} \phantom{i} 0i_2i_3 \cdots i_n = \bar{i}_{n+1}\bar{i}_{n+2}\cdots \bar{i}_{2n}	\\
	&	\textrm{and} \phantom{i} 0j_2j_3 \cdots j_n = j_{n+1}j_{n+2}\cdots j_{2n};	\\
	\phantom{\pm}
	\Big( \left|\mathcal{G}_p^+\right\rangle + \left|\mathcal{G}_p^-\right\rangle \Big)_{13}
	\Big( \left|\mathcal{G}_p^+\right\rangle - \left|\mathcal{G}_p^-\right\rangle \Big)_{24}
	\mp
	\Big( \left|\mathcal{G}_q^+\right\rangle + \left|\mathcal{G}_q^-\right\rangle \Big)_{13}
	\Big( \left|\mathcal{G}_q^+\right\rangle - \left|\mathcal{G}_q^-\right\rangle \Big)_{24}	\\
	\pm	
	\Big( \left|\mathcal{G}_q^+\right\rangle - \left|\mathcal{G}_q^-\right\rangle \Big) _{13}
	\Big( \left|\mathcal{G}_q^+\right\rangle + \left|\mathcal{G}_q^-\right\rangle \Big)_{24}	
	-
	\Big( \left|\mathcal{G}_p^+\right\rangle - \left|\mathcal{G}_p^-\right\rangle \Big) _{13}
	\Big( \left|\mathcal{G}_p^+\right\rangle + \left|\mathcal{G}_p^-\right\rangle \Big)_{24}
	& \textrm{if} \phantom{i} 0i_2i_3 \cdots i_n =\bar{i}_{n+1}\bar{i}_{n+2}\cdots \bar{i}_{2n}	\\
	&	\textrm{and} \phantom{i} 0j_2j_3 \cdots j_n = \bar{j}_{n+1}\bar{j}_{n+2}\cdots \bar{j}_{2n}
\end{cases}\notag\\
= &	
\begin{cases}
		\phantom{\pm}
		\left|\mathcal{G}_p^+\right\rangle_{13} \left|\mathcal{G}_p^-\right\rangle_{24} +
		\left|\mathcal{G}_p^-\right\rangle_{13} \left|\mathcal{G}_p^+\right\rangle_{24} \mp
		\left|\mathcal{G}_q^+\right\rangle_{13} \left|\mathcal{G}_q^-\right\rangle_{24} \mp
		\left|\mathcal{G}_q^-\right\rangle_{13} \left|\mathcal{G}_q^+\right\rangle_{24};
		\\
		\\
		\phantom{\pm}
 		\left|\mathcal{G}_p^+\right\rangle_{13} \left|\mathcal{G}_q^-\right\rangle_{24} +
		\left|\mathcal{G}_p^-\right\rangle_{13} \left|\mathcal{G}_q^+\right\rangle_{24} \mp
		\left|\mathcal{G}_q^+\right\rangle_{13} \left|\mathcal{G}_p^-\right\rangle_{24} \mp
		\left|\mathcal{G}_q^-\right\rangle_{13} \left|\mathcal{G}_p^+\right\rangle_{24};
		\\
		\\
-		\left|\mathcal{G}_p^+\right\rangle_{13} \left|\mathcal{G}_q^-\right\rangle_{24} +
		\left|\mathcal{G}_p^-\right\rangle_{13} \left|\mathcal{G}_q^+\right\rangle_{24} \pm
		\left|\mathcal{G}_q^+\right\rangle_{13} \left|\mathcal{G}_p^-\right\rangle_{24} \mp
		\left|\mathcal{G}_q^-\right\rangle_{13} \left|\mathcal{G}_p^+\right\rangle_{24};
		\\
		\\
-		\left|\mathcal{G}_p^+\right\rangle_{13} \left|\mathcal{G}_p^-\right\rangle_{24} +
		\left|\mathcal{G}_p^-\right\rangle_{13} \left|\mathcal{G}_p^+\right\rangle_{24} \pm
		\left|\mathcal{G}_q^+\right\rangle_{13} \left|\mathcal{G}_q^-\right\rangle_{24} \mp
		\left|\mathcal{G}_q^-\right\rangle_{13} \left|\mathcal{G}_q^+\right\rangle_{24},
\end{cases}
\end{align}
where
\begin{align}
\begin{cases}
\phantom{i} \left|\mathcal{G}_{p}^{\pm}\right\rangle =  \frac 1{\sqrt{2}}
\Big( 
\left|0\right\rangle \bigotimes_{k=2}^{n} \left|i_k\right\rangle \left|0\right\rangle \bigotimes_{k=2}^{n} \left|j_k\right\rangle \pm
\left|1\right\rangle \bigotimes_{k=2}^{n}  \left|\bar{i}_k\right\rangle \left|1\right\rangle \bigotimes_{k=2}^{n}  \left|\bar{j}_k\right\rangle 
\Big),
\\
\phantom{i} \left|\mathcal{G}_{q}^{\pm}\right\rangle =  \frac 1{\sqrt{2}}
\Big( 
\left|0\right\rangle \bigotimes_{k=2}^{n} \left|i_k\right\rangle \left|1\right\rangle \bigotimes_{k=2}^{n}  \left|\bar{j}_k\right\rangle \pm 
\left|1\right\rangle \bigotimes_{k=2}^{n}  \left|\bar{i}_k\right\rangle \left|0\right\rangle \bigotimes_{k=2}^{n} \left|j_k\right\rangle
\Big).
\end{cases}
\end{align}

From the above formulas, we can get a conclusion similar to that of entanglement swapping between two Bell states, 
that is, swapping two identical SGHZ states can generate two identical GHZ states, and vice versa.

Starting from the entanglement swapping of Bell states, we have considered the entanglement swapping
between two SGHZ states containing the same number of particles in the above entanglement swapping schemes.
Let us finally consider a more general case, that is, whether the conclusions obtained above are also valid
when the number of particles contained in each of the two SGHZ states is an arbitrary even number.
Let us assume that there are two initial SGHZ states containing $2m_1$ and $2m_2$ particles respectively,
and mark them by
\begin{align}
& \left|\mathcal{G}_{f}^{\pm}\right\rangle =  \frac 1{\sqrt{2}}
\Big(
\left|0\right\rangle \bigotimes_{k=2}^{2m_1} \left|i_k\right\rangle \pm 
\left|1\right\rangle \bigotimes_{k=2}^{2m_1}  \left|\bar{i}_k\right\rangle
\Big),
\notag \\
& \left|\mathcal{G}_{f'}^{\pm}\right\rangle =  \frac 1{\sqrt{2}}
\Big(
\left|0\right\rangle \bigotimes_{k=2}^{2m_2} \left|j_k\right\rangle \pm 
\left|1\right\rangle \bigotimes_{k=2}^{2m_2} \left|\bar{j}_k\right\rangle
\Big),
\end{align}
where $f = \sum_{k=2}^{2m_1} i_k\cdot2^{2m_1-k}$ and $f' = \sum_{k=2}^{2m_2} j_k\cdot2^{2m_1-k}$, then we have
\begin{align}
\label{two-GHZ-ES-general-case1}
&\left|\mathcal{G}_{f}^{\pm}\right\rangle_{12} \otimes 
	\left|\mathcal{G}_{f'}^{\pm}\right\rangle_{34}	\notag\\
= &\quad	\Big( \left|0\right\rangle \bigotimes_{k=2}^{2m_1} \left|i_k\right\rangle \pm
\left|1\right\rangle \bigotimes_{k=2}^{2m_1}  \left|\bar{i}_k\right\rangle \Big)_{12} \bigotimes
\Big( \left|0\right\rangle \bigotimes_{k=2}^{2m_2} \left|j_k\right\rangle \mp
\left|1\right\rangle \bigotimes_{k=2}^{2m_2}  \left|\bar{j}_k\right\rangle \Big)_{34}	\notag\\
= &\quad \Big( \left|0\right\rangle \bigotimes_{k=2}^{2m_1} \left|i_k\right\rangle \Big)_{12} \bigotimes
	\Big( \left|0\right\rangle \bigotimes_{k=2}^{2m_2} \left|j_k\right\rangle \Big)_{34}	
	\pm
	\Big( \left|0\right\rangle \bigotimes_{k=2}^{2m_1} \left|i_k\right\rangle \Big)_{12} \bigotimes
	\Big( \left|1\right\rangle \bigotimes_{k=2}^{2m_2}  \left|\bar{j}_k\right\rangle \Big)_{34}
	\notag\\
&	\pm 
	\Big( \left|1\right\rangle \bigotimes_{k=2}^{2m_1}  \left|\bar{i}_k\right\rangle \Big)_{12}	\bigotimes
	\Big( \left|0\right\rangle \bigotimes_{k=2}^{2m_2} \left|j_k\right\rangle \Big)_{34} 
	+
	 \Big( \left|1\right\rangle \bigotimes_{k=2}^{2m_1}  \left|\bar{i}_k\right\rangle \Big)_{12}	\bigotimes
	 \Big( \left|1\right\rangle \bigotimes_{k=2}^{2m_2}  \left|\bar{j}_k\right\rangle \Big)_{34}
	 \notag\\
= &\quad
	\Big( \left|0\right\rangle \bigotimes_{k=2}^{m_1} \left|i_k\right\rangle
	\left|0\right\rangle \bigotimes_{k=2}^{m_2} \left|j_k\right\rangle	\Big)_{13}	
	\bigotimes
	\Big( \bigotimes_{k=m_1+1}^{2m_1} \left|i_k\right\rangle
	\bigotimes_{k=m_2+1}^{2m_2} \left|j_k\right\rangle \Big)_{24}
	\pm	
	\Big( \left|0\right\rangle \bigotimes_{k=2}^{m_1} \left|i_k\right\rangle
	\left|1\right\rangle \bigotimes_{k=2}^{m_2} \left|\bar{j}_k\right\rangle	\Big)_{13}	
	\bigotimes
	\Big( \bigotimes_{k=m_1+1}^{2m_1} \left|i_k\right\rangle
	\bigotimes_{k=m_2+1}^{2m_2} \left|\bar{j}_k\right\rangle \Big)_{24}
	\notag\\
&	\pm 
	\Big( \left|1\right\rangle \bigotimes_{k=2}^{m_1} \left|\bar{i}_k\right\rangle
	\left|0\right\rangle \bigotimes_{k=2}^{m_2} \left|j_k\right\rangle	\Big)_{13}	
	\bigotimes
	\Big( \bigotimes_{k=m_1+1}^{2m_1} \left|\bar{i}_k\right\rangle
	\bigotimes_{k=m_2+1}^{2m_2} \left|j_k\right\rangle \Big)_{24}
	+
	\Big( \left|1\right\rangle \bigotimes_{k=2}^{m_1} \left|\bar{i}_k\right\rangle
	\left|1\right\rangle \bigotimes_{k=2}^{m_2} \left|\bar{j}_k\right\rangle	\Big)_{13}	
	\bigotimes
	\Big( \bigotimes_{k=m_1+1}^{2m_1} \left|\bar{i}_k\right\rangle
	\bigotimes_{k=m_2+1}^{2m_2} \left|\bar{j}_k\right\rangle \Big)_{24}
	\notag\\
= &	
\begin{cases}
	\phantom{\pm}
	\Big( \left|\mathcal{G}_{p'}^+\right\rangle + \left|\mathcal{G}_{p'}^-\right\rangle \Big)_{13}
	\Big( \left|\mathcal{G}_{p'}^+\right\rangle + \left|\mathcal{G}_{p'}^-\right\rangle \Big)_{24}	
	\pm
	\Big( \left|\mathcal{G}_{q'}^+\right\rangle + \left|\mathcal{G}_{q'}^-\right\rangle \Big)_{13}
	\Big( \left|\mathcal{G}_{q'}^+\right\rangle + \left|\mathcal{G}_{q'}^-\right\rangle \Big)_{24}	\\
	\pm	
	\Big( \left|\mathcal{G}_{q'}^+\right\rangle - \left|\mathcal{G}_{q'}^-\right\rangle \Big)_{13}
	\Big( \left|\mathcal{G}_{q'}^+\right\rangle - \left|\mathcal{G}_{q'}^-\right\rangle \Big)_{24}	
	+
	\Big( \left|\mathcal{G}_{p'}^+\right\rangle - \left|\mathcal{G}_{p'}^-\right\rangle \Big)_{13}
	\Big( \left|\mathcal{G}_{p'}^+\right\rangle - \left|\mathcal{G}_{p'}^-\right\rangle \Big)_{24}
	& \textrm{if} \phantom{i} 0i_2i_3 \cdots i_{m_1} = i_{m_1+1}i_{m_1+2}\cdots i_{2m_1}	\\
	&	\textrm{and} \phantom{i} 0j_2j_3 \cdots j_{m_2} = j_{m_2+1}j_{m_2+2}\cdots j_{2m_2};	\\
	\phantom{\pm}
	\Big( \left|\mathcal{G}_{p'}^+\right\rangle + \left|\mathcal{G}_{p'}^-\right\rangle \Big)_{13}
	\Big( \left|\mathcal{G}_{q'}^+\right\rangle + \left|\mathcal{G}_{q'}^-\right\rangle \Big)_{24}	
	\pm
	\Big( \left|\mathcal{G}_{q'}^+\right\rangle + \left|\mathcal{G}_{q'}^-\right\rangle \Big)_{13}
	\Big( \left|\mathcal{G}_{p'}^+\right\rangle + \left|\mathcal{G}_{p'}^-\right\rangle \Big)_{24}	\\
	\pm	
	\Big( \left|\mathcal{G}_{q'}^+\right\rangle - \left|\mathcal{G}_{q'}^-\right\rangle \Big)_{13}
	\Big( \left|\mathcal{G}_{p'}^+\right\rangle - \left|\mathcal{G}_{p'}^-\right\rangle \Big)_{24}	
	+
	\Big( \left|\mathcal{G}_{p'}^+\right\rangle - \left|\mathcal{G}_{p'}^-\right\rangle \Big)_{13}
	\Big( \left|\mathcal{G}_{p'}^+\right\rangle - \left|\mathcal{G}_{p'}^-\right\rangle \Big)_{24}
	& \textrm{if} \phantom{i} 0i_2i_3 \cdots i_{m_1} = i_{m_1+1}i_{m_1+2}\cdots i_{2m_1}	\\
	&	\textrm{and} \phantom{i} 0j_2j_3 \cdots j_{m_2} = \bar{j}_{m_2+1}\bar{j}_{m_2+2}\cdots \bar{j}_{2m_2};	\\
	\phantom{\pm}
	\Big( \left|\mathcal{G}_{p'}^+\right\rangle + \left|\mathcal{G}_{p'}^-\right\rangle \Big)_{13}
	\Big( \left|\mathcal{G}_{q'}^+\right\rangle - \left|\mathcal{G}_{q'}^-\right\rangle \Big)_{24}	
	\pm
	\Big( \left|\mathcal{G}_{q'}^+\right\rangle + \left|\mathcal{G}_{q'}^-\right\rangle \Big)_{13}
	\Big( \left|\mathcal{G}_{p'}^+\right\rangle - \left|\mathcal{G}_{p'}^-\right\rangle \Big)_{24}	\\
	\pm	
	\Big( \left|\mathcal{G}_{q'}^+\right\rangle - \left|\mathcal{G}_{q'}^-\right\rangle \Big) _{13}
	\Big( \left|\mathcal{G}_{p'}^+\right\rangle + \left|\mathcal{G}_{p'}^-\right\rangle \Big)_{24}	
	+
	\Big( \left|\mathcal{G}_{p'}^+\right\rangle - \left|\mathcal{G}_{p'}^-\right\rangle \Big) _{13}
	\Big( \left|\mathcal{G}_{q'}^+\right\rangle + \left|\mathcal{G}_{q'}^-\right\rangle \Big)_{24}
	& \textrm{if} \phantom{i} 0i_2i_3 \cdots i_{m_1} = \bar{i}_{m_1+1}\bar{i}_{m_1+2}\cdots \bar{i}_{2m_1}	\\
	&	\textrm{and} \phantom{i} 0j_2j_3 \cdots j_{m_2} = j_{m_2+1}j_{m_2+2}\cdots j_{2m_2};	\\
	\phantom{\pm}
	\Big( \left|\mathcal{G}_{p'}^+\right\rangle + \left|\mathcal{G}_{p'}^-\right\rangle \Big)_{13}
	\Big( \left|\mathcal{G}_{p'}^+\right\rangle - \left|\mathcal{G}_{p'}^-\right\rangle \Big)_{24}	
	\pm
	\Big( \left|\mathcal{G}_{q'}^+\right\rangle + \left|\mathcal{G}_{q'}^-\right\rangle \Big)_{13}
	\Big( \left|\mathcal{G}_{q'}^+\right\rangle - \left|\mathcal{G}_{q'}^-\right\rangle \Big)_{24}	\\
	\pm	
	\Big( \left|\mathcal{G}_{q'}^+\right\rangle - \left|\mathcal{G}_{q'}^-\right\rangle \Big) _{13}
	\Big( \left|\mathcal{G}_{q'}^+\right\rangle + \left|\mathcal{G}_{q'}^-\right\rangle \Big)_{24}	
	+
	\Big( \left|\mathcal{G}_{p'}^+\right\rangle - \left|\mathcal{G}_{p'}^-\right\rangle \Big) _{13}
	\Big( \left|\mathcal{G}_{p'}^+\right\rangle + \left|\mathcal{G}_{p'}^-\right\rangle \Big)_{24}
	& \textrm{if} \phantom{i} 0i_2i_3 \cdots i_{m_1} =\bar{i}_{m_1+1}\bar{i}_{m_1+2}\cdots \bar{i}_{2m_1}	\\
	&	\textrm{and} \phantom{i} 0j_2j_3 \cdots j_{m_2} = \bar{j}_{m_2+1}\bar{j}_{m_2+2}\cdots \bar{j}_{2m_2}
\end{cases}\notag\\
= &	
\begin{cases}
		\phantom{\pm}
		\left|\mathcal{G}_{p'}^+\right\rangle_{13} \left|\mathcal{G}_{p'}^+\right\rangle_{24} +
		\left|\mathcal{G}_{p'}^-\right\rangle_{13} \left|\mathcal{G}_{p'}^-\right\rangle_{24} \pm
		\left|\mathcal{G}_{q'}^+\right\rangle_{13} \left|\mathcal{G}_{q'}^+\right\rangle_{24} \pm
		\left|\mathcal{G}_{q'}^-\right\rangle_{13} \left|\mathcal{G}_{q'}^-\right\rangle_{24};
		\\
		\\
		\phantom{\pm}
		\left|\mathcal{G}_{p'}^+\right\rangle_{13} \left|\mathcal{G}_{q'}^+\right\rangle_{24} +
		\left|\mathcal{G}_{p'}^-\right\rangle_{13} \left|\mathcal{G}_{q'}^-\right\rangle_{24} \pm
		\left|\mathcal{G}_{q'}^+\right\rangle_{13} \left|\mathcal{G}_{p'}^+\right\rangle_{24} \pm
		\left|\mathcal{G}_{q'}^-\right\rangle_{13} \left|\mathcal{G}_{p'}^-\right\rangle_{24};
		\\
		\\
		\phantom{\pm}
		\left|\mathcal{G}_{p'}^+\right\rangle_{13} \left|\mathcal{G}_{q'}^+\right\rangle_{24} -
		\left|\mathcal{G}_{p'}^-\right\rangle_{13} \left|\mathcal{G}_{q'}^-\right\rangle_{24} \pm
		\left|\mathcal{G}_{q'}^+\right\rangle_{13} \left|\mathcal{G}_{p'}^+\right\rangle_{24} \mp
		\left|\mathcal{G}_{q'}^-\right\rangle_{13} \left|\mathcal{G}_{p'}^-\right\rangle_{24};
		\\
		\\
		\phantom{\pm}
		\left|\mathcal{G}_{p'}^+\right\rangle_{13} \left|\mathcal{G}_{p'}^+\right\rangle_{24} -
		\left|\mathcal{G}_{p'}^-\right\rangle_{13} \left|\mathcal{G}_{p'}^-\right\rangle_{24} \pm
		\left|\mathcal{G}_{q'}^+\right\rangle_{13} \left|\mathcal{G}_{q'}^+\right\rangle_{24} \mp
		\left|\mathcal{G}_{q'}^-\right\rangle_{13} \left|\mathcal{G}_{q'}^-\right\rangle_{24},
\end{cases}
\end{align}
and 
\begin{align}
\label{two-GHZ-ES-general-case1}
&\left|\mathcal{G}_{f}^{\pm}\right\rangle_{12} \otimes 
	\left|\mathcal{G}_{f^{\prime}}^{\mp}\right\rangle_{34}	\notag\\
= &\quad	\Big( \left|0\right\rangle \bigotimes_{k=2}^{2m_1} \left|i_k\right\rangle \pm
\left|1\right\rangle \bigotimes_{k=2}^{2m_1}  \left|\bar{i}_k\right\rangle \Big)_{12} 
\bigotimes
\Big( \left|0\right\rangle \bigotimes_{k=2}^{2m_2} \left|j_k\right\rangle \mp
\left|1\right\rangle \bigotimes_{k=2}^{2m_2}  \left|\bar{j}_k\right\rangle \Big)_{34}	\notag\\
= &\quad \Big( \left|0\right\rangle \bigotimes_{k=2}^{2m_1} \left|i_k\right\rangle \Big)_{12} \bigotimes
	\Big( \left|0\right\rangle \bigotimes_{k=2}^{2m_2} \left|j_k\right\rangle \Big)_{34}	
	\mp
	\Big( \left|0\right\rangle \bigotimes_{k=2}^{2m_1} \left|i_k\right\rangle \Big)_{12} \bigotimes
	\Big( \left|1\right\rangle \bigotimes_{k=2}^{2m_2}  \left|\bar{j}_k\right\rangle \Big)_{34}
	\notag\\
&	\pm 
	\Big( \left|1\right\rangle \bigotimes_{k=2}^{2m_1}  \left|\bar{i}_k\right\rangle \Big)_{12}	\bigotimes
	\Big( \left|0\right\rangle \bigotimes_{k=2}^{2m_2} \left|j_k\right\rangle \Big)_{34} 
	-
	 \Big( \left|1\right\rangle \bigotimes_{k=2}^{2m_1}  \left|\bar{i}_k\right\rangle \Big)_{12}	\bigotimes
	 \Big( \left|1\right\rangle \bigotimes_{k=2}^{2m_2}  \left|\bar{j}_k\right\rangle \Big)_{34}
	 \notag\\
= &\quad
	\Big( \left|0\right\rangle \bigotimes_{k=2}^{m_1} \left|i_k\right\rangle
	\left|0\right\rangle \bigotimes_{k=2}^{m_2} \left|j_k\right\rangle	\Big)_{13}	\bigotimes
	\Big( \bigotimes_{k=n+1}^{2m_1} \left|i_k\right\rangle
	\bigotimes_{k=n+1}^{2m_2} \left|j_k\right\rangle \Big)_{24}
	\mp	
	\Big( \left|0\right\rangle \bigotimes_{k=2}^{m_1} \left|i_k\right\rangle
	\left|1\right\rangle \bigotimes_{k=2}^{m_2} \left|\bar{j}_k\right\rangle	\Big)_{13}	\bigotimes
	\Big( \bigotimes_{k=n+1}^{2m_1} \left|i_k\right\rangle
	\bigotimes_{k=n+1}^{2m_2} \left|\bar{j}_k\right\rangle \Big)_{24}
	\notag\\
&	\pm 
	\Big( \left|1\right\rangle \bigotimes_{k=2}^{m_1} \left|\bar{i}_k\right\rangle
	\left|0\right\rangle \bigotimes_{k=2}^{m_2} \left|j_k\right\rangle	\Big)_{13}	\bigotimes
	\Big( \bigotimes_{k=n+1}^{2m_1} \left|\bar{i}_k\right\rangle
	\bigotimes_{k=n+1}^{2m_2} \left|j_k\right\rangle \Big)_{24}
	-
	\Big( \left|1\right\rangle \bigotimes_{k=2}^{m_1} \left|\bar{i}_k\right\rangle
	\left|1\right\rangle \bigotimes_{k=2}^{m_2} \left|\bar{j}_k\right\rangle	\Big)_{13}	\bigotimes
	\Big( \bigotimes_{k=n+1}^{2m_1} \left|\bar{i}_k\right\rangle
	\bigotimes_{k=n+1}^{2m_2} \left|\bar{j}_k\right\rangle \Big)_{24}
	\notag\\
= &	
\begin{cases}
	\phantom{\pm}
	\Big( \left|\mathcal{G}_{p'}^+\right\rangle + \left|\mathcal{G}_{p'}^-\right\rangle \Big)_{13}
	\Big( \left|\mathcal{G}_{p'}^+\right\rangle + \left|\mathcal{G}_{p'}^-\right\rangle \Big)_{24}	
	\mp
	\Big( \left|\mathcal{G}_{q'}^+\right\rangle + \left|\mathcal{G}_{q'}^-\right\rangle \Big)_{13}
	\Big( \left|\mathcal{G}_{q'}^+\right\rangle + \left|\mathcal{G}_{q'}^-\right\rangle \Big)_{24}	\\
	\pm	
	\Big( \left|\mathcal{G}_{q'}^+\right\rangle - \left|\mathcal{G}_{q'}^-\right\rangle \Big)_{13}
	\Big( \left|\mathcal{G}_{q'}^+\right\rangle - \left|\mathcal{G}_{q'}^-\right\rangle \Big)_{24}	
	-
	\Big( \left|\mathcal{G}_{p'}^+\right\rangle - \left|\mathcal{G}_{p'}^-\right\rangle \Big)_{13}
	\Big( \left|\mathcal{G}_{p'}^+\right\rangle - \left|\mathcal{G}_{p'}^-\right\rangle \Big)_{24}
	& \textrm{if} \phantom{i} 0i_2i_3 \cdots i_{m_1} = i_{m_1+1}i_{m_1+2}\cdots i_{2m_1}	\\
	&	\textrm{and} \phantom{i} 0j_2j_3 \cdots j_{m_2} = j_{m_2+1}j_{m_2+2}\cdots j_{2m_2};	\\
	\phantom{\pm}
	\Big( \left|\mathcal{G}_{p'}^+\right\rangle + \left|\mathcal{G}_{p'}^-\right\rangle \Big)_{13}
	\Big( \left|\mathcal{G}_{q'}^+\right\rangle + \left|\mathcal{G}_{q'}^-\right\rangle \Big)_{24}	
	\mp
	\Big( \left|\mathcal{G}_{q'}^+\right\rangle + \left|\mathcal{G}_{q'}^-\right\rangle \Big)_{13}
	\Big( \left|\mathcal{G}_{p'}^+\right\rangle + \left|\mathcal{G}_{p'}^-\right\rangle \Big)_{24}	\\
	\pm	
	\Big( \left|\mathcal{G}_{q'}^+\right\rangle - \left|\mathcal{G}_{q'}^-\right\rangle \Big)_{13}
	\Big( \left|\mathcal{G}_{p'}^+\right\rangle - \left|\mathcal{G}_{p'}^-\right\rangle \Big)_{24}	
	-
	\Big( \left|\mathcal{G}_{p'}^+\right\rangle - \left|\mathcal{G}_{p'}^-\right\rangle \Big)_{13}
	\Big( \left|\mathcal{G}_{p'}^+\right\rangle - \left|\mathcal{G}_{p'}^-\right\rangle \Big)_{24}
	& \textrm{if} \phantom{i} 0i_2i_3 \cdots i_{m_1} = i_{m_1+1}i_{m_1+2}\cdots i_{2m_1}	\\
	&	\textrm{and} \phantom{i} 0j_2j_3 \cdots j_{m_2} = \bar{j}_{m_2+1}\bar{j}_{m_2+2}\cdots \bar{j}_{2m_2};	\\
	\phantom{\pm}
	\Big( \left|\mathcal{G}_{p'}^+\right\rangle + \left|\mathcal{G}_{p'}^-\right\rangle \Big)_{13}
	\Big( \left|\mathcal{G}_{q'}^+\right\rangle - \left|\mathcal{G}_{q'}^-\right\rangle \Big)_{24}	
	\mp
	\Big( \left|\mathcal{G}_{q'}^+\right\rangle + \left|\mathcal{G}_{q'}^-\right\rangle \Big)_{13}
	\Big( \left|\mathcal{G}_{p'}^+\right\rangle - \left|\mathcal{G}_{p'}^-\right\rangle \Big)_{24}	\\
	\pm	
	\Big( \left|\mathcal{G}_{q'}^+\right\rangle - \left|\mathcal{G}_{q'}^-\right\rangle \Big) _{13}
	\Big( \left|\mathcal{G}_{p'}^+\right\rangle + \left|\mathcal{G}_{p'}^-\right\rangle \Big)_{24}	
	-
	\Big( \left|\mathcal{G}_{p'}^+\right\rangle - \left|\mathcal{G}_{p'}^-\right\rangle \Big) _{13}
	\Big( \left|\mathcal{G}_{q'}^+\right\rangle + \left|\mathcal{G}_{q'}^-\right\rangle \Big)_{24}
	& \textrm{if} \phantom{i} 0i_2i_3 \cdots i_{m_1} = \bar{i}_{m_1+1}\bar{i}_{m_1+2}\cdots \bar{i}_{2m_1}	\\
	&	\textrm{and} \phantom{i} 0j_2j_3 \cdots j_{m_2} = j_{m_2+1}j_{m_2+2}\cdots j_{2m_2};	\\
	\phantom{\pm}
	\Big( \left|\mathcal{G}_{p'}^+\right\rangle + \left|\mathcal{G}_{p'}^-\right\rangle \Big)_{13}
	\Big( \left|\mathcal{G}_{p'}^+\right\rangle - \left|\mathcal{G}_{p'}^-\right\rangle \Big)_{24}
	\mp
	\Big( \left|\mathcal{G}_{q'}^+\right\rangle + \left|\mathcal{G}_{q'}^-\right\rangle \Big)_{13}
	\Big( \left|\mathcal{G}_{q'}^+\right\rangle - \left|\mathcal{G}_{q'}^-\right\rangle \Big)_{24}	\\
	\pm	
	\Big( \left|\mathcal{G}_{q'}^+\right\rangle - \left|\mathcal{G}_{q'}^-\right\rangle \Big) _{13}
	\Big( \left|\mathcal{G}_{q'}^+\right\rangle + \left|\mathcal{G}_{q'}^-\right\rangle \Big)_{24}	
	-
	\Big( \left|\mathcal{G}_{p'}^+\right\rangle - \left|\mathcal{G}_{p'}^-\right\rangle \Big) _{13}
	\Big( \left|\mathcal{G}_{p'}^+\right\rangle + \left|\mathcal{G}_{p'}^-\right\rangle \Big)_{24}
	& \textrm{if} \phantom{i} 0i_2i_3 \cdots i_{m_1} =\bar{i}_{m_1+1}\bar{i}_{m_1+2}\cdots \bar{i}_{2m_1}	\\
	&	\textrm{and} \phantom{i} 0j_2j_3 \cdots j_{m_2} = \bar{j}_{m_2+1}\bar{j}_{m_2+2}\cdots \bar{j}_{2m_2}
\end{cases}\notag\\
= &	
\begin{cases}
		\phantom{\pm}
		\left|\mathcal{G}_{p'}^+\right\rangle_{13} \left|\mathcal{G}_{p'}^-\right\rangle_{24} +
		\left|\mathcal{G}_{p'}^-\right\rangle_{13} \left|\mathcal{G}_{p'}^+\right\rangle_{24} \mp
		\left|\mathcal{G}_{q'}^+\right\rangle_{13} \left|\mathcal{G}_{q'}^-\right\rangle_{24} \mp
		\left|\mathcal{G}_{q'}^-\right\rangle_{13} \left|\mathcal{G}_{q'}^+\right\rangle_{24};
		\\
		\\
		\phantom{\pm}
 		\left|\mathcal{G}_{p'}^+\right\rangle_{13} \left|\mathcal{G}_{q'}^-\right\rangle_{24} +
		\left|\mathcal{G}_{p'}^-\right\rangle_{13} \left|\mathcal{G}_{q'}^+\right\rangle_{24} \mp
		\left|\mathcal{G}_{q'}^+\right\rangle_{13} \left|\mathcal{G}_{p'}^-\right\rangle_{24} \mp
		\left|\mathcal{G}_{q'}^-\right\rangle_{13} \left|\mathcal{G}_{p'}^+\right\rangle_{24};
		\\
		\\
-		\left|\mathcal{G}_{p'}^+\right\rangle_{13} \left|\mathcal{G}_{q'}^-\right\rangle_{24} +
		\left|\mathcal{G}_{p'}^-\right\rangle_{13} \left|\mathcal{G}_{q'}^+\right\rangle_{24} \pm
		\left|\mathcal{G}_{q'}^+\right\rangle_{13} \left|\mathcal{G}_{p'}^-\right\rangle_{24} \mp
		\left|\mathcal{G}_{q'}^-\right\rangle_{13} \left|\mathcal{G}_{p'}^+\right\rangle_{24};
		\\
		\\
-		\left|\mathcal{G}_{p'}^+\right\rangle_{13} \left|\mathcal{G}_{p'}^-\right\rangle_{24} +
		\left|\mathcal{G}_{p'}^-\right\rangle_{13} \left|\mathcal{G}_{p'}^+\right\rangle_{24} \pm
		\left|\mathcal{G}_{q'}^+\right\rangle_{13} \left|\mathcal{G}_{q'}^-\right\rangle_{24} \mp
		\left|\mathcal{G}_{q'}^-\right\rangle_{13} \left|\mathcal{G}_{q'}^+\right\rangle_{24},
\end{cases}
\end{align}
where
\begin{align}
\begin{cases}
\phantom{i} \left|\mathcal{G}_{p'}^{\pm}\right\rangle =  \frac 1{\sqrt{2}}
\Big( 
\left|0\right\rangle \bigotimes_{k=2}^{m_1} \left|i_k\right\rangle \left|0\right\rangle \bigotimes_{k=2}^{m_2} \left|j_k\right\rangle \pm
\left|1\right\rangle \bigotimes_{k=2}^{m_1}  \left|\bar{i}_k\right\rangle \left|1\right\rangle \bigotimes_{k=2}^{m_2}  \left|\bar{j}_k\right\rangle 
\Big),
\\
\phantom{i} \left|\mathcal{G}_{q'}^{\pm}\right\rangle =  \frac 1{\sqrt{2}}
\Big( 
\left|0\right\rangle \bigotimes_{k=2}^{m_1} \left|i_k\right\rangle \left|1\right\rangle \bigotimes_{k=2}^{m_2}  \left|\bar{j}_k\right\rangle \pm 
\left|1\right\rangle \bigotimes_{k=2}^{m_1}  \left|\bar{i}_k\right\rangle \left|0\right\rangle \bigotimes_{k=2}^{m_2} \left|j_k\right\rangle
\Big).
\end{cases}
\end{align}
In all the above entanglement swapping schemes,
we only consider the case of swapping the particles 2 and 3 (i.e., measuring the first $n$ particles in each state),
but not the case of swapping the particles 2 and 4.
In fact, the latter case can be deduced similarly as the former. For simplicity,
we would not like to repeat it in this paper, but directly summarize our results as follows.

\begin{theorem}
\label{theorem_1}

Suppose that there are two entangled states containing $m_1$ and $m_2$ particles, and they are in one of states
\begin{align}
\label{Bell_SGHZ}
\left|\mathscr{X}^{\pm}\right\rangle = \frac 1{\sqrt{2}}
\Big( \left|0\right\rangle \bigotimes_{k=2}^{2m} \left|i_k\right\rangle \pm
\left|1\right\rangle \bigotimes_{k=2}^{2m}  \left|\bar{i}_k\right\rangle \Big),
\end{align}
respectively,
where $ i_1,i_2,\ldots,i_{2m} \in \{ 0, 1 \}$
and $m \in \rm N_+$ (i.e., when $m = 1$,
$\left|\mathscr{X}^{\pm}\right\rangle$ are in Bell states; when $m > 1$,
they are in SGHZ states). Let us mark the two entangled states by
$\left|\mathscr{G}_1\right\rangle_{12}$ and $\left|\mathscr{G}_2\right\rangle_{34}$, where
$\left|\mathscr{G}_1\right\rangle_{12} = \frac 1{\sqrt{2}}
\Big( \left|0\right\rangle \bigotimes_{k=2}^{2m_1} \left|i_k^1\right\rangle \pm
\left|1\right\rangle \bigotimes_{k=2}^{2m_1}  \left|\bar{i}_k^1\right\rangle \Big)$
and
$\left|\mathscr{G}_2\right\rangle_{34} = \frac 1{\sqrt{2}}
\Big( \left|0\right\rangle \bigotimes_{k=2}^{2m_2} \left|i_k^2\right\rangle \pm
\left|1\right\rangle \bigotimes_{k=2}^{2m_2}  \left|\bar{i}_k^2\right\rangle \Big)$.
Note here that the subscripts $(1,3)$ outside the Dirac notations denote the first $m_1$ particles
and the first $m_2$ particles in the two states, and $(2,4)$ the last $m_1$ particles 
and the last $m_2$ particles, respectively.
Suppose that a GHZ (Bell) measurement is performed on the particles (1,3) [or (1,4)],
where the measurement is a Bell measurement iff $m = 1$.
Let us denote the measurement result as
$\left|\mathcal{G}_a\right\rangle$,
and the state that the remaining particles collapse into as
$\left|\mathcal{G}_b\right\rangle$.
If $\left|\mathscr{G}_1\right\rangle_{12}$ and $\left|\mathscr{G}_2\right\rangle_{34}$
meet the following two conditions:
\begin{enumerate}

\item $0i_2^h i_3^h \cdots i_{m}^h = i_{k+1}^h i_{k+2}^h \cdots i_{2m}^h
\phantom{l} or \phantom{l}
0i_2^h i_3^h \cdots i_{m}^h$  $= \bar{i}_{l+1}^h \bar{i}_{k+2}^h \cdots \bar{i}_{2m}^h 
\phantom{l} \forall m=m_1,m_2 \phantom{l} and \phantom{l} h = 1,2$.

\item Both $\left|\mathscr{G}_1\right\rangle_{12}$ and
$\left|\mathscr{G}_2\right\rangle_{34}$ are in $\left|\mathscr{X}^+\right\rangle$ or $\left|\mathscr{X}^-\right\rangle$, 
that is, the symbol of the second term in the expressions of the two states is the same,

\end{enumerate}
then $\left|\mathcal{G}_a\right\rangle$ and $\left|\mathcal{G}_b\right\rangle$ are the same,
otherwise they are different.

\end{theorem}


\section{Multi-state entanglement swapping in qubit systems}

We have shown above that two identical Bell (GHZ) states can be created
by entanglement swapping when two initial Bell (GHZ) states
satisfy the conditions given in Theorem \ref{theorem_1}.
A natural question is whether two identical GHZ states can be
created by swapping more than two Bell (SGHZ) states.
Before addressing this question, we would like to concentrate our attention on 
entanglement swapping between any number of Bell (GHZ) states.
Since Bell states can be obtained when $m = 2$ in Eq. \ref{GHZ2}, 
we only consider entanglement swapping between multi-particle GHZ states.
Suppose that there are $n(n>2)$ multi-particle GHZ states each composed of $m_h$ particles
where $m_h > 2$ and $h = 1,2,\dots,n$, 
in which case the following composite system can be constructed,
\begin{align}
\label{composite-GHZ-system}
\left|\mathscr{C}(M)\right\rangle = \bigotimes_{h=1}^{n} \left|\psi(m_h)\right\rangle,
\end{align}
where $M = \sum_{h=1}^n m_h$ and
\begin{align}
\label{each-GHZ-state}
\left|\psi(m_h)\right\rangle = \frac {1}{\sqrt{2}}
\Big( \left|0\right\rangle \bigotimes_{k=2}^{m_h} \left|i_k^h\right\rangle
\pm
\left|1\right\rangle \bigotimes_{k=2}^{m_h} \left|\bar{i}_k^h\right\rangle \Big).
\end{align}
Now suppose that the first $l_h (1\le l_h \le m_h)$ particles are selected from each GHZ state,
then these particles can be measured by using the complete orthogonal basis 
constructed by the $L$-particle ($L = \sum_{h=1}^n l_h$) GHZ states.
Such a basis is as follows:
\begin{align}
\label{L-GHZ-basis}
\left|\mathscr{B}(L)\right\rangle = \frac 1{\sqrt{2}}
\Big( \left|0\right\rangle \bigotimes_{k=2}^{l_1} \left|i_k^1\right\rangle
\bigotimes_{h=2}^{n} \bigotimes_{k=1}^{l_h} \left|i_k^h\right\rangle
\pm
\left|1\right\rangle \bigotimes_{k=2}^{l_1} \left|\bar{i}_k^1\right\rangle \bigotimes_{h=2}^{n}
\bigotimes_{k=1}^{l_h} \left|\bar{i}_k^h\right\rangle \Big).
\end{align}
With this basis, the measurements on the selected $L$ particles will project them to this basis.
At the same time, the remaining particles in system 
$\left|\mathscr{C}(M)\right\rangle$ are projected onto
\begin{align}
\label{remaining_state}
\left|\mathscr{B}(M-L)\right\rangle = \frac 1{\sqrt{2}}
\Big( \left|0\right\rangle \bigotimes_{k=l_1+2}^{m_1} \left|i_k^1\right\rangle
\bigotimes_{h=2}^{n} \bigotimes_{k=l_h+1}^{m_h} \left|i_k^h\right\rangle
\pm
\left|1\right\rangle \bigotimes_{k=l_1+2}^{m_1} \left|\bar{i}_k^1\right\rangle \bigotimes_{h=2}^{n}
\bigotimes_{k=l_h+1}^{m_h} \left|\bar{i}_k^h\right\rangle \Big).
\end{align}

Let us now move on to the issue raised above.
As before, we first consider the entanglement swapping between any number of SGHZ states
containing the same number of particles.
Assume that there are $n$ entangled states 
each in one of the states $\left|\mathscr{X}^{\pm}\right\rangle$ (see Eq. \ref{Bell_SGHZ}, similarly hereinafter),
denoted as
\begin{align}
\left|\mathscr{G}_1\right\rangle_{12}, \left|\mathscr{G}_2\right\rangle_{34}, \ldots,
\left|\mathscr{G}_n\right\rangle_{(2n-1)2n},
\end{align}
where the subscripts $(1,3,\ldots,2n-1)$ and $(2,4,\ldots,2n)$ outside the Dirac notations denote the first $l$ particles
and the last $l$ particles in these states, respectively, and 
$
0i_2^h i_3^h \cdots i_l^h = i_{l+1}^h i_{l+2}^h \cdots i_{2l}^h	 \quad \text{or} \quad
0i_2^h i_3^h \cdots i_l^h = \bar{i}_{l+1}^h \bar{i}_{l+2}^h \cdots \bar{i}_{2l}^h \phantom{i} \forall h = 1,2,\ldots,n.
$
Let us consider the first case where each state is in one of the states
$\left|\mathscr{X}^{+}\right\rangle$. Let us mark these states by 
$\left|\mathscr{G}_1^+\right\rangle_{12}, \left|\mathscr{G}_2^+\right\rangle_{34}, \ldots,
\left|\mathscr{G}_n^+\right\rangle_{(2n-1)2n}$, respectively.
Suppose that a GHZ measurement is performed on
the first $l$ particles in each state, we can arrive at
\begin{align}
\label{all_in_positive_ES}
& 	\bigotimes_{h=1}^{n} \left|\mathscr{G}_h^+\right\rangle_{2h-1,2h}
	\notag\\
= & \quad
	\Big( \bigotimes_{h=1}^n \left|0\right\rangle \bigotimes_{k=2}^{2l} \left|i_k^h\right\rangle \Big)^1_{1,2,3, \dots, 2n}
	+
	\Big( \bigotimes_{h=1}^{n-1} \left|0\right\rangle \bigotimes_{k=2}^{2l} \left|i_k^h\right\rangle
	\left|1\right\rangle \bigotimes_{k=2}^{2l} \left|\bar{i}_k^n\right\rangle \Big)^2_{1,2,3, \dots, 2n}
	\notag\\
&	+
	\Big( \bigotimes_{h=1}^{n-2} \left|0\right\rangle \bigotimes_{k=2}^{2l} \left|i_k^h\right\rangle
	\left|1\right\rangle \bigotimes_{k=2}^{2l} \left|\bar{i}_k^{n-1}\right\rangle 
	\left|0\right\rangle \bigotimes_{k=2}^{2l} \left|i_k^n\right\rangle	\Big)^3_{1,2,3,\dots, 2n}
	+ \dots 
	+
	\Big( \bigotimes_{h=1}^n \left|1\right\rangle \bigotimes_{k=2}^{2l} \left|\bar{i}_k^h\right\rangle \Big)^{2^n}_{1,2,3, \dots, 2n}
	\notag\\
= & \quad
	\Big(
	\bigotimes_{h=1}^n \left|0\right\rangle \bigotimes_{k=2}^{l} \left|i_k^h\right\rangle
	\bigotimes_{h=1}^n \bigotimes_{k=l+1}^{2l} \left|i_k^h\right\rangle
	\Big)^1_{1,3,\dots, 2n-1,2,4,\dots,2n}
	+
	\Big(
	\bigotimes_{h=1}^{n-1} \left|0\right\rangle \bigotimes_{k=2}^{l} \left|i_k^h\right\rangle
	\left|1\right\rangle \bigotimes_{k=2}^{l} \left|\bar{i}_k^n\right\rangle
	\bigotimes_{h=1}^{n-1} \bigotimes_{k=l+1}^{2l} \left|i_k^h\right\rangle
	\bigotimes_{k=l+1}^{2l} \left|\bar{i}_k^n\right\rangle
	\Big)^2_{1,3,\dots, 2n-1,2,4,\dots,2n}
	\notag\\
&	+
	\Big(
	\bigotimes_{h=1}^{n-2} \left|0\right\rangle \bigotimes_{k=2}^{l} \left|i_k^h\right\rangle
	\left|1\right\rangle \bigotimes_{k=2}^{l} \left|\bar{i}_k^{n-1}\right\rangle
	\left|0\right\rangle \bigotimes_{k=2}^{l} \left|i_k^n\right\rangle
	\bigotimes_{h=1}^{n-2} \bigotimes_{k=l+1}^{2l} \left|i_k^h\right\rangle
	\bigotimes_{k=l+1}^{2l} \left|\bar{i}_k^{n-1}\right\rangle
	\bigotimes_{k=l+1}^{2l} \left|i_k^n\right\rangle
	\Big)^3_{1,3,\dots,2n-1,2,4,\dots,2n}
	\notag\\
&	+ \cdots +
	\Big(
	\bigotimes_{h=1}^{n} \left|1\right\rangle \bigotimes_{k=2}^{l} \left|\bar{i}_k^h\right\rangle
	\bigotimes_{h=1}^{n} \bigotimes_{k=l+1}^{2l} \left|\bar{i}_k^h\right\rangle
	\Big)^{2^n}_{1,3,\dots, 2n-1,2,4,\dots,2n}
	\notag\\
= &	
\begin{dcases}
	\sum_{\begin{subarray}{c}
	a_k^j \in \{0,1\},\phantom{i} j = 1,2,\ldots,2^{n-1} \\
	0 a_2^j a_3^j \cdots a_{nl}^j = a_{nl+1}^j a_{nl+2}^j \cdots a_{2nl}^j
	\end{subarray}}
	\Big(
	\left|0\right\rangle \bigotimes_{k=2}^{2nl} \left|a_k^j\right\rangle
	+
	\left|1\right\rangle \bigotimes_{k=2}^{2nl} \left|\bar{a}_k^{2^n-j+1}\right\rangle
	\Big)
	\phantom{i}\textrm{if} \phantom{i} 0i_2^h i_3^h \cdots i_l^h = i_{l+1}^h i_{l+2}^h \cdots i_{2l}^h;
	\\
	\sum_{\begin{subarray}{c}
	b_k^j \in \{0,1\},\phantom{i} j = 1,2,\ldots,2^{n-1} \\
	0 b_2^j b_3^j \cdots b_{nl}^j = \bar{b}_{nl+1}^j \bar{b}_{nl+2}^j \cdots \bar{b}_{2nl}^j
	\end{subarray}}
	\Big(
	\left|0\right\rangle \bigotimes_{k=2}^{2nl} \left|b_k^j\right\rangle
	+
	\left|1\right\rangle \bigotimes_{k=2}^{2nl} \left|\bar{b}_k^{2^n-j+1}\right\rangle
	\Big)
	\phantom{i}\textrm{if} \phantom{i} 0i_2^h i_3^h \cdots i_l^h
	=\bar{i}_{l+1}^h \bar{i}_{l+2}^h \cdots \bar{i}_{2l}^h,
\end{dcases}\notag\\
= &	
\begin{dcases}
	\phantom{i}
	\sum_{p}
	\Big[
	\Big( \left|\mathcal{G}_p^{+}\right\rangle + \left|\mathcal{G}_p^{-}\right\rangle	\Big)
	\Big( \left|\mathcal{G}_p^{+}\right\rangle + \left|\mathcal{G}_p^{-}\right\rangle	\Big)	+
	\Big( \left|\mathcal{G}_p^{+}\right\rangle - \left|\mathcal{G}_p^{-}\right\rangle	\Big)
	\Big( \left|\mathcal{G}_p^{+}\right\rangle - \left|\mathcal{G}_p^{-}\right\rangle	\Big)
	\Big];
	\\
	\phantom{i}
	\sum_{q}
	\Big( \left|\mathcal{G}_q^{+}\right\rangle + \left|\mathcal{G}_q^{-}\right\rangle	\Big)
	\Big( \left|\mathcal{G}_q^{+}\right\rangle - \left|\mathcal{G}_q^{-}\right\rangle	\Big)	+
	\Big( \left|\mathcal{G}_q^{+}\right\rangle - \left|\mathcal{G}_q^{-}\right\rangle	\Big)
	\Big( \left|\mathcal{G}_q^{+}\right\rangle + \left|\mathcal{G}_q^{-}\right\rangle	\Big)
	\Big],
\end{dcases}\notag\\
= &	
\begin{dcases}
	\phantom{i}
	\sum_{p}
	\Big( \left|\mathcal{G}_p^{+}\right\rangle  \left|\mathcal{G}_p^{+}\right\rangle +
	\left|\mathcal{G}_p^{-}\right\rangle \left|\mathcal{G}_p^{-}\right\rangle	\Big);	
	\\
	\phantom{i}
	\sum_{q}
	\Big( \left|\mathcal{G}_q^{+}\right\rangle  \left|\mathcal{G}_q^{+}\right\rangle -
	\left|\mathcal{G}_q^{-}\right\rangle \left|\mathcal{G}_q^{-}\right\rangle	\Big),
\end{dcases}
\end{align}
where $p = \sum_{k=2}^{nl} a_k \cdot 2^{nl-k} \phantom{i} \text{and} \phantom{i} q = \sum_{k=2}^{nl} b_k \cdot 2^{nl-k}$.
Note that in the first step of the above formula, 
we obtain the polynomial with $2^n$ terms and mark their orders by the superscripts $1, 2,\dots,2^n$ respectively.
Then we add the $j$-th term and the $(2^n-j+1)$-th term
of the polynomial in turn in the third step, such that we realize the elimination in the last step (similarly hereinafter).
Next is the case where each state is in one of the states $\left|\mathscr{X}^{-}\right\rangle$.
Marking them by $\left|\mathscr{G}_1^-\right\rangle_{12}, \left|\mathscr{G}_2^-\right\rangle_{34},\ldots,
\left|\mathscr{G}_n^-\right\rangle_{(2n-1)2n}$, we can get
\begin{align}
\label{all_in_negative_ES}
& 	\bigotimes_{h=1}^{n} \left|\mathscr{G}_h^-\right\rangle_{2h-1,2h}
	\notag\\
= & \quad
	(-1)^{c_1}
	\Big( \bigotimes_{h=1}^n \left|0\right\rangle \bigotimes_{k=2}^{2l} \left|i_k^h\right\rangle \Big)^1_{1,2,3, \dots, 2n}
	+(-1)^{c_2}
	\Big( \bigotimes_{h=1}^{n-1} \left|0\right\rangle \bigotimes_{k=2}^{2l} \left|i_k^h\right\rangle
	\left|1\right\rangle \bigotimes_{k=2}^{2l} \left|\bar{i}_k^n\right\rangle \Big)^2_{1,2,3, \dots, 2n}
	\notag\\
&	+(-1)^{c_3}
	\Big( \bigotimes_{h=1}^{n-2} \left|0\right\rangle \bigotimes_{k=2}^{2l} \left|i_k^h\right\rangle
	\left|1\right\rangle \bigotimes_{k=2}^{2l} \left|\bar{i}_k^{n-1}\right\rangle 
	\left|0\right\rangle \bigotimes_{k=2}^{2l} \left|i_k^n\right\rangle	\Big)^3_{1,2,3,\dots, 2n}
	+ \dots 
	+(-1)^{c_{2^n}}
	\Big( \bigotimes_{h=1}^n \left|1\right\rangle \bigotimes_{k=2}^{2l} \left|\bar{i}_k^h\right\rangle \Big)^{2^n}_{1,2,3, \dots, 2n}
	\notag\\
= & \quad
	(-1)^{c_1}
	\Big(
	\bigotimes_{h=1}^n \left|0\right\rangle \bigotimes_{k=2}^{l} \left|i_k^h\right\rangle
	\bigotimes_{h=1}^n \bigotimes_{k=l+1}^{2l} \left|i_k^h\right\rangle
	\Big)^1_{1,3,\dots, 2n-1,2,4,\dots,2n}
	\notag\\
&	+(-1)^{c_2}
	\Big(
	\bigotimes_{h=1}^{n-1} \left|0\right\rangle \bigotimes_{k=2}^{l} \left|i_k^h\right\rangle
	\left|1\right\rangle \bigotimes_{k=2}^{l} \left|\bar{i}_k^n\right\rangle
	\bigotimes_{h=1}^{n-1} \bigotimes_{k=l+1}^{2l} \left|i_k^h\right\rangle
	\bigotimes_{k=l+1}^{2l} \left|\bar{i}_k^n\right\rangle
	\Big)^2_{1,3,\dots, 2n-1,2,4,\dots,2n}
	\notag\\
&	+(-1)^{c_3}
	\Big(
	\bigotimes_{h=1}^{n-2} \left|0\right\rangle \bigotimes_{k=2}^{l} \left|i_k^h\right\rangle
	\left|1\right\rangle \bigotimes_{k=2}^{l} \left|\bar{i}_k^{n-1}\right\rangle
	\left|0\right\rangle \bigotimes_{k=2}^{l} \left|i_k^n\right\rangle
	\bigotimes_{h=1}^{n-2} \bigotimes_{k=l+1}^{2l} \left|i_k^h\right\rangle
	\bigotimes_{k=l+1}^{2l} \left|\bar{i}_k^{n-1}\right\rangle
	\bigotimes_{k=l+1}^{2l} \left|i_k^n\right\rangle
	\Big)^3_{1,3,\dots,2n-1,2,4,\dots,2n}
	\notag\\
&	+ \cdots +(-1)^{c_{2^n}}
	\Big(
	\bigotimes_{h=1}^{n} \left|1\right\rangle \bigotimes_{k=2}^{l} \left|\bar{i}_k^h\right\rangle
	\bigotimes_{h=1}^{n} \bigotimes_{k=l+1}^{2l} \left|\bar{i}_k^h\right\rangle
	\Big)^{2^n}_{1,3,\dots, 2n-1,2,4,\dots,2n}
	\notag\\
= &	
\begin{dcases}
	\sum_{\begin{subarray}{c}
	a_k^j \in \{0,1\},\phantom{i} j = 1,2,\ldots,2^{n-1} \\
	0 a_2^j a_3^j \cdots a_{nl}^j = a_{nl+1}^j a_{nl+2}^j \cdots a_{2nl}^j
	\end{subarray}}
	\Big[
	(-1)^{c_j}
	\left|0\right\rangle \bigotimes_{k=2}^{2nl} \left|a_k^j\right\rangle
	+
	(-1)^{n-c_j}
	\left|1\right\rangle \bigotimes_{k=2}^{2nl} \left|\bar{a}_k^{2^n-j+1}\right\rangle
	\Big]
	\phantom{i}\textrm{if} \phantom{i} 0i_2^h i_3^h \cdots i_l^h = i_{l+1}^h i_{l+2}^h \cdots i_{2l}^h;
	\\
	\sum_{\begin{subarray}{c}
	b_k^j \in \{0,1\},\phantom{i} j = 1,2,\ldots,2^{n-1} \\
	0 b_2^j b_3^j \cdots b_{nl}^j = \bar{b}_{nl+1}^j \bar{b}_{nl+2}^j \cdots \bar{b}_{2nl}^j
	\end{subarray}}
	\Big[
	(-1)^{c_j}
	\left|0\right\rangle \bigotimes_{k=2}^{2nl} \left|b_k^j\right\rangle
	+
	(-1)^{n-c_j}
	\left|1\right\rangle \bigotimes_{k=2}^{2nl} \left|\bar{b}_k^{2^n-j+1}\right\rangle
	\Big]
	\phantom{i}\textrm{if} \phantom{i} 0i_2^h i_3^h \cdots i_l^h
	=\bar{i}_{l+1}^h \bar{i}_{l+2}^h \cdots \bar{i}_{2l}^h,
\end{dcases}\notag\\
= &	
\begin{dcases}
\quad
\begin{dcases}
	\sum_p
	(-1)^{c_j}
	\Big[
	\Big( \left|\mathcal{G}_p^{+}\right\rangle + \left|\mathcal{G}_p^{-}\right\rangle	\Big)
	\Big( \left|\mathcal{G}_p^{+}\right\rangle + \left|\mathcal{G}_p^{-}\right\rangle	\Big)	
	-
	\Big( \left|\mathcal{G}_p^{+}\right\rangle - \left|\mathcal{G}_p^{-}\right\rangle	\Big)
	\Big( \left|\mathcal{G}_p^{+}\right\rangle - \left|\mathcal{G}_p^{-}\right\rangle	\Big)
	\Big]
	\quad\textrm{if} \phantom{i} n \phantom{i}	\textrm{is an odd number};
	\notag\\
	\sum_p
	(-1)^{c_j}
	\Big[
	\Big( \left|\mathcal{G}_p^{+}\right\rangle + \left|\mathcal{G}_p^{-}\right\rangle	\Big)
	\Big( \left|\mathcal{G}_p^{+}\right\rangle + \left|\mathcal{G}_p^{-}\right\rangle	\Big)	
	+
	\Big( \left|\mathcal{G}_p^{+}\right\rangle - \left|\mathcal{G}_p^{-}\right\rangle	\Big)
	\Big( \left|\mathcal{G}_p^{+}\right\rangle - \left|\mathcal{G}_p^{-}\right\rangle	\Big)
	\Big]
	\quad\textrm{if} \phantom{i} n \phantom{i}	\textrm{is an even number};
\end{dcases}
\\
\quad
\begin{dcases}
	\sum_q
	(-1)^{c_j}
	\Big[
	\Big( \left|\mathcal{G}_q^{+}\right\rangle + \left|\mathcal{G}_q^{-}\right\rangle	\Big)
	\Big( \left|\mathcal{G}_q^{+}\right\rangle - \left|\mathcal{G}_q^{-}\right\rangle	\Big)	
	-
	\Big( \left|\mathcal{G}_q^{+}\right\rangle - \left|\mathcal{G}_q^{-}\right\rangle	\Big)
	\Big( \left|\mathcal{G}_q^{+}\right\rangle + \left|\mathcal{G}_q^{-}\right\rangle	\Big)
	\Big]
	\quad\textrm{if} \phantom{i} n \phantom{i}	\textrm{is an odd number};
	\notag\\
	\sum_q
	(-1)^{c_j}
	\Big[
	\Big( \left|\mathcal{G}_q^{+}\right\rangle + \left|\mathcal{G}_q^{-}\right\rangle	\Big)
	\Big( \left|\mathcal{G}_q^{+}\right\rangle - \left|\mathcal{G}_q^{-}\right\rangle	\Big)	
	+
	\Big( \left|\mathcal{G}_q^{+}\right\rangle - \left|\mathcal{G}_q^{-}\right\rangle	\Big)
	\Big( \left|\mathcal{G}_q^{+}\right\rangle + \left|\mathcal{G}_q^{-}\right\rangle	\Big)
	\Big]
	\quad\textrm{if} \phantom{i} n \phantom{i}	\textrm{is an even number},
\end{dcases}
\end{dcases}
\\
= &	
\begin{dcases}
\quad
\begin{dcases}
	\sum_p
	(-1)^{c_j}
	\Big( \left|\mathcal{G}_p^{+}\right\rangle  \left|\mathcal{G}_p^{-}\right\rangle	+
	\left|\mathcal{G}_p^{-}\right\rangle \left|\mathcal{G}_p^{+}\right\rangle	\Big);
	\\
	\sum_p
	(-1)^{c_j}
	\Big( \left|\mathcal{G}_p^{+}\right\rangle  \left|\mathcal{G}_p^{+}\right\rangle +
	\left|\mathcal{G}_p^{-}\right\rangle \left|\mathcal{G}_p^{-}\right\rangle	\Big);
\end{dcases}
\\
\quad
\begin{dcases}
	\sum_q
	(-1)^{c_j}
	\Big( \left|\mathcal{G}_q^{-}\right\rangle  \left|\mathcal{G}_q^{+}\right\rangle	-
	\left|\mathcal{G}_q^{+}\right\rangle \left|\mathcal{G}_q^{-}\right\rangle	\Big);
	\\
	\sum_q 
	(-1)^{c_j}
	\Big( \left|\mathcal{G}_q^{+}\right\rangle  \left|\mathcal{G}_q^{+}\right\rangle	-
	\left|\mathcal{G}_q^{-}\right\rangle \left|\mathcal{G}_q^{-}\right\rangle	\Big),
\end{dcases}
\end{dcases}
\end{align}
where $c_j (j=1,2,\ldots,2^{n-1})$ denote the number of $1\bar{i}_2\bar{i}_3 \cdots \bar{i}_{2l}$
in the item
$\left|0\right\rangle \bigotimes_{k=2}^{2nl} \left|a_k^j\right\rangle$ (similarly hereinafter).
In the case described in Eq. \ref{all_in_positive_ES}, 
two identical GHZ states can be obtained by entanglement swapping, 
while in the case described in Eq. \ref{all_in_negative_ES}, 
two identical GHZ states can be obtained when $n$ is even.
More generally, let us assume that there are $n$ entangled states, 
each of which consists of $2l$ particles,
and among which $m$ states are in $\left|\mathscr{X}^{-}\right\rangle$
while the rest are in $\left|\mathscr{X}^{+}\right\rangle$,
then we can arrive at
\begin{align}
\label{general_ES}
& 	\bigotimes_{h=1}^{n} \left|\mathscr{G}_h\right\rangle_{2h-1,2h}
	\notag\\
= & \quad
	(-1)^{c_1}
	\Big( \bigotimes_{h=1}^n \left|0\right\rangle \bigotimes_{k=2}^{2l} \left|i_k^h\right\rangle \Big)^1_{1,2,3, \dots, 2n}
	+(-1)^{c_2}
	\Big( \bigotimes_{h=1}^{n-1} \left|0\right\rangle \bigotimes_{k=2}^{2l} \left|i_k^h\right\rangle
	\left|1\right\rangle \bigotimes_{k=2}^{2l} \left|\bar{i}_k^n\right\rangle \Big)^2_{1,2,3, \dots, 2n}
	\notag\\
&	+(-1)^{c_3}
	\Big( \bigotimes_{h=1}^{n-2} \left|0\right\rangle \bigotimes_{k=2}^{2l} \left|i_k^h\right\rangle
	\left|1\right\rangle \bigotimes_{k=2}^{2l} \left|\bar{i}_k^{n-1}\right\rangle 
	\left|0\right\rangle \bigotimes_{k=2}^{2l} \left|i_k^n\right\rangle	\Big)^3_{1,2,3,\dots, 2n}
	+ \dots 
	+(-1)^{c_{2^n}}
	\Big( \bigotimes_{h=1}^n \left|1\right\rangle \bigotimes_{k=2}^{2l} \left|\bar{i}_k^h\right\rangle \Big)^{2^n}_{1,2,3, \dots, 2n}
	\notag\\
= & \quad
	(-1)^{c_1}
	\Big(
	\bigotimes_{h=1}^n \left|0\right\rangle \bigotimes_{k=2}^{l} \left|i_k^h\right\rangle
	\bigotimes_{h=1}^n \bigotimes_{k=l+1}^{2l} \left|i_k^h\right\rangle
	\Big)^1_{1,3,\dots, 2n-1,2,4,\dots,2n}
	\notag\\
&	+(-1)^{c_2}
	\Big(
	\bigotimes_{h=1}^{n-1} \left|0\right\rangle \bigotimes_{k=2}^{l} \left|i_k^h\right\rangle
	\left|1\right\rangle \bigotimes_{k=2}^{l} \left|\bar{i}_k^n\right\rangle
	\bigotimes_{h=1}^{n-1} \bigotimes_{k=l+1}^{2l} \left|i_k^h\right\rangle
	\bigotimes_{k=l+1}^{2l} \left|\bar{i}_k^n\right\rangle
	\Big)^2_{1,3,\dots, 2n-1,2,4,\dots,2n}
	\notag\\
&	+(-1)^{c_3}
	\Big(
	\bigotimes_{h=1}^{n-2} \left|0\right\rangle \bigotimes_{k=2}^{l} \left|i_k^h\right\rangle
	\left|1\right\rangle \bigotimes_{k=2}^{l} \left|\bar{i}_k^{n-1}\right\rangle
	\left|0\right\rangle \bigotimes_{k=2}^{l} \left|i_k^n\right\rangle
	\bigotimes_{h=1}^{n-2} \bigotimes_{k=l+1}^{2l} \left|i_k^h\right\rangle
	\bigotimes_{k=l+1}^{2l} \left|\bar{i}_k^{n-1}\right\rangle
	\bigotimes_{k=l+1}^{2l} \left|i_k^n\right\rangle
	\Big)^3_{1,3,\dots,2n-1,2,4,\dots,2n}
	\notag\\
&	+ \cdots + (-1)^{c_{2^n}}
	\Big(
	\bigotimes_{h=1}^{n} \left|1\right\rangle \bigotimes_{k=2}^{l} \left|\bar{i}_k^h\right\rangle
	\bigotimes_{h=1}^{n} \bigotimes_{k=l+1}^{2l} \left|\bar{i}_k^h\right\rangle
	\Big)^{2^n}_{1,3,\dots, 2n-1,2,4,\dots,2n}
	\notag\\
= &	
\begin{dcases}
	\sum_{\begin{subarray}{c}
	a_k^j \in \{0,1\},\phantom{i} j = 1,2,\ldots,2^{n-1} \\
	0 a_2^j a_3^j \cdots a_{nl}^j = a_{nl+1}^j a_{nl+2}^j \cdots a_{2nl}^j
	\end{subarray}}
	\Big[
	(-1)^{c_j}
	\left|0\right\rangle \bigotimes_{k=2}^{2nl} \left|a_k^j\right\rangle
	+
	(-1)^{m-c_j}
	\left|1\right\rangle \bigotimes_{k=2}^{2nl} \left|\bar{a}_k^{2^n-j+1}\right\rangle
	\Big]
	\phantom{i}\textrm{if} \phantom{i} 0i_2^h i_3^h \cdots i_l^h = i_{l+1}^h i_{l+2}^h \cdots i_{2l}^h;
	\\
	\sum_{\begin{subarray}{c}
	b_k^j \in \{0,1\},\phantom{i} j = 1,2,\ldots,2^{n-1} \\
	0 b_2^j b_3^j \cdots b_{nl}^j = \bar{b}_{nl+1}^j \bar{b}_{nl+2}^j \cdots \bar{b}_{2nl}^j
	\end{subarray}}
	\Big[
	(-1)^{c_j}
	\left|0\right\rangle \bigotimes_{k=2}^{2nl} \left|b_k^j\right\rangle
	+
	(-1)^{m-c_j}
	\left|1\right\rangle \bigotimes_{k=2}^{2nl} \left|\bar{b}_k^{2^n-j+1}\right\rangle
	\Big]
	\phantom{i}\textrm{if} \phantom{i} 0i_2^h i_3^h \cdots i_l^h
	=\bar{i}_{l+1}^h \bar{i}_{l+2}^h \cdots \bar{i}_{2l}^h,
\end{dcases}\notag\\
= &	
\begin{dcases}
\quad
\begin{dcases}
	\sum_p
	(-1)^{c_j}
	\Big[
	\Big( \left|\mathcal{G}_p^{+}\right\rangle + \left|\mathcal{G}_p^{-}\right\rangle	\Big)
	\Big( \left|\mathcal{G}_p^{+}\right\rangle + \left|\mathcal{G}_p^{-}\right\rangle	\Big)	
	-
	\Big( \left|\mathcal{G}_p^{+}\right\rangle - \left|\mathcal{G}_p^{-}\right\rangle	\Big)
	\Big( \left|\mathcal{G}_p^{+}\right\rangle - \left|\mathcal{G}_p^{-}\right\rangle	\Big)
	\Big]
	\quad\textrm{if} \phantom{i} m \phantom{i}	\textrm{is an odd number};
	\notag\\
	\sum_p
	(-1)^{c_j}
	\Big[
	\Big( \left|\mathcal{G}_p^{+}\right\rangle + \left|\mathcal{G}_p^{-}\right\rangle	\Big)
	\Big( \left|\mathcal{G}_p^{+}\right\rangle + \left|\mathcal{G}_p^{-}\right\rangle	\Big)	
	+
	\Big( \left|\mathcal{G}_p^{+}\right\rangle - \left|\mathcal{G}_p^{-}\right\rangle	\Big)
	\Big( \left|\mathcal{G}_p^{+}\right\rangle - \left|\mathcal{G}_p^{-}\right\rangle	\Big)
	\Big]
	\quad\textrm{if} \phantom{i} m \phantom{i}	\textrm{is an even number};
\end{dcases}
\\
\quad
\begin{dcases}
	\sum_q
	(-1)^{c_j}
	\Big[
	\Big( \left|\mathcal{G}_q^{+}\right\rangle + \left|\mathcal{G}_q^{-}\right\rangle	\Big)
	\Big( \left|\mathcal{G}_q^{+}\right\rangle - \left|\mathcal{G}_q^{-}\right\rangle	\Big)	
	-
	\Big( \left|\mathcal{G}_q^{+}\right\rangle - \left|\mathcal{G}_q^{-}\right\rangle	\Big)
	\Big( \left|\mathcal{G}_q^{+}\right\rangle + \left|\mathcal{G}_q^{-}\right\rangle	\Big)
	\Big]
	\quad\textrm{if} \phantom{i} m \phantom{i}	\textrm{is an odd number};
	\notag\\
	\sum_q
	(-1)^{c_j}
	\Big[
	\Big( \left|\mathcal{G}_q^{+}\right\rangle + \left|\mathcal{G}_q^{-}\right\rangle	\Big)
	\Big( \left|\mathcal{G}_q^{+}\right\rangle - \left|\mathcal{G}_q^{-}\right\rangle	\Big)	
	+
	\Big( \left|\mathcal{G}_q^{+}\right\rangle - \left|\mathcal{G}_q^{-}\right\rangle	\Big)
	\Big( \left|\mathcal{G}_q^{+}\right\rangle + \left|\mathcal{G}_q^{-}\right\rangle	\Big)
	\Big]
	\quad\textrm{if} \phantom{i} m \phantom{i}	\textrm{is an even number},
\end{dcases}
\end{dcases}
\\
= &	
\begin{dcases}
\quad
\begin{dcases}
	\sum_p
	(-1)^{c_j}
	\Big( \left|\mathcal{G}_p^{+}\right\rangle  \left|\mathcal{G}_p^{-}\right\rangle	+
	\left|\mathcal{G}_p^{-}\right\rangle \left|\mathcal{G}_p^{+}\right\rangle	\Big);
	\\
	\sum_p
	(-1)^{c_j}
	\Big( \left|\mathcal{G}_p^{+}\right\rangle  \left|\mathcal{G}_p^{+}\right\rangle +
	\left|\mathcal{G}_p^{-}\right\rangle \left|\mathcal{G}_p^{-}\right\rangle	\Big);
\end{dcases}
\\
\quad
\begin{dcases}
	\sum_q
	(-1)^{c_j}
	\Big( \left|\mathcal{G}_q^{-}\right\rangle  \left|\mathcal{G}_q^{+}\right\rangle	-
	\left|\mathcal{G}_q^{+}\right\rangle \left|\mathcal{G}_q^{-}\right\rangle	\Big);
	\\
	\sum_q 
	(-1)^{c_j}
	\Big( \left|\mathcal{G}_q^{+}\right\rangle  \left|\mathcal{G}_q^{+}\right\rangle-
	\left|\mathcal{G}_q^{-}\right\rangle \left|\mathcal{G}_q^{-}\right\rangle	\Big).
\end{dcases}
\end{dcases}
\end{align}

As before, let us finally consider the case where the number of particles 
contained in each SGHZ state is an arbitrary even number.
Suppose there are $n$ entangled states containing $2m_1,2m_2,\dots,2m_n$ particles respectively,
and among which $m$ states are in $\left|\mathscr{X}^{-}\right\rangle$
while the rest are in $\left|\mathscr{X}^{+}\right\rangle$.
Let us mark them by 
$\left|\mathbb{G}_{1}\right\rangle_{1,2},\left|\mathbb{G}_{2}\right\rangle_{3,4},\dots,\left|\mathbb{G}_{n}\right\rangle_{2n-1,2n}$,
then we can arrive at
\begin{align}
\label{most-general-ES}
& 	\bigotimes_{h=1}^{n} \left|\mathbb{G}_h\right\rangle_{2h-1,2h}
	\notag\\
= & \quad
	(-1)^{c_1}
	\Big( \bigotimes_{h=1}^n \left|0\right\rangle \bigotimes_{k=2}^{2m_h} \left|i_k^h\right\rangle \Big)^1_{1,2,3, \dots, 2n}
	+(-1)^{c_2}
	\Big( \bigotimes_{h=1}^{n-1} \left|0\right\rangle \bigotimes_{k=2}^{2m_h} \left|i_k^h\right\rangle
	\left|1\right\rangle \bigotimes_{k=2}^{2m_n} \left|\bar{i}_k^n\right\rangle \Big)^2_{1,2,3, \dots, 2n}
	\notag\\
&	+(-1)^{c_3}
	\Big( \bigotimes_{h=1}^{n-2} \left|0\right\rangle \bigotimes_{k=2}^{2m_h} \left|i_k^h\right\rangle
	\left|1\right\rangle \bigotimes_{k=2}^{2m_{n-1}} \left|\bar{i}_k^{n-1}\right\rangle 
	\left|0\right\rangle \bigotimes_{k=2}^{2m_n} \left|i_k^n\right\rangle	\Big)^3_{1,2,3,\dots, 2n}
	+ \dots 
	+(-1)^{c_{2^n}}
	\Big( \bigotimes_{h=1}^n \left|1\right\rangle \bigotimes_{k=2}^{2m_h} \left|\bar{i}_k^h\right\rangle \Big)^{2^n}_{1,2,3, \dots, 2n}
	\notag\\
= & \quad
	(-1)^{c_1}
	\Big(
	\bigotimes_{h=1}^n \left|0\right\rangle \bigotimes_{k=2}^{m_h} \left|i_k^h\right\rangle
	\bigotimes_{h=1}^n \bigotimes_{k=m_h+1}^{2m_h} \left|i_k^h\right\rangle
	\Big)^1_{1,3,\dots, 2n-1,2,4,\dots,2n}
	\notag\\
&	+(-1)^{c_2}
	\Big(
	\bigotimes_{h=1}^{n-1} \left|0\right\rangle \bigotimes_{k=2}^{m_h} \left|i_k^h\right\rangle
	\left|1\right\rangle \bigotimes_{k=2}^{m_n} \left|\bar{i}_k^n\right\rangle
	\bigotimes_{h=1}^{n-1} \bigotimes_{k=m_h+1}^{2m_h} \left|i_k^h\right\rangle
	\bigotimes_{k=m_n+1}^{2m_n} \left|\bar{i}_k^n\right\rangle
	\Big)^2_{1,3,\dots, 2n-1,2,4,\dots,2n}
	\notag\\
&	+(-1)^{c_3}
	\Big(
	\bigotimes_{h=1}^{n-2} \left|0\right\rangle \bigotimes_{k=2}^{m_h} \left|i_k^h\right\rangle
	\left|1\right\rangle \bigotimes_{k=2}^{m_{n-1}} \left|\bar{i}_k^{n-1}\right\rangle
	\left|0\right\rangle \bigotimes_{k=2}^{m_n} \left|i_k^n\right\rangle
	\bigotimes_{h=1}^{n-2} \bigotimes_{k=m_h+1}^{2m_h} \left|i_k^h\right\rangle
	\bigotimes_{k=m_{n-1}+1}^{2m_{n-1}} \left|\bar{i}_k^{n-1}\right\rangle
	\bigotimes_{k=m_n+1}^{2m_n} \left|i_k^n\right\rangle
	\Big)^3_{1,3,\dots,2n-1,2,4,\dots,2n}
	\notag\\
&	+ \cdots + (-1)^{c_{2^n}}
	\Big(
	\bigotimes_{h=1}^{n} \left|1\right\rangle \bigotimes_{k=2}^{m_h} \left|\bar{i}_k^h\right\rangle
	\bigotimes_{h=1}^{n} \bigotimes_{k=m_h+1}^{2m_h} \left|\bar{i}_k^h\right\rangle
	\Big)^{2^n}_{1,3,\dots, 2n-1,2,4,\dots,2n}
	\notag\\
= &	
\begin{dcases}
	\sum_{\begin{subarray}{c}
	a_k^j \in \{0,1\},\phantom{i} j = 1,2,\ldots,2^{n-1} \\
	0 a_2^j a_3^j \cdots a_{M}^j = a_{M+1}^j a_{M+2}^j \cdots a_{2M}^j
	\end{subarray}}
	\Big[
	(-1)^{c_j}
	\left|0\right\rangle \bigotimes_{k=2}^{2M} \left|a_k^j\right\rangle
	+
	(-1)^{m-c_j}
	\left|1\right\rangle \bigotimes_{k=2}^{2M} \left|\bar{a}_k^{2^n-j+1}\right\rangle
	\Big]
	\phantom{i}\textrm{if} \phantom{i} 0i_2^h i_3^h \cdots i_l^h = i_{l+1}^h i_{l+2}^h \cdots i_{2l}^h;
	\\
	\sum_{\begin{subarray}{c}
	b_k^j \in \{0,1\},\phantom{i} j = 1,2,\ldots,2^{n-1} \\
	0 b_2^j b_3^j \cdots b_{M}^j = \bar{b}_{M+1}^j \bar{b}_{M+2}^j \cdots \bar{b}_{2M}^j
	\end{subarray}}
	\Big[
	(-1)^{c_j}
	\left|0\right\rangle \bigotimes_{k=2}^{2M} \left|b_k^j\right\rangle
	+
	(-1)^{m-c_j}
	\left|1\right\rangle \bigotimes_{k=2}^{2M} \left|\bar{b}_k^{2^n-j+1}\right\rangle
	\Big]
	\phantom{i}\textrm{if} \phantom{i} 0i_2^h i_3^h \cdots i_l^h
	=\bar{i}_{l+1}^h \bar{i}_{l+2}^h \cdots \bar{i}_{2l}^h,
\end{dcases}\notag\\
= &	
\begin{dcases}
\quad
\begin{dcases}
	\sum_p
	(-1)^{c_j}
	\Big[
	\Big( \left|\mathcal{G}_p^{+}\right\rangle + \left|\mathcal{G}_p^{-}\right\rangle	\Big)
	\Big( \left|\mathcal{G}_p^{+}\right\rangle + \left|\mathcal{G}_p^{-}\right\rangle	\Big)	
	-
	\Big( \left|\mathcal{G}_p^{+}\right\rangle - \left|\mathcal{G}_p^{-}\right\rangle	\Big)
	\Big( \left|\mathcal{G}_p^{+}\right\rangle - \left|\mathcal{G}_p^{-}\right\rangle	\Big)
	\Big]
	\quad\textrm{if} \phantom{i} m \phantom{i}	\textrm{is an odd number};
	\notag\\
	\sum_p
	(-1)^{c_j}
	\Big[
	\Big( \left|\mathcal{G}_p^{+}\right\rangle + \left|\mathcal{G}_p^{-}\right\rangle	\Big)
	\Big( \left|\mathcal{G}_p^{+}\right\rangle + \left|\mathcal{G}_p^{-}\right\rangle	\Big)	
	+
	\Big( \left|\mathcal{G}_p^{+}\right\rangle - \left|\mathcal{G}_p^{-}\right\rangle	\Big)
	\Big( \left|\mathcal{G}_p^{+}\right\rangle - \left|\mathcal{G}_p^{-}\right\rangle	\Big)
	\Big]
	\quad\textrm{if} \phantom{i} m \phantom{i}	\textrm{is an even number};
\end{dcases}
\\
\quad
\begin{dcases}
	\sum_q
	(-1)^{c_j}
	\Big[
	\Big( \left|\mathcal{G}_q^{+}\right\rangle + \left|\mathcal{G}_q^{-}\right\rangle	\Big)
	\Big( \left|\mathcal{G}_q^{+}\right\rangle - \left|\mathcal{G}_q^{-}\right\rangle	\Big)	
	-
	\Big( \left|\mathcal{G}_q^{+}\right\rangle - \left|\mathcal{G}_q^{-}\right\rangle	\Big)
	\Big( \left|\mathcal{G}_q^{+}\right\rangle + \left|\mathcal{G}_q^{-}\right\rangle	\Big)
	\Big]
	\quad\textrm{if} \phantom{i} m \phantom{i}	\textrm{is an odd number};
	\notag\\
	\sum_q
	(-1)^{c_j}
	\Big[
	\Big( \left|\mathcal{G}_q^{+}\right\rangle + \left|\mathcal{G}_q^{-}\right\rangle	\Big)
	\Big( \left|\mathcal{G}_q^{+}\right\rangle - \left|\mathcal{G}_q^{-}\right\rangle	\Big)	
	+
	\Big( \left|\mathcal{G}_q^{+}\right\rangle - \left|\mathcal{G}_q^{-}\right\rangle	\Big)
	\Big( \left|\mathcal{G}_q^{+}\right\rangle + \left|\mathcal{G}_q^{-}\right\rangle	\Big)
	\Big]
	\quad\textrm{if} \phantom{i} m \phantom{i}	\textrm{is an even number},
\end{dcases}
\end{dcases}
\\
= &	
\begin{dcases}
\quad
\begin{dcases}
	\sum_p
	(-1)^{c_j}
	\Big( \left|\mathcal{G}_p^{+}\right\rangle  \left|\mathcal{G}_p^{-}\right\rangle	+
	\left|\mathcal{G}_p^{-}\right\rangle \left|\mathcal{G}_p^{+}\right\rangle	\Big);
	\\
	\sum_p
	(-1)^{c_j}
	\Big( \left|\mathcal{G}_p^{+}\right\rangle  \left|\mathcal{G}_p^{+}\right\rangle +
	\left|\mathcal{G}_p^{-}\right\rangle \left|\mathcal{G}_p^{-}\right\rangle	\Big);
\end{dcases}
\\
\quad
\begin{dcases}
	\sum_q
	(-1)^{c_j}
	\Big( \left|\mathcal{G}_q^{-}\right\rangle  \left|\mathcal{G}_q^{+}\right\rangle	-
	\left|\mathcal{G}_q^{+}\right\rangle \left|\mathcal{G}_q^{-}\right\rangle	\Big);
	\\
	\sum_q 
	(-1)^{c_j}
	\Big( \left|\mathcal{G}_q^{+}\right\rangle  \left|\mathcal{G}_q^{+}\right\rangle-
	\left|\mathcal{G}_q^{-}\right\rangle \left|\mathcal{G}_q^{-}\right\rangle	\Big),
\end{dcases}
\end{dcases}
\end{align}
where $M = \sum_{h=1}^n m_h$, $p = \sum_{k=2}^{M} a_k \cdot 2^{M-k}$ and 
$q = \sum_{k=2}^{M} b_k \cdot 2^{M-k}$.

 We can now summarize above results as follows.

\begin{theorem}
\label{theorem_2}

Suppose that there are $n (n \ge 2)$ entangled states
containing $2m_1,2m_2,\dots,2m_n$ particles respectively,
in which each state is in one of the states $\left|\mathscr{X}^{\pm}\right\rangle$,
and $m$ of them are in $\left|\mathscr{X}^{-}\right\rangle$
while the rest are in $\left|\mathscr{X}^{+}\right\rangle$.
Let us denote these states as
$\left|\mathscr{G}_{1}\right\rangle,\left|\mathscr{G}_{2}\right\rangle,\dots,\left|\mathscr{G}_{n}\right\rangle$,
where
$\left|\mathscr{G}_h\right\rangle = \frac 1{\sqrt{2}}
\Big( \left|0\right\rangle \bigotimes_{k=2}^{2m_h} \left|i_k^h\right\rangle \pm
\left|1\right\rangle \bigotimes_{k=2}^{2m_h}  \left|\bar{i}_k^h\right\rangle \Big), h=1,2,\ldots,n$.
Suppose that a GHZ (Bell) measurement is performed on the first $m_h$ particles in each state
(the measurement is a Bell measurement iff $n=2$ and $m_h = 1$).
Let us denote the measurement result as $\left|\mathcal{G}_a\right\rangle$,
and the state that the remaining particles collapse into as $\left|\mathcal{G}_b\right\rangle$.
If the $n$ entangled states meet the following two conditions:
\begin{enumerate}

\item $m$ is an even number,

\item
$0i_2^h i_3^h \cdots i_{m_h}^h = i_{m_h+1}^h i_{m_h+2}^h \cdots i_{2m_h}^h$ or
$0i_2^h i_3^h \cdots i_{m_h}^h = \bar{i}_{m_h+1}^h \bar{i}_{m_h+2}^h \cdots \bar{i}_{2m_h}^h$ \phantom{i},

\end{enumerate}
then $\left|\mathcal{G}_a\right\rangle$
and $\left|\mathcal{G}_b\right\rangle$ are the same.

\end{theorem}

We can easily get the following two corollaries:

\begin{corollary}
\label{corollary_1}

Suppose that all the entangled states in Theorem \ref{theorem_2} satisfy the second condition.
If each of the entangled states is in one of the states $\left|\mathscr{X}^{+}\right\rangle$,
then $\left|\mathcal{G}_a\right\rangle$ and $\left|\mathcal{G}_b\right\rangle$ 
obtained by entanglement swapping are the same.
If each one is in one of $\left|\mathscr{X}^{-}\right\rangle$,
then $\left|\mathcal{G}_a\right\rangle$ and $\left|\mathcal{G}_b\right\rangle$
are the same when $n$ is even, and they are different when $n$ is odd.

\end{corollary}

\begin{corollary}
\label{corollary_2}

Suppose that the $n$ entangled states in Theorem \ref{theorem_2} are the same.
If they are in one of the states $\left|\mathscr{X}^{+}\right\rangle$,
then $\left|\mathcal{G}_a\right\rangle$ and $\left|\mathcal{G}_b\right\rangle$ obtained by entanglement swapping
are the same.
If they are in one of the states $\left|\mathscr{X}^{-}\right\rangle$,
then $\left|\mathcal{G}_a\right\rangle$ and $\left|\mathcal{G}_b\right\rangle$
are the same when $n$ is even, and they are different when $n$ is odd.

\end{corollary}

In what follows we would like to give a few simple examples to verify the correctness of 
Eqs. \ref{all_in_positive_ES}-\ref{general_ES}, 
which may also be beneficial to their clarity.
Let us take the entanglement swapping scheme of three identical 
Bell states and of four identical Bell states as examples, respectively, to verify the correctness.
Entanglement swapping between three identical Bell states is given by
\begin{align}
\label{ES_three_identical_Bell1}
& 	\Big( \left|0i\right\rangle + \left|1\bar{i}\right\rangle \Big)_{12}	\otimes
	\Big( \left|0i\right\rangle + \left|1\bar{i}\right\rangle \Big)_{34}	\otimes
	\Big( \left|0i\right\rangle + \left|1\bar{i}\right\rangle \Big)_{56}	\notag\\
= & \quad	 	\left|0i0i0i\right\rangle_{123456} +
	\left|0i0i1\bar{i}\right\rangle_{123456}+
	\left|0i1\bar{i}0i\right\rangle_{123456} +
	\left|0i1\bar{i}1\bar{i}\right\rangle_{123456}		\notag\\
   & 	+	\left|1\bar{i}0i0i\right\rangle_{123456} + 
   \left|1\bar{i}0i1\bar{i}\right\rangle_{123456} +
	\left|1\bar{i}1\bar{i}0i\right\rangle_{123456} + 
	\left|1\bar{i}1\bar{i}1\bar{i}\right\rangle_{123456}		\notag\\
= & \quad	 	\left|000iii\right\rangle_{13572468} + 
	\left|001ii\bar{i}\right\rangle_{135246} +
	\left|010i\bar{i}i\right\rangle_{135246} + 
	\left|011i\bar{i}\bar{i}\right\rangle_{135246}		\notag\\
   & 	+	\left|100\bar{i}ii\right\rangle_{135246} + 
   \left|101\bar{i}i\bar{i}\right\rangle_{135246} +
	\left|110\bar{i}\bar{i}i\right\rangle_{135246} \pm 
	\left|111\bar{i}\bar{i}\bar{i}\right\rangle_{135246}		\notag\\
= &	
\begin{cases}
	\phantom{\pm}
	\left|\mathcal{G}_0^{+}\right\rangle \left|\mathcal{G}_0^{+}\right\rangle +
	\left|\mathcal{G}_0^{-}\right\rangle \left|\mathcal{G}_0^{-}\right\rangle +
	\left|\mathcal{G}_1^{+}\right\rangle \left|\mathcal{G}_1^{+}\right\rangle +
	\left|\mathcal{G}_1^{-}\right\rangle \left|\mathcal{G}_1^{-}\right\rangle +
	\left|\mathcal{G}_2^{+}\right\rangle \left|\mathcal{G}_2^{+}\right\rangle +
	\left|\mathcal{G}_2^{-}\right\rangle \left|\mathcal{G}_2^{-}\right\rangle +	
	\left|\mathcal{G}_3^{+}\right\rangle \left|\mathcal{G}_3^{+}\right\rangle +
	\left|\mathcal{G}_3^{-}\right\rangle \left|\mathcal{G}_3^{-}\right\rangle
	& \textrm{if} \phantom{i} i = 0;	\\
	\\
	\phantom{\pm}
	\left|\mathcal{G}_0^{+}\right\rangle \left|\mathcal{G}_0^{+}\right\rangle -
	\left|\mathcal{G}_0^{-}\right\rangle \left|\mathcal{G}_0^{-}\right\rangle +
	\left|\mathcal{G}_1^{+}\right\rangle \left|\mathcal{G}_1^{+}\right\rangle -
	\left|\mathcal{G}_1^{-}\right\rangle \left|\mathcal{G}_1^{-}\right\rangle +
	\left|\mathcal{G}_2^{+}\right\rangle \left|\mathcal{G}_2^{+}\right\rangle -
	\left|\mathcal{G}_2^{-}\right\rangle \left|\mathcal{G}_2^{-}\right\rangle +	
	\left|\mathcal{G}_3^{+}\right\rangle \left|\mathcal{G}_3^{+}\right\rangle -
	\left|\mathcal{G}_3^{-}\right\rangle \left|\mathcal{G}_3^{-}\right\rangle
	& \textrm{if} \phantom{i} i = 1,
\end{cases}
\end{align}
and
\begin{align}
\label{ES_three_identical_Bell2}
& 	\Big( \left|0i\right\rangle - \left|1\bar{i}\right\rangle \Big)_{12}	\otimes
	\Big( \left|0i\right\rangle - \left|1\bar{i}\right\rangle \Big)_{34}	\otimes
	\Big( \left|0i\right\rangle - \left|1\bar{i}\right\rangle \Big)_{56}	\notag\\
= & \quad	 	\left|0i0i0i\right\rangle_{123456} - 
	\left|0i0i1\bar{i}\right\rangle_{123456} -
	\left|0i1\bar{i}0i\right\rangle_{123456} + 
	\left|0i1\bar{i}1\bar{i}\right\rangle_{123456}		\notag\\
   & 	-	\left|1\bar{i}0i0i\right\rangle_{123456} + 
   \left|1\bar{i}0i1\bar{i}\right\rangle_{123456} +
	\left|1\bar{i}1\bar{i}0i\right\rangle_{123456} - 
	\left|1\bar{i}1\bar{i}1\bar{i}\right\rangle_{123456}		\notag\\
= & \quad	 	\left|000iii\right\rangle_{135246} - 
	\left|001ii\bar{i}\right\rangle_{135246} -
	\left|010i\bar{i}i\right\rangle_{135246} + 
	\left|011i\bar{i}\bar{i}\right\rangle_{135246}		\notag\\
   & 	-	\left|100\bar{i}ii\right\rangle_{135246} + 
   \left|101\bar{i}i\bar{i}\right\rangle_{135246} +
	\left|110\bar{i}\bar{i}i\right\rangle_{135246} - 
	\left|111\bar{i}\bar{i}\bar{i}\right\rangle_{135246}		\notag\\
= &	
\begin{cases}
	\phantom{\pm}
	\left|\mathcal{G}_0^{+}\right\rangle \left|\mathcal{G}_0^{-}\right\rangle +
	\left|\mathcal{G}_0^{-}\right\rangle \left|\mathcal{G}_0^{+}\right\rangle -
	\left|\mathcal{G}_1^{+}\right\rangle \left|\mathcal{G}_1^{-}\right\rangle -
	\left|\mathcal{G}_1^{-}\right\rangle \left|\mathcal{G}_1^{+}\right\rangle -
	\left|\mathcal{G}_2^{+}\right\rangle \left|\mathcal{G}_2^{-}\right\rangle -
	\left|\mathcal{G}_2^{-}\right\rangle \left|\mathcal{G}_2^{+}\right\rangle +	
	\left|\mathcal{G}_3^{+}\right\rangle \left|\mathcal{G}_3^{-}\right\rangle +
	\left|\mathcal{G}_3^{-}\right\rangle \left|\mathcal{G}_3^{+}\right\rangle
	& \textrm{if} \phantom{i} i = 0;	\\
	\\
-	\left|\mathcal{G}_0^{+}\right\rangle \left|\mathcal{G}_0^{-}\right\rangle +
	\left|\mathcal{G}_0^{-}\right\rangle \left|\mathcal{G}_0^{+}\right\rangle +
	\left|\mathcal{G}_1^{+}\right\rangle \left|\mathcal{G}_1^{-}\right\rangle -
	\left|\mathcal{G}_1^{-}\right\rangle \left|\mathcal{G}_1^{+}\right\rangle +
	\left|\mathcal{G}_2^{+}\right\rangle \left|\mathcal{G}_2^{-}\right\rangle -
	\left|\mathcal{G}_2^{-}\right\rangle \left|\mathcal{G}_2^{+}\right\rangle -	
	\left|\mathcal{G}_3^{+}\right\rangle \left|\mathcal{G}_3^{-}\right\rangle +
	\left|\mathcal{G}_3^{-}\right\rangle \left|\mathcal{G}_3^{+}\right\rangle
	& \textrm{if} \phantom{i} i = 1.
\end{cases}
\end{align}
Entanglement swapping between four identical Bell states is given by
\begin{align}
\label{ES_identical_four_Bell}
& 	\Big( \left|0i\right\rangle \pm \left|1\bar{i}\right\rangle \Big)_{12}	\otimes
	\Big( \left|0i\right\rangle \pm \left|1\bar{i}\right\rangle \Big)_{34}	\otimes
	\Big( \left|0i\right\rangle \pm \left|1\bar{i}\right\rangle \Big)_{56}	\otimes
	\Big( \left|0i\right\rangle \pm \left|1\bar{i}\right\rangle \Big)_{78} \notag\\
= & \quad	 	\left|0i0i0i0i\right\rangle_{12345678} \pm \left|0i0i0i1\bar{i}\right\rangle_{12345678}	\pm
	\left|0i0i1\bar{i}0i\right\rangle_{12345678} + \left|0i0i1\bar{i}1\bar{i}\right\rangle_{12345678}		\notag\\
   & 	\pm	\left|0i1\bar{i}0i0i\right\rangle_{12345678} + \left|0i1\bar{i}0i1\bar{i}\right\rangle_{12345678}	+
	\left|0i1\bar{i}1\bar{i}0i\right\rangle_{12345678} \pm \left|0i1\bar{i}1\bar{i}1\bar{i}\right\rangle_{12345678}		\notag\\
   & 	\pm	\left|1\bar{i}0i0i0i\right\rangle_{12345678} + \left|1\bar{i}0i0i1\bar{i}\right\rangle_{12345678}	+
	\left|1\bar{i}0i1\bar{i}0i\right\rangle_{12345678} \pm \left|1\bar{i}0i1\bar{i}1\bar{i}\right\rangle_{12345678}		\notag\\
   & 	+	\left|1\bar{i}1\bar{i}0i0i\right\rangle_{12345678} \pm \left|1\bar{i}1\bar{i}0i1\bar{i}\right\rangle_{12345678}	\pm
	\left|1\bar{i}1\bar{i}1\bar{i}0i\right\rangle_{12345678} + \left|1\bar{i}1\bar{i}1\bar{i}1\bar{i}\right\rangle_{12345678}	\notag\\
= & \quad	 	\left|0000iiii\right\rangle_{13572468} \pm \left|0001iii\bar{i}\right\rangle_{13572468}	\pm
	\left|0010ii\bar{i}i\right\rangle_{13572468} + \left|0011ii\bar{i}\bar{i}\right\rangle_{13572468}		\notag\\
   & 	\pm	\left|0100i\bar{i}ii\right\rangle_{12345678} + \left|0101i\bar{i}i\bar{i}\right\rangle_{13572468}	+
	\left|0110i\bar{i}\bar{i}i\right\rangle_{13572468} \pm \left|0111i\bar{i}\bar{i}\bar{i}\right\rangle_{13572468}		\notag\\
   & 	\pm	\left|1000\bar{i}iii\right\rangle_{13572468} + \left|1001\bar{i}ii\bar{i}\right\rangle_{13572468}	+
	\left|1010\bar{i}i\bar{i}i\right\rangle_{13572468} \pm \left|1011\bar{i}i\bar{i}\bar{i}\right\rangle_{13572468}		\notag\\
   & 	+	\left|1100\bar{i}\bar{i}ii\right\rangle_{13572468} \pm \left|1101\bar{i}\bar{i}i\bar{i}\right\rangle_{13572468}	\pm
	\left|1110\bar{i}\bar{i}\bar{i}i\right\rangle_{13572468} + \left|1111\bar{i}\bar{i}\bar{i}\bar{i}\right\rangle_{13572468}	\notag\\
= &	
\begin{cases}
	\phantom{\pm}
	\left|\mathcal{G}_0^{+}\right\rangle \left|\mathcal{G}_0^{+}\right\rangle +
	\left|\mathcal{G}_0^{-}\right\rangle \left|\mathcal{G}_0^{-}\right\rangle \pm
	\left|\mathcal{G}_1^{+}\right\rangle \left|\mathcal{G}_1^{+}\right\rangle \pm
	\left|\mathcal{G}_1^{-}\right\rangle \left|\mathcal{G}_1^{-}\right\rangle \pm
	\left|\mathcal{G}_2^{+}\right\rangle \left|\mathcal{G}_2^{+}\right\rangle \pm
	\left|\mathcal{G}_2^{-}\right\rangle \left|\mathcal{G}_2^{-}\right\rangle +	
	\left|\mathcal{G}_3^{+}\right\rangle \left|\mathcal{G}_3^{+}\right\rangle +
	\left|\mathcal{G}_3^{-}\right\rangle \left|\mathcal{G}_3^{-}\right\rangle 	\\
\pm \left|\mathcal{G}_4^{+}\right\rangle \left|\mathcal{G}_4^{+}\right\rangle \pm
	\left|\mathcal{G}_4^{-}\right\rangle \left|\mathcal{G}_4^{-}\right\rangle +
	\left|\mathcal{G}_5^{+}\right\rangle \left|\mathcal{G}_5^{+}\right\rangle +
	\left|\mathcal{G}_5^{-}\right\rangle \left|\mathcal{G}_5^{-}\right\rangle +
	\left|\mathcal{G}_6^{+}\right\rangle \left|\mathcal{G}_6^{+}\right\rangle +
	\left|\mathcal{G}_6^{-}\right\rangle \left|\mathcal{G}_6^{-}\right\rangle \pm
	\left|\mathcal{G}_7^{+}\right\rangle \left|\mathcal{G}_7^{+}\right\rangle \pm
	\left|\mathcal{G}_7^{-}\right\rangle \left|\mathcal{G}_7^{-}\right\rangle,
	& \textrm{if} \phantom{i} i = 0;	\\
	\\
	\phantom{\pm}
	\left|\mathcal{G}_0^{+}\right\rangle \left|\mathcal{G}_0^{+}\right\rangle -
	\left|\mathcal{G}_0^{-}\right\rangle \left|\mathcal{G}_0^{-}\right\rangle \pm
	\left|\mathcal{G}_1^{+}\right\rangle \left|\mathcal{G}_1^{+}\right\rangle \mp
	\left|\mathcal{G}_1^{-}\right\rangle \left|\mathcal{G}_1^{-}\right\rangle \pm
	\left|\mathcal{G}_2^{+}\right\rangle \left|\mathcal{G}_2^{+}\right\rangle \mp
	\left|\mathcal{G}_2^{-}\right\rangle \left|\mathcal{G}_2^{-}\right\rangle +	
	\left|\mathcal{G}_3^{+}\right\rangle \left|\mathcal{G}_3^{+}\right\rangle -
	\left|\mathcal{G}_3^{-}\right\rangle \left|\mathcal{G}_3^{-}\right\rangle 	\\
\pm \left|\mathcal{G}_4^{+}\right\rangle \left|\mathcal{G}_4^{+}\right\rangle \mp
	\left|\mathcal{G}_4^{-}\right\rangle \left|\mathcal{G}_4^{-}\right\rangle +
	\left|\mathcal{G}_5^{+}\right\rangle \left|\mathcal{G}_5^{+}\right\rangle -
	\left|\mathcal{G}_5^{-}\right\rangle \left|\mathcal{G}_5^{-}\right\rangle +
	\left|\mathcal{G}_6^{+}\right\rangle \left|\mathcal{G}_6^{+}\right\rangle -
	\left|\mathcal{G}_6^{-}\right\rangle \left|\mathcal{G}_6^{-}\right\rangle \pm
	\left|\mathcal{G}_7^{+}\right\rangle \left|\mathcal{G}_7^{+}\right\rangle \mp
	\left|\mathcal{G}_7^{-}\right\rangle \left|\mathcal{G}_7^{-}\right\rangle,
	& \textrm{if} \phantom{i} i = 1.
\end{cases}
\end{align}
From Eqs. \ref{ES_three_identical_Bell1} and \ref{ES_three_identical_Bell2},
two GHZ states generated by entanglement swapping are the same
if the three initial Bell states are the same and in $\left|0i\right\rangle + \left|1\bar{i}\right\rangle$,
which the two generated GHZ states are different
when three initial Bell states are in $\left|0i\right\rangle - \left|1\bar{i}\right\rangle$.
From Eq. \ref{ES_identical_four_Bell}, when the number of initial Bell states is four,
whether they are in $\left|0i\right\rangle + \left|1\bar{i}\right\rangle$
or $\left|0i\right\rangle - \left|1\bar{i}\right\rangle$,
the two GHZ states generated are the same.

Next we discuss the entanglement exchange between four GHZ states that are not all the same.
Let us first focus on the following example, where three states are the same and the other is different from them,
\begin{align}
\label{ES_different_four_Bell_1}
& \Big( \left|0i\right\rangle \pm \left|1\bar{i}\right\rangle \Big)_{12}	\otimes
	\Big( \left|0i\right\rangle \pm \left|1\bar{i}\right\rangle \Big)_{34}	\otimes
	\Big( \left|0i\right\rangle \pm \left|1\bar{i}\right\rangle \Big)_{56}	\otimes
	\Big( \left|0i\right\rangle \mp \left|1\bar{i}\right\rangle \Big)_{78}	\notag\\
= & \quad	 	
	\left|0i0i0i0i\right\rangle_{12345678} \mp
	\left|0i0i0i1\bar{i}\right\rangle_{12345678} \pm
	\left|0i0i1\bar{i}0i\right\rangle_{12345678} -
	\left|0i0i1\bar{i}1\bar{i}\right\rangle_{12345678}	\notag\\
   & 	\pm	\left|0i1\bar{i}0i0i\right\rangle_{12345678} -
   \left|0i1\bar{i}0i1\bar{i}\right\rangle_{12345678} +
	\left|0i1\bar{i}1\bar{i}0i\right\rangle_{12345678} \mp
	\left|0i1\bar{i}1\bar{i}1\bar{i}\right\rangle_{12345678}		\notag\\
   & 	\pm	\left|1\bar{i}0i0i0i\right\rangle_{12345678} -
   \left|1\bar{i}0i0i1\bar{i}\right\rangle_{12345678} +
	\left|1\bar{i}0i1\bar{i}0i\right\rangle_{12345678} \mp
	\left|1\bar{i}0i1\bar{i}1\bar{i}\right\rangle_{12345678}		\notag\\
   & 	+	\left|1\bar{i}1\bar{i}0i0i\right\rangle_{12345678} \mp
   \left|1\bar{i}1\bar{i}0i1\bar{i}\right\rangle_{12345678}	\pm
	\left|1\bar{i}1\bar{i}1\bar{i}0i\right\rangle_{12345678} -
	\left|1\bar{i}1\bar{i}1\bar{i}1\bar{i}\right\rangle_{12345678}	\notag\\
= & \quad	 	\left|0000iiii\right\rangle_{13572468} \mp 
	\left|0001iii\bar{i}\right\rangle_{13572468}	\pm
	\left|0010ii\bar{i}i\right\rangle_{13572468} - 
	\left|0011ii\bar{i}\bar{i}\right\rangle_{13572468}		\notag\\
   & 	\pm	\left|0100i\bar{i}ii\right\rangle_{12345678} -
   \left|0101i\bar{i}i\bar{i}\right\rangle_{13572468}	 +
	\left|0110i\bar{i}1\bar{i}i\right\rangle_{13572468} \mp
	\left|0111i\bar{i}\bar{i}\bar{i}\right\rangle_{13572468}		\notag\\
   & 	\pm	\left|1000\bar{i}iii\right\rangle_{13572468} -
   \left|1001\bar{i}ii\bar{i}\right\rangle_{13572468}	+
	\left|1010\bar{i}i\bar{i}i\right\rangle_{13572468} \mp 
	\left|1011\bar{i}i\bar{i}\bar{i}\right\rangle_{13572468}		\notag\\
   & 	+	\left|1100\bar{i}\bar{i}ii\right\rangle_{13572468} \mp
   \left|1101\bar{i}\bar{i}i\bar{i}\right\rangle_{13572468}	\pm
	\left|1110\bar{i}\bar{i}\bar{i}i\right\rangle_{13572468} -
	\left|1111\bar{i}\bar{i}\bar{i}\bar{i}\right\rangle_{13572468}	\notag\\
= &	
\begin{cases}
	\phantom{\pm}
	\left|\mathcal{G}_0^{+}\right\rangle \left|\mathcal{G}_0^{-}\right\rangle +
	\left|\mathcal{G}_0^{-}\right\rangle \left|\mathcal{G}_0^{+}\right\rangle \mp
	\left|\mathcal{G}_1^{+}\right\rangle \left|\mathcal{G}_1^{-}\right\rangle \mp
	\left|\mathcal{G}_1^{-}\right\rangle \left|\mathcal{G}_1^{+}\right\rangle \pm
	\left|\mathcal{G}_2^{+}\right\rangle \left|\mathcal{G}_2^{-}\right\rangle \pm
	\left|\mathcal{G}_2^{-}\right\rangle \left|\mathcal{G}_2^{+}\right\rangle -	
	\left|\mathcal{G}_3^{+}\right\rangle \left|\mathcal{G}_3^{-}\right\rangle -
	\left|\mathcal{G}_3^{-}\right\rangle \left|\mathcal{G}_3^{+}\right\rangle 	\\
\pm \left|\mathcal{G}_4^{+}\right\rangle \left|\mathcal{G}_4^{-}\right\rangle \pm
	\left|\mathcal{G}_4^{-}\right\rangle \left|\mathcal{G}_4^{+}\right\rangle -
	\left|\mathcal{G}_5^{+}\right\rangle \left|\mathcal{G}_5^{-}\right\rangle -
	\left|\mathcal{G}_5^{-}\right\rangle \left|\mathcal{G}_5^{+}\right\rangle +
	\left|\mathcal{G}_6^{+}\right\rangle \left|\mathcal{G}_6^{-}\right\rangle +
	\left|\mathcal{G}_6^{-}\right\rangle \left|\mathcal{G}_6^{+}\right\rangle \mp
	\left|\mathcal{G}_7^{+}\right\rangle \left|\mathcal{G}_7^{-}\right\rangle \mp
	\left|\mathcal{G}_7^{-}\right\rangle \left|\mathcal{G}_7^{+}\right\rangle,
	& \textrm{if} \phantom{i} i = 0;	\\
	\\
-  \left|\mathcal{G}_0^{+}\right\rangle \left|\mathcal{G}_0^{-}\right\rangle +
	\left|\mathcal{G}_0^{-}\right\rangle \left|\mathcal{G}_0^{+}\right\rangle \pm
	\left|\mathcal{G}_1^{+}\right\rangle \left|\mathcal{G}_1^{-}\right\rangle \mp
	\left|\mathcal{G}_1^{-}\right\rangle \left|\mathcal{G}_1^{+}\right\rangle \mp
	\left|\mathcal{G}_2^{+}\right\rangle \left|\mathcal{G}_2^{-}\right\rangle \pm
	\left|\mathcal{G}_2^{-}\right\rangle \left|\mathcal{G}_2^{+}\right\rangle +	
	\left|\mathcal{G}_3^{+}\right\rangle \left|\mathcal{G}_3^{-}\right\rangle -
	\left|\mathcal{G}_3^{-}\right\rangle \left|\mathcal{G}_3^{+}\right\rangle 	\\
\mp \left|\mathcal{G}_4^{+}\right\rangle \left|\mathcal{G}_4^{-}\right\rangle \pm
	\left|\mathcal{G}_4^{-}\right\rangle \left|\mathcal{G}_4^{+}\right\rangle +
	\left|\mathcal{G}_5^{+}\right\rangle \left|\mathcal{G}_5^{-}\right\rangle -
	\left|\mathcal{G}_5^{-}\right\rangle \left|\mathcal{G}_5^{+}\right\rangle -
	\left|\mathcal{G}_6^{+}\right\rangle \left|\mathcal{G}_6^{-}\right\rangle +
	\left|\mathcal{G}_6^{-}\right\rangle \left|\mathcal{G}_6^{+}\right\rangle \pm
	\left|\mathcal{G}_7^{+}\right\rangle \left|\mathcal{G}_7^{-}\right\rangle \mp
	\left|\mathcal{G}_7^{-}\right\rangle \left|\mathcal{G}_7^{+}\right\rangle,
	& \textrm{if} \phantom{i} i = 1,
\end{cases}
\end{align}
Let us then focus on the following example, where two states are the same and the other two are also the same,
\begin{align}
\label{ES_different_four_Bell_2}
& 	\Big( \left|0i\right\rangle \pm \left|1\bar{i}\right\rangle \Big)_{12}	\otimes
	\Big( \left|0i\right\rangle \pm \left|1\bar{i}\right\rangle \Big)_{34}	\otimes
	\Big( \left|0i\right\rangle \mp \left|1\bar{i}\right\rangle \Big)_{56}	\otimes
	\Big( \left|0i\right\rangle \mp \left|1\bar{i}\right\rangle \Big)_{78}	\notag\\
= & \quad	 	\left|0i0i0i0i\right\rangle_{12345678} \mp \left|0i0i0i1\bar{i}\right\rangle_{12345678}	\mp
	\left|0i0i1\bar{i}0i\right\rangle_{12345678} + \left|0i0i1\bar{i}1\bar{i}\right\rangle_{12345678}	\notag\\
   & 	\pm	\left|0i1\bar{i}0i0i\right\rangle_{12345678} - \left|0i1\bar{i}0i1\bar{i}\right\rangle_{12345678}	 -
	\left|0i1\bar{i}1\bar{i}0i\right\rangle_{12345678} \pm \left|0i1\bar{i}1\bar{i}1\bar{i}\right\rangle_{12345678}		\notag\\
   & 	\pm	\left|1\bar{i}0i0i0i\right\rangle_{12345678} - \left|1\bar{i}0i0i1\bar{i}\right\rangle_{12345678}	 -
	\left|1\bar{i}0i1\bar{i}0i\right\rangle_{12345678} \pm \left|1\bar{i}0i1\bar{i}1\bar{i}\right\rangle_{12345678}		\notag\\
   & 	+	\left|1\bar{i}1\bar{i}0i0i\right\rangle_{12345678} \mp \left|1\bar{i}1\bar{i}0i1\bar{i}\right\rangle_{12345678}	\mp
	\left|1\bar{i}1\bar{i}1\bar{i}0i\right\rangle_{12345678} + \left|1\bar{i}1\bar{i}1\bar{i}1\bar{i}\right\rangle_{12345678}	\notag\\
= & \quad	 	\left|0000iiii\right\rangle_{13572468} \mp \left|0001iii\bar{i}\right\rangle_{13572468}	\mp
	\left|0010ii\bar{i}i\right\rangle_{13572468} + \left|0011ii\bar{i}\bar{i}\right\rangle_{13572468}		\notag\\
   & 	\pm	\left|0100i\bar{i}ii\right\rangle_{13572468} - \left|0101i\bar{i}i\bar{i}\right\rangle_{13572468}	 -
	\left|0110i\bar{i}1\bar{i}i\right\rangle_{13572468} \pm \left|0111i\bar{i}\bar{i}\bar{i}\right\rangle_{13572468}		\notag\\
   & 	\pm	\left|1000\bar{i}iii\right\rangle_{13572468} + \left|1001\bar{i}ii\bar{i}\right\rangle_{13572468}	+
	\left|1010\bar{i}i\bar{i}i\right\rangle_{13572468} \pm \left|1011\bar{i}i\bar{i}\bar{i}\right\rangle_{13572468}		\notag\\
   & 	+	\left|1100\bar{i}\bar{i}ii\right\rangle_{13572468} \mp \left|1101\bar{i}\bar{i}i\bar{i}\right\rangle_{13572468}	\mp
	\left|1110\bar{i}\bar{i}\bar{i}i\right\rangle_{13572468} + \left|1111\bar{i}\bar{i}\bar{i}\bar{i}\right\rangle_{13572468}	\notag\\
= &	
\begin{cases}
	\phantom{\pm}
	\left|\mathcal{G}_0^{+}\right\rangle \left|\mathcal{G}_0^{+}\right\rangle +
	\left|\mathcal{G}_0^{-}\right\rangle \left|\mathcal{G}_0^{-}\right\rangle \mp
	\left|\mathcal{G}_1^{+}\right\rangle \left|\mathcal{G}_1^{+}\right\rangle \mp
	\left|\mathcal{G}_1^{-}\right\rangle \left|\mathcal{G}_1^{-}\right\rangle \mp
	\left|\mathcal{G}_2^{+}\right\rangle \left|\mathcal{G}_2^{+}\right\rangle \mp
	\left|\mathcal{G}_2^{-}\right\rangle \left|\mathcal{G}_2^{-}\right\rangle +	
	\left|\mathcal{G}_3^{+}\right\rangle \left|\mathcal{G}_3^{+}\right\rangle +
	\left|\mathcal{G}_3^{-}\right\rangle \left|\mathcal{G}_3^{-}\right\rangle 	\\
\pm \left|\mathcal{G}_4^{+}\right\rangle \left|\mathcal{G}_4^{+}\right\rangle \pm
	\left|\mathcal{G}_4^{-}\right\rangle \left|\mathcal{G}_4^{-}\right\rangle -
	\left|\mathcal{G}_5^{+}\right\rangle \left|\mathcal{G}_5^{+}\right\rangle -
	\left|\mathcal{G}_5^{-}\right\rangle \left|\mathcal{G}_5^{-}\right\rangle -
	\left|\mathcal{G}_6^{+}\right\rangle \left|\mathcal{G}_6^{+}\right\rangle -
	\left|\mathcal{G}_6^{-}\right\rangle \left|\mathcal{G}_6^{-}\right\rangle \pm
	\left|\mathcal{G}_7^{+}\right\rangle \left|\mathcal{G}_7^{+}\right\rangle \pm
	\left|\mathcal{G}_7^{-}\right\rangle \left|\mathcal{G}_7^{-}\right\rangle,
	& \textrm{if} \phantom{i} i = 0;	\\
	\\
	\phantom{\pm}
	\left|\mathcal{G}_0^{+}\right\rangle \left|\mathcal{G}_0^{+}\right\rangle -
	\left|\mathcal{G}_0^{-}\right\rangle \left|\mathcal{G}_0^{-}\right\rangle \mp
	\left|\mathcal{G}_1^{+}\right\rangle \left|\mathcal{G}_1^{+}\right\rangle \pm
	\left|\mathcal{G}_1^{-}\right\rangle \left|\mathcal{G}_1^{-}\right\rangle \mp
	\left|\mathcal{G}_2^{+}\right\rangle \left|\mathcal{G}_2^{+}\right\rangle \pm
	\left|\mathcal{G}_2^{-}\right\rangle \left|\mathcal{G}_2^{-}\right\rangle +
	\left|\mathcal{G}_3^{+}\right\rangle \left|\mathcal{G}_3^{+}\right\rangle -
	\left|\mathcal{G}_3^{-}\right\rangle \left|\mathcal{G}_3^{-}\right\rangle 	\\
\pm \left|\mathcal{G}_4^{+}\right\rangle \left|\mathcal{G}_4^{+}\right\rangle \mp
	\left|\mathcal{G}_4^{-}\right\rangle \left|\mathcal{G}_4^{-}\right\rangle -
	\left|\mathcal{G}_5^{+}\right\rangle \left|\mathcal{G}_5^{+}\right\rangle +
	\left|\mathcal{G}_5^{-}\right\rangle \left|\mathcal{G}_5^{-}\right\rangle -
	\left|\mathcal{G}_6^{+}\right\rangle \left|\mathcal{G}_6^{+}\right\rangle +
	\left|\mathcal{G}_6^{-}\right\rangle \left|\mathcal{G}_6^{-}\right\rangle \pm
	\left|\mathcal{G}_7^{+}\right\rangle \left|\mathcal{G}_7^{+}\right\rangle \mp
	\left|\mathcal{G}_7^{-}\right\rangle \left|\mathcal{G}_7^{-}\right\rangle,
	& \textrm{if} \phantom{i} i = 1.
\end{cases}
\end{align}
Obviously, in each case described in Eq. \ref{ES_different_four_Bell_1},
the two GHZ states obtained by entanglement swapping are different,
while the two GHZ states obtained are the same in each case described in Eq. \ref{ES_different_four_Bell_2}.
It can be seen from these examples that the conclusions put forward above have been verified. 
For simplicity, we use Bell states as the initial states in these examples. 
As for the examples of entanglement swapping between SGHZ states, we will not present them here.


\section{Applications}

It has been shown that the entanglement swapping of Bell states and GHZ states have important applications not only
in quantum repeaters \cite{XuP119172017} and entanglement purification \cite{BoseS6011998},
but also in quantum information processing such as quantum cryptography
\cite{CabelloA6151999,LeeJ7032004,ZhangZ7232005,
ZXMan392006,KangMS2492015,ZhouN254462005}, 
quantum teleportation \cite{HongLu276562000}, and quantum computation \cite{LiQ8942014}.
In this section, by proposing three quantum protocols,
we describe the applications of the entanglement swapping schemes proposed above in
quantum key distribution, quantum private comparison, and quantum secret sharing.
For simplicity, in what follows, we wold only like to briefly describe the steps of the quantum protocols.
The detailed discussion of the quantum protocols, such as security analysis and efficiency analysis, is beyond the scope of this paper.

\subsection{Quantum key distribution}

Alice and Bob can prepare a set of shared keys through the following steps.

\begin{enumerate}

\item Alice (Bob) prepare randomly $n$ quantum states, 
each of which is in one of the states
$\left|\mathscr{X}^{\pm}\right\rangle$ (see Eq. \ref{Bell_SGHZ}),
and marks them by
$
\left|\mathcal{X}(p_1^h,p_2^h,\ldots,p_{2l}^h)\right\rangle
\Big(
\left|\mathcal{X}(q_1^h,q_2^h,\ldots,q_{2l}^h)\right\rangle
\Big),
$
where 
\begin{align}
&\left|\mathcal{X}(p_1^h,p_2^h,\ldots,p_{2l}^h)\right\rangle = \frac 1{\sqrt{2}}
\Big(
\left|0\right\rangle \bigotimes_{k=2}^{2l} \left|a_k^h\right\rangle \pm \left|1\right\rangle \bigotimes_{k=2}^{2l} \left|\bar{a}_k^h\right\rangle 
\Big) \notag \\
&\Big[\left|\mathcal{X}(q_1^h,q_2^h,\ldots,q_{2l}^h)\right\rangle = \frac 1{\sqrt{2}}
\Big(
\left|0\right\rangle \bigotimes_{k=2}^{2l} \left|b_k^h\right\rangle \pm \left|1\right\rangle \bigotimes_{k=2}^{2l} \left|\bar{b}_k^h\right\rangle 
\Big) \Big],
\end{align}
where $p_1^h,p_2^h,\ldots,p_{2l}^h(q_1^h,q_2^h,\ldots,q_{2l}^h)$
denote $2l$ particles in the $h$-th state, and
$0a_2^h a_3^h \cdots a_{l}^h = a_{l+1}^h a_{l+2}^h \cdots a_{2l}^h 
(0b_2^h b_3^h \cdots b_{l}^h = b_{l+1}^h b_{l+2}^h \cdots b_{2l}^h)$
or
$0a_2^h a_3^h \cdots a_{l}^h = \bar{a}_{l+1}^h \bar{a}_{l+2}^h \cdots \bar{a}_{2l}^h$
($0b_2^h b_3^h $ $ \cdots $ $b_{l}^h$ $= \bar{b}_{l+1}^h \bar{b}_{l+2}^h \cdots \bar{b}_{2l}^h$)
$\phantom{l} \forall h=1,2,\dots,n$.

\item Alice (Bob) takes the last (first) $l$ particles out from
$\left|\mathcal{X}(p_1^h,p_2^h,\ldots,p_{2l}^h\right\rangle$ 
\big($\left|\mathcal{X}(q_1^h,q_2^h,\ldots,q_{2l}^h\right\rangle$\big)
to construct the sequence 
\begin{align}
&p_{l+1}^1,p_{l+2}^1,\ldots,p_{2l}^1,p_{l+1}^2,p_{l+2}^2,\ldots,p_{2l}^2,\ldots,
p_{l+1}^n,p_{l+2}^n,\ldots,p_{2l}^n \notag \\
&(q_1^1,q_2^1,\ldots,q_l^1,q_1^2,q_2^2,\ldots,q_l^2,\ldots,q_1^n,q_2^n,\ldots,q_l^n),
\end{align}
and marks it by $S_a(S_b)$. The remaining particles construct another new sequence
\begin{align}
&p_1^1,p_2^1,\ldots,p_{l}^1,p_1^2,p_2^2,\ldots,p_{l}^2,\ldots,
p_1^n,p_2^n,\ldots,p_{l}^n \notag \\
&(q_{l+1}^1,q_{l+2}^1,\ldots,q_{2l}^1,q_{l+1}^2,q_{l+2}^2,\ldots,
q_{2l}^2,\ldots,q_{l+1}^n,q_{l+2}^n,\ldots,q_{2l}^n).
\end{align}

\item Alice and Bob sends $S_a$ and $S_b$ to each other.
Then Alice measures in turn the particles marked by
$p_1^h,p_2^h,\ldots,p_{l}^h$ and $q_1^h,q_2^h,\ldots,q_{l}^h$
(if $l=1$, Alice performs Bell measurements, 
otherwise she performs GHZ measurements).
Likewise, Bob measures in turn the particles marked by
$p_{l+1}^h,p_{l+2}^h,\ldots,p_{2l}^h$ and $q_{l+1}^h,q_{l+2}^h,\ldots,q_{2l}^h$.
Let us denote the measurement results of Alice (Bob) as
$\left|\mathcal{X}_a^h\right\rangle \big(\left|\mathcal{X}_b^h\right\rangle\big)$, where
\begin{align}
\left|\mathcal{X}_a^h\right\rangle = \frac 1{\sqrt{2}}
\Big(
\left|0\right\rangle \bigotimes_{k=2}^{2l} \left|i_k^h\right\rangle \pm \left|1\right\rangle \bigotimes_{k=2}^{2l} \left|\bar{i}_k^h\right\rangle \Big),
\left|\mathcal{X}_b^h\right\rangle = \frac 1{\sqrt{2}}
\Big(
\left|0\right\rangle \bigotimes_{k=2}^{2l} \left|j_k^h\right\rangle \pm \left|1\right\rangle \bigotimes_{k=2}^{2l} \left|\bar{j}_k^h\right\rangle \Big).
\end{align}

\item Alice and Bob publish the quantum states that they prepare in step 1 to each other.
If the two states prepared are different,
Alice and Bob use them for eavesdropping checking.
Otherwise, they can generate a shared key sequence in many ways
(according to Theorem~\ref{theorem_1}, 
the measurement results of Alice and Bob are the same;
Let us assume that there are enough states prepared by Alice and Bob 
to generate a set of keys).
For example, they can calculate $0 \oplus_{k=2}^{2l} i_k^h$
and $0 \oplus_{k=2}^{2l} j_k^h$ respectively, and take the calculation results as a set of keys,
where the symbol $\oplus$ denotes the module 2 operation (i.e., XOR operator).
They can also calculate 
$\sum_{k=2}^{2l} i_k^h \cdot 2^{2l-k}$
and 
$\sum_{k=2}^{2l} j_k^h \cdot 2^{2l-k}$ to generate keys.
Note here that 
$0 \oplus_{k=2}^{2l} i_k^h = 0 \oplus_{k=2}^{2l} j_k^h$
and
$\sum_{k=2}^{2l} i_k^h \cdot 2^{2l-k} = \sum_{k=2}^{2l} j_k^h \cdot 2^{2l-k}$.

\end{enumerate}


\subsection{Quantum private comparison}

Let us first give two prerequisites of the protocol.

\begin{enumerate}

\item Suppose that Alice and Bob
have the secret data $X$ and $Y$, respectively, and that the binary representations of
$X$ and $Y$ are $\left(  x_{1},x_{2},\ldots,x_{n} \right)$ and
$\left(  y_{1}, y_{2},\ldots,y_{n}  \right)$, respectively, where $n \in \rm N_+$,
$ x_{h}, y_{h} \in \{ 0, 1 \} \phantom{l} \forall h=1,2,\ldots,n$.
Alice and Bob want to judge whether $X = Y$
with the help of a semi-honest third party (usually called TP in quantum private comparison protocols);
TP is assumed to be faithful to execute the protocol processes
and not to conspire with Alice or Bob,
but he can record the calculation results generated in the protocol,
from which he may attempt to deduce the participants' data \cite{ZXJi1249112019,JiZX19112019}.

\item Alice, Bob and TP agree on the following coding rules:
$\left|0\right\rangle \leftrightarrow 0$ and $\left|1\right\rangle \leftrightarrow 1$.

\end{enumerate}


The steps of the protocol are as follows.

\begin{enumerate}

\item According to the value of $x_{h} (y_{h})$, 
Alice (Bob) prepares the Bell states
\begin{align}
\left|B(p_1^h,p_2^h)\right\rangle = 
\frac 1{\sqrt{2}}
\big( \left|0i_2^h\right\rangle +
\left|1\bar{i}_2^h\right\rangle \big)
\quad
\Big[
\left|B(q_1^h,q_2^h)\right\rangle = 
\frac 1{\sqrt{2}}
\big( \left|0j_2^h\right\rangle +
\left|1\bar{j}_2^h\right\rangle \big)
\Big],
\end{align}
where the particles in the state are marked by
$p_1^h,p_2^h$ ($q_1^h,q_2^h$),
the value of $i_2^h$ $(j_2^h)$ is the same as that of $x_{h} (y_{h})$.

\item Alice (Bob) takes all the particles out from
$\left|B(p_1^h,p_2^h)\right\rangle
\Big(\left|B(q_1^h,q_2^h)\right\rangle \Big)$
to construct the sequence
\begin{align}
p_1^1,p_2^1,p_1^2,p_2^2,\ldots,p_1^n,p_2^n \phantom{j}
\big(q_1^1,q_2^1,q_1^2,q_2^2,\ldots,q_1^n,q_2^n\big),
\end{align}
and denotes it as $S_a(S_b)$.

\item Alice(Bob) uses the decoy photon technology to send $S_a (S_b)$ to TP, 
where the decoy photon technology is used for eavesdropping checking
\cite{JiZX19112019,JiZX20122020}.
If an eavesdropper is detected, they terminate the protocol and start over.
Otherwise, TP performs Bell measurements on the first particle in $\left|B(p_1^h,p_2^h)\right\rangle$
and the first particle in $\left|B(q_1^h,q_2^h)\right\rangle$,
and marks the measurement results by
$\left|B_1^h\right\rangle$.
Similarly, he then measures the remaining particles in 
$\left|B(p_1^h,p_2^h)\right\rangle$ and $\left|B(q_1^h,q_2^h)\right\rangle$,
and marks the measurement results by $\left|B_2^h\right\rangle$.
According to Theorem \ref{theorem_1}, 
if $\left|B_1^h\right\rangle = \left|B_2^h\right\rangle \phantom{j} \forall h = 1,2,\ldots,n$,
TP can conclude that $X = Y$, otherwise $X \ne Y$.
Finally, TP announces publicly the comparison result to Alice and Bob.

\end{enumerate}

In this protocol, if TP attempts to perform single-particle measurements on
the Bell states that he receives at the last step of the protocol,
he can guess Alice's data $X$ with the successful probability of $1/2^n$.
Obviously, $1/2^n$ decreases with the increase in $n$,
and when the value of $n$ is small, there are many ways to increase it.
For example, Alice and Bob can agree in advance on a secret positive integer $M$
(or generate it using a quantum key distribution protocol),
and then they can calculate $X \times M$, or $X^M$ ($Y \times M$, or $Y^M$).
For Alice and Bob, they can only steal each other's data through quantum channels
because there is no particle exchanged between them.
In this case, they will not succeed and will be caught by eavesdropping checking \cite{JiZX19112019}.


\subsection{Quantum secret sharing}

Suppose that Alice wants to share a secret with 
$\textrm{Bob}_1,\textrm{Bob}_2,\ldots,\textrm{Bob}_n$ ($n \ge 2$),
in which the secret can only be deduced
by the collaboration of $\textrm{Bob}_1,\textrm{Bob}_2,\ldots,\textrm{Bob}_n$.
Alice, $\textrm{Bob}_1,\textrm{Bob}_2,\ldots,\textrm{Bob}_n$ agree on the coding rules:
$\left|0\right\rangle \rightarrow 0$ and $\left|1\right\rangle \rightarrow 1$.
They can achieve this through the following steps:

\begin{enumerate}

\item	Alice prepares $n$ copies of the Bell state
\begin{equation}
\left|\phi^{+}(p_1,p_2)\right\rangle = \frac 1{\sqrt{2}}
( \left|00\right\rangle + \left|11\right\rangle ),
\end{equation}
and marks them by
\begin{align}
\left|\phi^{+}(p_1^1,p_2^1)\right\rangle,
\left|\phi^{+}(p_1^2,p_2^2)\right\rangle,\ldots,
\left|\phi^{+}(p_1^n,p_2^n)\right\rangle,
\end{align}
in turn to construct an ordered sequence, where the superscripts $1,2,\ldots,n$
denote the order of the Bell states in the sequence,
and the subscripts 1,2 denote two particles in each state.

\item Alice takes the first particle marked by $p_1^h (h=1,2,\ldots,n)$ out from
$\left|\phi^{+}(p_1^h,p_2^h)\right\rangle$
to construct the new sequence $p_1^1,p_1^2,\ldots,p_1^n$.
The remaining particles construct another new sequence
$p_2^1,p_2^2,\ldots,p_2^n$.
Then Alice sends the particle marked by $p_1^h$ to $\textrm{Bob}_h$.

\item Alice performs GHZ (Bell) measurements (the measurement is a Bell measurement if $n = 2$)
on the articles marked by $p_2^1,p_2^2,\ldots,p_2^n$, and then $\textrm{Bob}_i$
performs single-particle measurements on his particle marked by $p_1^h$.
Let us denote the measurement result of Alice as
$\left|\mathcal{X}_a\right\rangle$, then
\begin{align}
\left|\mathcal{X}_a\right\rangle = \frac 1{\sqrt{2}}
\Big( 
\left|0\right\rangle \bigotimes_{k=2}^{n} \left|i_k\right\rangle \pm \left|1\right\rangle \bigotimes_{k=2}^{n} \left|\bar{i}_k\right\rangle 
\Big).
\end{align}
Alice calculates
$\sum_{k=2}^{n} i_k \cdot 2^{n-k}$,
and takes the calculation result as a key;
this key is unknown to $\textrm{Bob}_1,\textrm{Bob}_2,\ldots,\textrm{Bob}_n$, 
unless they share their measurement results with each other.

\end{enumerate}

For eavesdropping checking, Alice can prepare $n+mn$ copies of the Bell state
$\left|\phi^{+}(p_1,p_2)\right\rangle$ in the first step.
Then she takes the first particle in each state,
and send the $m+1$ particles to $\textrm{Bob}_i$.
Alice and $\textrm{Bob}_i$ use $m$ Bell states to
check whether there is eavesdropping in the quantum channel between them,
and the remaining Bell state to complete secret sharing.

Let us provide a simple example to check the correctness of the protocol.
Suppose that Alice's measurement result in the last step of the protocol is
$\left|\mathcal{X}_a\right\rangle = \frac 1{\sqrt{2}}
\Big( \left|0110\right\rangle + \left|1001\right\rangle \Big)$,
then Alice calculates $1\cdot2^{2}+1\cdot2^{1}+0\cdot2^0 = 6$ and takes it as the secret.
According to Corollary \ref{corollary_2}, 
the state that the particles of $\textrm{Bob}_1,\textrm{Bob}_2,\textrm{Bob}_3$ and $\textrm{Bob}_4$ 
collapses into are the same as $\left|\mathcal{X}_a\right\rangle$.
Therefore, after they perform single-particle measurements, 
the binary number corresponding to the measurement results is either 0110 or 1001.
If the binary number is 1001, they flip it to 0110.
In this way and by sharing the measurement results, they can know Alice's secret.

\section{Discussion}

We have proposed a series of entanglement swapping schemes in qubit systems,
and described quantum protocols for quantum key distribution,
quantum private comparison and quantum secret sharing based on entanglement swapping, respectively.
At present,  a lot of quantum information procession protocols have been proposed
based on the entanglement swapping of Bell states and GHZ states.
Compared with the other protocols without entanglement swapping, 
entanglement swapping makes these protocols have some obvious advantages in security and efficiency. 
The three proposed protocols are also based on the entanglement swapping of Bell states and GHZ states, 
thus they naturally inherit some of these advantages.
Below we list the main advantages of the proposed protocols:

\begin{enumerate}

\item In the processes of preparing keys in the quantum key distribution protocol and generating shared secrets
in the quantum secret sharing protocol, participants in the protocols (e.g., Alice and Bob)
do not need to choose between alternative measurements,
which can improve the rate of the bits generated by the transmitted qubit \cite{CabelloA6151999,KarimipourV652002}.

\item In the proposed quantum private comparison protocol, the Bell states are prepared by Alice and Bob rather than by TP,
which can improve the security and efficiency of the protocol \cite{JiZX19112019}.
For example, in order to achieve the purpose of private comparison,
Alice and Bob will not prepare false quantum states,
hence there is no need to adopt additional means to verify the authenticity of the Bell states in the protocol.
In addition, when the length of binary representations of $X$ and $Y$ is large (i.e., the value of n is large),
entanglement swapping makes the protocol avoid using a quantum key distribution protocol
to generate keys, thus the protocol has advantages in efficiency compared with the existing protocols
using quantum key distribution protocols \cite{ZXJi1249112019,JiZX19112019,JiZX181682019}.

\item Entanglement swapping can expand the distance of quantum communication and
make large-scale long-distance quantum communication possible.
Therefore, entanglement swapping makes the quantum protocols feasible
when the distance between participants is relatively long \cite{NicolasS8312011,ZhouN254462005}.

\end{enumerate}

In what follows we would like to briefly review the existing entanglement swapping schemes for multi-particle systems,
and then make a comparison between our proposed schemes and these schemes.
Let us take a look at Bouda and Buzek's schemes \cite{BoudaJ34202001} first. Indeed, in their proposed schemes, 
all the entanglement swapping cases of d-level maximally entangled states are considered, that is,
swapping any number of particles in any number of multi-particle maximally entangled states.
Now let us show the latter directly, because it is the most general case \cite{BoudaJ34202001}.
The expression of a d-level $(m+1)$-particle ($m \ge 2$) maximally entangled state 
(also known as cat states, see Ref. \cite{KarimipourV652002}) is given by
\begin{align}
\label{one-maximally-entangled-state}
\left|\phi(u_0,u_1,\dots,u_m)\right\rangle = \frac {1}{\sqrt{d}} 
\sum_{l=0}^{d-1} \zeta^{l u_0} \left|l,l \oplus u_1,l \oplus u_2,\dots,l \oplus u_m\right\rangle,
\end{align}
where $u_0,u_1,\dots,u_m \in \{0,1,\dots,d-1\}$, $ \zeta = e^{2\pi i/d} $,
and the symbol $\oplus$ denotes the addition modulo $d$.
Obviously, all the cat states are complete and orthonormal \cite{BoudaJ34202001,KarimipourV652002,JiZX181682019}, 
\begin{align}
\begin{cases}
\phantom{i} \langle\phi(u_0,u_1,\dots,u_m)|\phi(u_0',u_1',\dots,u_m')\rangle = 
\delta_{u_0,u_0'} \delta_{u_1,u_1'} \cdots \delta_{u_m,u_m'}, \\
\phantom{i} \left|u_0,u_1,\dots,u_m\right\rangle = \frac {1}{\sqrt{d}} 
\sum_{l=0}^{d-1} \zeta^{-l u_0} \left|\phi(l,u_1 \ominus u_0,u_2 \ominus u_0,\dots,u_m \ominus u_0)\right\rangle,
\end{cases}
\end{align}
where the symbol $\ominus$ denotes subtraction modulo $d$,
thus they can form a set of complete orthogonal basis in the Hilbert space 
$\mathcal{H} \otimes \mathcal{H} \otimes \cdots \otimes \mathcal{H}$
composed of $m+1$ d-level Hilbert space $\mathcal{H}$ \cite{BoudaJ34202001}.
A composite system composed of $n$ cat states can be expressed as
\begin{align}
\label{composite-maximally-system}
\left|\varPhi(M)\right\rangle = \bigotimes_{h=1}^{n} \left|\phi(u_0^h,u_1^h,\dots,u_{m_h}^h)\right\rangle,
\end{align}
where $M= \sum_{h=1}^n m_h + n$ and
the $h$-th cat state contains $m_h+1$ particles, specifically,
\begin{align}
\label{each-maximally-state}
\left|\phi(u_0^h,u_1^h,\dots,u_{m_h}^h)\right\rangle = \frac {1}{\sqrt{d}}
\sum_{l=0}^{d-1} \zeta^{l u_0^h} \left|l,l \oplus u_1^h,l \oplus u_2^h,\dots,l \oplus u_{m_h}^h\right\rangle,
\quad h =1,2,\dots,n.
\end{align}
Suppose that $k_h (1 \le k_h \le m_h)$ particles are selected from the $h$-th cat state,
then measuring them with the basis constructed by the $K$-particle ($K = \sum_{h=1}^n k_h$) cat states,
where the basis is given by
\begin{align}
\label{K-particle-cat-basis}
\left|\phi(v^0,v_2^1,v_3^1,\dots,v_{k_1}^1,v_1^2,v_2^2,\dots,v_{k_2}^2,\dots,v_1^n,v_2^n,\dots,v_{k_n}^n)\right\rangle,
\end{align}
such that the measurements on the selected $K$ particles will project them to this basis.
The state that the remaining unmeasured $M-K$ particles in $\left|\varPhi(M)\right\rangle$ project onto is marked by
\begin{align}
\label{remaining-cat-state}
\left|\phi(w^0,w_2^1,w_3^1,\dots,w_{m_1-k_1+1}^1,w_1^2,w_2^2,\dots,w_{m_2-k_2+1}^2,\dots,
w_1^n,w_2^n,\dots,w_{m_n-k_n+1}^n)\right\rangle,
\end{align}
where 
\begin{align}
\begin{cases}
\phantom{i} w_i^h = u_i^h \oplus l^h \ominus l^1  \quad \text{for} \quad i  \le m_h - k_h, \\
\phantom{i} w_{m_h - k_h + 1} = l^{h+1} \ominus l^1,
\end{cases}
\end{align}
By looking for the states which are in the following form
\begin{align}
\left|l^1,l^1 \oplus u_1^1,\dots,l^1 \oplus u_{m_1-k_1+1}^1,\dots,l^1 \oplus u_{m_1}^1,\dots,
l^n,l^n \oplus u_1^n,\dots,l^n \oplus u_{m_n-k_n+1}^n,\dots,l^n \oplus u_{m_n}^n\right\rangle,
\end{align}
one can get
\begin{align}
l^h = l^1 \oplus v_{k_{h-1}}^{h-1} \oplus u_{m_1-k_1+1}^1 \ominus u_{m_h - k_h + 1}^h, \quad h = 2,3,\dots,n.
\end{align}
To deduce $w^0$, let
\begin{align}
\prod_{h=1}^n \zeta^{l^h u_0^h} = \zeta^{(l^1 + u_{m_1-k_1+1}^1)v^0} \zeta^{l^1 w^0} \zeta^x,
\end{align}
where $\zeta^x$ is part of phase shift of 
$\left|\phi(w^0,w_1^1,w_2^1,\dots,w_{m_1-k_1+1}^1,w_1^2,w_2^2,\dots,w_{m_2-k_2+1}^2,\dots,
w_1^n,w_2^n,\dots,w_{m_n-k_n+1}^n)\right\rangle$, then one can get
\begin{align}
\big(\sum_{h=1}^n u_0^h - v^0 - w^0 \big) \cdot l^1 \phantom{i} \text{mod} \phantom{i} d  \equiv  
\big[\sum_{h=2}^n (u_{m_1-k_1+1}^1 + v_{k_{h-1}}^{h-1} - u_{m_h-k_h+1}^{h}) u_0^h + u_{m_1-k_1+1}^1 v^0 + x \big]
\phantom{i} \text{mod} \phantom{i} d ,
\end{align}
where the right side of the equation is equal to 0, which yields
\begin{align}
w^0 = \big(\sum_{h=1}^n u_0^h - v^0 \big) \phantom{i} \text{mod} \phantom{i} d.
\end{align}

The above parameters are determined by comparing the initial cat states
with the ones after measurements under the condition that the composite system
after entanglement swapping is determined in advance. For more details see Ref. \cite{BoudaJ34202001}.
Compared with the results given by Bouda and Buzek, the results of Karimipour et at. \cite{KarimipourV652002}
and Hardy et al. \cite{HardyL6252000} looks clearer, or more acceptable.
Therefore, we would now like to sort out some of Bouda and Buzek's main results
and describe them in the way that Karimipour et at. presents their results.
Let us start by showing a basic case, that is, swapping two particles in two cat states
by performing measurements using the d-level Bell basis given by \cite{BoudaJ34202001,KarimipourV652002}
\begin{align}
\label{d-level-Bell-state}
\left|\phi(u_0,u_1)\right\rangle = \frac {1}{\sqrt{d}} 
\sum_{l=0}^{d-1} \zeta^{l u_0} \left|l,l \oplus u_1\right\rangle.
\end{align}
Indeed, these Bell states are two-particle forms of cat states, such that they form a complete orthogonal basis
in the d-level Hilbert space $\mathcal{H} \otimes \mathcal{H}$.
Let us denote the two initial entangled states as 
$\left|\phi(u_0^1,u_1^1,\dots,u_{m_1}^1)\right\rangle$ containing $m_1+1$ particles
and $\left|\phi(u_0^2,u_1^2,\dots,u_{m_2}^2)\right\rangle$ containing $m_2+1$ particles, then one can arrive at
\begin{align}
\label{two-d-level-cat-states}
& \bigotimes_{h=1}^{2} \left|\phi(u_0^h,u_1^h,\dots,u_{m_h}^h)\right\rangle \notag \\
= & \frac {1}{d} \sum_{i,j=0}^{d-1} \zeta^{u_0^1 i \oplus u_0^2 j} 
\left|i,i \oplus u_1^1,\dots,i \oplus u_p^1,\dots,i \oplus u_{m_1}^1\right\rangle 
\left|j,j \oplus u_1^2,\dots,j \oplus u_q^2,\dots,j \oplus u_{m_2}^2\right\rangle \notag \\
= & \frac {1}{d} \sum_{i,j=0}^{d-1} \zeta^{u_0^1 i \oplus u_0^2 j} 
\left|i,i \oplus u_1^1,\dots,i \oplus u_{p-1}^1,i \oplus u_{p+1}^1,\dots,i \oplus u_{m_1}^1,
j,j \oplus u_1^2,\dots,j \oplus u_{q-1}^2,j \oplus u_{q+1}^2,\dots,j \oplus u_{m_2}^2\right\rangle  
\left|i \oplus u_p^1, j \oplus u_q^2\right\rangle \notag \\
= & \frac {1}{d^2} \sum_{i,j,l,k=0}^{d-1} \zeta^{\big[ u_0^1 i + u_0^2 j - il - (i + u_p^1)k \big] \phantom{i} \text{mod} \phantom{i} d}
\left|\phi(l,u_1^1,\dots,u_{p-1}^1,u_{p+1}^1,\dots,u_{m_1}^1,
j \ominus i,j \oplus u_1^2 \ominus i,\dots,j \oplus u_{q-1}^2 \ominus i,j \oplus u_{q+1}^2 \ominus i,\dots,
j \oplus u_{m_2}^2 \ominus i)\right\rangle \notag \\
& \quad \otimes \left|\phi(k,j \oplus u_q^2 \ominus i \ominus u_p^1)\right\rangle \notag \\
= & \frac {1}{d} \sum_{v_0,v_1=0}^{d-1}
\zeta^{\big[(u_p^1 - u_q^2 + v_1)u_0^2 - u_p^1 v_0\big] \phantom{i} \text{mod} \phantom{i} d}
|\phi(u_0^1 \oplus u_0^2 \ominus v_0,u_1^1,u_2^1,\dots,u_{p-1}^1,u_{p+1}^1,\dots,u_{m_1}^1, 
u_p^1 \ominus u_q^2 \oplus v_1,u_p^1 \ominus u_q^2 \oplus v_1 \oplus u_1^2,\dots \notag \\
& \quad
\dots,
u_p^1 \ominus u_q^2 \oplus v_1 \oplus u_{q-1}^2,u_p^1 \ominus u_q^2 \oplus v_1 \oplus u_{q+1}^2,\dots,
u_p^1 \ominus u_q^2 \oplus v_1 \oplus u_{m_2}^2)\rangle
\otimes \left|\phi(v_0,v_1)\right\rangle
\end{align}
where the particles are swapped at the second step in the above formula as before.
Next is the most general case. Without losing generality, 
let us select the first $k_h+1$ particles from each cat state and measure the last $m_h - k_h$ particles, 
then the entanglement swapping case can be described as follows:
\begin{align}
\label{n-d-level-cat-states-ES}
& \bigotimes_{h=1}^{n} \left|\phi(u_0^h,u_1^h,\dots,u_{m_h}^h)\right\rangle \notag \\
= & \frac {1}{d^{n/2}} \sum_{l^1,l^2,\dots,l^n=0}^{d-1} \zeta^{\sum_{h=1}^{n} l^h u_0^h} 
\bigotimes_{t=1}^{n} \left|l^h, l^h \oplus u_1^h,\dots,l^h \oplus u_{m_h}^h\right\rangle \notag \\
= & \frac {1}{d^{n/2}} \sum_{l^1,l^2,\dots,l^n=0}^{d-1} \zeta^{\sum_{h=1}^{n} l^h u_0^h \phantom{i} \text{mod} \phantom{i} d} 
\bigotimes_{h=1}^{n} \left|l^h, l^h \oplus u_1^h,\dots,l^h \oplus u_{k_h}^h\right\rangle
\bigotimes_{h=1}^{n} \left|l^h \oplus u_{k_h+1}^h, l^h \oplus u_{k_h+2}^h,\dots,l^h \oplus u_{m_h}^h\right\rangle \notag \\
= & \frac {1}{d^{n/2+1}} \sum_{l^1,l^2,\dots,l^n,i,j=0}^{d-1} 
\zeta^{\big(\sum_{h=1}^{n} l^h u_0^h - l^1 i - (l^1 + u_{k_1+1})j \big)\phantom{i} \text{mod} \phantom{i} d} 
|\phi(i,u_1^1,u_2^1,\dots,u_{k_1}^1,
l^2 \ominus l^1,l^2 \ominus l^1 \oplus u_1^2,\dots,l^2 \ominus l^1 \oplus u_{k_2}^2,\dots \notag \\
& \quad \dots,
l^n \ominus l^1,,l^n \ominus l^1 \oplus u_1^2,\dots,l^n \ominus l^1 \oplus u_{k_n}^n\rangle
\otimes |\phi(j,u_{k_1+2}^1 \ominus u_{k_1+1}^1,u_{k_1+3}^1 \ominus u_{k_1+1}^1,\dots,u_{m_1}^1 \ominus u_{k_1+1}^1, \notag \\
& \quad
l^2 \ominus l^1 \oplus u_{k_2+1} \ominus u_{k_1+1},\dots,l^2 \ominus l^1 \oplus u_{m_2+2}^2 \ominus u_{k_1+1}^1,\dots,
l^n \ominus l^1 \oplus u_{k_n+1}^n \ominus u_{k_1+1}^1,\dots,l^n \ominus l^1 \oplus u_{m_n+2}^n \ominus u_{k_1+1})\rangle \notag \\
= & \frac {1}{d^{n/2}} \sum_{v^0,v_1^2,\dots,v_1^n=0}^{d-1}  
|\phi\big(\oplus_{h=1}^{n} u_0^h \ominus v^0,u_1^1,u_2^1,\dots,u_{k_1}^1,
v_1^2 \oplus u_{k_1+1}^1 \ominus u_{k_2+1}^2,v_1^2 \oplus u_{k_1+1}^1 \ominus u_{k_2+1}^2 \oplus u_1^2,\dots \notag \\
& \quad \dots,v_1^2 \oplus u_{k_1+1}^1 \ominus u_{k_2+1}^2 \oplus u_{k_2}^2,\dots,
v_1^n \oplus u_{k_1+1}^1 \ominus u_{k_n+1}^n,v_1^n \oplus u_{k_1+1}^1 \ominus u_{k_n+1}^n \oplus u_1^n,\dots,
v_1^n \oplus u_{k_1+1}^1 \ominus u_{k_n+1}^n \oplus u_{k_n}^n \big)\rangle \notag \\
& \quad \otimes |\phi(v^0,v_1^1,v_2^1,\dots,v_{m_1-k_1-1}^1,v_1^2,v_2^2,\dots,v_{m_2-k_2}^2,\dots,
v_1^n,v_2^n,\dots,v_{m_n-k_n}^n)\rangle,
\end{align}

Let us then introduce the entanglement swapping scheme proposed by Bose et al. \cite{BoseS5721998}.
The expression of a $m$-particle ($m \ge 2$) 2-level cat state is given by
\begin{align}
\label{one-cat-state}
\left|\mathcal{C}(m)\right\rangle = \frac {1}{\sqrt{2}}
\Big( \bigotimes_{k=1}^{m} \left|i_k\right\rangle
\pm
\bigotimes_{k=1}^{m} \left|\bar{i}_k\right\rangle \Big),
\end{align}
where $i_k \in \{0,1\}$.
The following composite system can be constructed with $n$ 2-level cat states
\begin{align}
\label{composite-cat-system}
\bigotimes_{h=1}^{n} \left|\mathcal{C}(m_h)\right\rangle,
\end{align}
where each state is composed of $m_h (m_h \ge 2 \phantom{i} \forall n = 1,2,\dots,n)$ particles and can be expressed as
\begin{align}
\label{one-cat-state}
\left|\mathcal{C}(m_h)\right\rangle = \frac {1}{\sqrt{2}}
\Big( \bigotimes_{k=1}^{m_h} \left|i_k^h\right\rangle
\pm
\bigotimes_{k=1}^{m_h} \left|\bar{i}_k^h\right\rangle \Big).
\end{align}
Suppose that the first $l_h (1\le l_h \le m_h)$ particles are selected from each cat state,
then one can perform a joint measurement on these particles with the complete orthogonal basis
constructed by $L$-particle ($L = \sum_{h=1}^n l_h$) cat states,
where the basis is given by
\begin{align}
\label{L-cat-basis}
\left|\mathcal{B}(L)\right\rangle = \frac 1{\sqrt{2}}
\Big(
\bigotimes_{h=1}^{n} 
\bigotimes_{k=1}^{l_h} \left|i_k^h\right\rangle
\pm
\bigotimes_{h=1}^{n}
\bigotimes_{k=1}^{l_h} \left|\bar{i}_k^h\right\rangle 
\Big).
\end{align}
With this basis, the measurements performed on the selected $L$ particles will project them to this basis,
and the remaining particles in system $\left|\mathcal{C}(S)\right\rangle$ are projected onto
\begin{align}
\label{remaining-cat-state}
\left|\mathcal{B}(\sum_{h=1}^n m_h - L)\right\rangle = \frac 1{\sqrt{2}}
\Big(
\bigotimes_{h=1}^{n}
\bigotimes_{k=l_h+1}^{m_h} \left|i_k^h\right\rangle
\pm
\bigotimes_{h=1}^{n}
\bigotimes_{k=l_h+1}^{m_h} \left|\bar{i}_k^h\right\rangle 
\Big).
\end{align}
To sum up, the above entanglement swapping process can be denoted by the following schematic expression:
\begin{align}
\bigotimes_{h=1}^{n} \left|\mathcal{C}(m_h)\right\rangle \rightarrow 
\left|\mathcal{B}(L)\right\rangle \otimes \left|\mathcal{B}(\sum_{h=1}^n m_h - L)\right\rangle.
\end{align}

Finally, let us take a look at Karimipour et al.'s results, which are introduced in Ref. \cite{KarimipourV652002}.
Let one of the two initial cat states in Eq. \ref{d-level-Bell-state} be a d-level Bell state, 
and suppose that another cat state contains $m+1$ particles.
Swapping the particle with the mark $u_k^1$ in the cat state and the particle with the mark $u_1^2$ in the Bell state,
one can get
\begin{align}
\label{cat-bell-ES}
&\left|\phi(u_0^1,u_1^1,\dots,u_m^1)\right\rangle \otimes \left|\phi(u_0^2,u_1^2)\right\rangle \notag \\
= & \frac {1}{d} \sum_{v_0,v_1=0}^{d-1} \zeta^{(u_k^1 \ominus v_1)(u_0^2 \ominus v_0)}
\left|\phi(u_0^1 \oplus u_0^2 \ominus v_0,u_1^1,u_2^1,\dots,u_1^2 \oplus u_k^1 \ominus v_1,\dots,u_m^1)\right\rangle
\otimes \left|\phi(v_0,v_1)\right\rangle,
\end{align}
let $u_k^1 \ominus v_1=l_1$ and $u_0^2 \ominus v_0 = l_2$, one can get 
\begin{align}
&\left|\phi(u_0^1,u_1^1,\dots,u_{m_1}^1)\right\rangle \otimes \left|\phi(u_0^2,u_1^2)\right\rangle \notag \\
= & \frac {1}{d} \sum_{l_1, l_2=0}^{d-1} \zeta^{l_1 l_2}
\left|\phi(u_0^1 \oplus l_2,u_1^1,u_2^1,\dots,u_1^2 \oplus l_1,\dots,u_m^1)\right\rangle
\otimes \left|\phi(v_0 \ominus l_2,v_1 \ominus l_1)\right\rangle.
\end{align}
This is the result obtained by Karimipour et al. \cite{KarimipourV652002}.
Indeed, strictly speaking, Karimipour et al.'s results are included in the ones considered by Bouda and Buzek \cite{BoudaJ34202001}.
Nevertheless, we argue that Karimipour et al.'s result is clear and can be perceived directly.
Further, as far as the application of entanglement swapping in quantum information processing is concerned,
Karimipour et al.'s result plays a more positive role. Based on their work,
some quantum information processing protocols were proposed, such as
the quantum secret sharing protocol \cite{KarimipourV652002} and the quantum secure multi-party summation
protocols introduced by us recently \cite{JiZX181682019}.

We would now like to make a brief comparison among our entanglement swapping scheme,
Bouda and Buzek's scheme \cite{BoudaJ34202001}, and Bose et al.'s scheme \cite{BoseS5721998} 
to show the difference and connection between our work and their work.
For clarity, our schemes and their schemes are listed in Table \ref{table}.
From Eq. \ref{one-maximally-entangled-state}, one can get the Bell states
\begin{align}
\label{two-level-Bell-state}
&\left|\phi(0,0)\right\rangle = \frac 1{\sqrt{2}}
\Big(   \left|00\right\rangle + \left|11\right\rangle  \Big),
\quad
\left|\phi(0,1)\right\rangle = \frac 1{\sqrt{2}}
\Big(   \left|01\right\rangle + \left|10\right\rangle  \Big), \notag\\
&\left|\phi(1,0)\right\rangle = \frac 1{\sqrt{2}}
\Big(   \left|00\right\rangle - \left|11\right\rangle  \Big),
\quad
\left|\phi(1,1)\right\rangle = \frac 1{\sqrt{2}}
\Big(   \left|01\right\rangle - \left|10\right\rangle  \Big),
\end{align}
which is also shown in Eq. \ref{Bell_states}. Likewise, one can get any multi-particle GHZ states
such as the familiar three-particle GHZ states
\begin{align}
\label{two-level-GHZ-state}
&\left|\phi(0,0,0)\right\rangle = \frac 1{\sqrt{2}}
\Big(   \left|000\right\rangle + \left|111\right\rangle  \Big),
\quad
\left|\phi(0,0,1)\right\rangle = \frac 1{\sqrt{2}}
\Big(   \left|001\right\rangle + \left|110\right\rangle  \Big), \notag\\
&\left|\phi(0,1,0)\right\rangle = \frac 1{\sqrt{2}}
\Big(   \left|010\right\rangle + \left|101\right\rangle  \Big),
\quad
\left|\phi(0,1,1)\right\rangle = \frac 1{\sqrt{2}}
\Big(   \left|011\right\rangle + \left|100\right\rangle  \Big), \notag\\
&\left|\phi(1,0,0)\right\rangle = \frac 1{\sqrt{2}}
\Big(   \left|000\right\rangle - \left|111\right\rangle  \Big),
\quad
\left|\phi(1,0,1)\right\rangle = \frac 1{\sqrt{2}}
\Big(   \left|001\right\rangle - \left|110\right\rangle  \Big),	\notag\\
&\left|\phi(1,1,0)\right\rangle = \frac 1{\sqrt{2}}
\Big(   \left|010\right\rangle - \left|101\right\rangle  \Big),
\quad
\left|\phi(1,1,1)\right\rangle = \frac 1{\sqrt{2}}
\Big(   \left|011\right\rangle - \left|100\right\rangle  \Big).
\end{align}
Therefore, the most general entanglement swapping scheme shown in Eq. \ref{n-d-level-cat-states-ES}
includes our scheme shown in Eqs. \ref{L-GHZ-basis} and \ref{remaining_state}.
Let us then discuss whether Eqs. \ref{L-GHZ-basis} and \ref{remaining_state} can be derived from
Bose et al.'s results shown in  Eqs. \ref{L-cat-basis} and \ref{remaining-cat-state}.
In fact, the two-level cat states introduced by Bose et al. contains the following three types of states:
\begin{align}
\begin{cases}
\left|\mathscr{S}_1(m)\right\rangle = \frac {1}{\sqrt{2}}
\Big(\left|0\right\rangle \bigotimes_{k=2}^m \left|i_{k}\right\rangle
+
\left|1\right\rangle \bigotimes_{k=2}^m \left|\bar{i}_{k}\right\rangle \Big), \\
\left|\mathscr{S}_2(m)\right\rangle = \frac {1}{\sqrt{2}}
\Big(\left|0\right\rangle \bigotimes_{k=2}^m \left|i_{k}\right\rangle
-
\left|1\right\rangle \bigotimes_{k=2}^m \left|\bar{i}_{k}\right\rangle \Big), \\
\left|\mathscr{S}_3(m)\right\rangle = \frac {1}{\sqrt{2}}
\Big(\left|1\right\rangle \bigotimes_{k=2}^m \left|i_{k}\right\rangle
-
\left|0\right\rangle \bigotimes_{k=2}^m \left|\bar{i}_{k}\right\rangle \Big).
\end{cases}
\end{align}
Obviously, the two-level cat states are essentially different from the GHZ states shown in Eq. \ref{GHZ2},
because the GHZ states do not contain the states $\left|\mathscr{S}_3(m)\right\rangle$.
Therefore, the results shown in Eqs. \ref{L-GHZ-basis} and \ref{remaining_state} 
cannot be directly derived from Bose et al.'s results unless the state of 
the first particle is limited to the state $\frac {1}{\sqrt{2}}(\left|0\right\rangle \pm \left|1\right\rangle)$.
Of course, Bouda and Buzek's results do not include Bose et al.'s results due to
the d-level cat states introduced by Bouda and Buzek do not contain the states
$\left|\mathscr{S}_3(m)\right\rangle$ as well.
Admittedly, from the perspective of these differences and connections,
our results shown in Eqs. \ref{L-GHZ-basis} and \ref{remaining_state}
seem to be less important, but such more transparent results are
enlightening to find their applications in many fields such as quantum information processing.
Regardless of this, this paper is mainly devoted to studying the entanglement swapping of Bell states and SGHZ states.
The proposed schemes in this paper extend the properties of the entanglement swapping for
Bell states and multi-particle GHZ states. In other words, our main work is to further
consider the entanglement swapping of Bell states and the multi-particle GHZ states in a detailed way,
rather than limited to the general rules of their entanglement swapping.

\begin{table}[!h]
\centering
\setlength{\tabcolsep}{18pt}

\caption{Entanglement swapping schemes}
\label{table}
\begin{tabular}{cc}

\noalign{\smallskip}\hline\noalign{\smallskip}

Bose et al.'s schemes  \cite{BoseS5721998}
& Entanglement swapping between any number of two-level cat states \\

\noalign{\bigskip}

Bouda and Buzek's shemes \cite{BoudaJ34202001}
& Entanglement swapping between any number of d-level cat states \\

\noalign{\bigskip}

Karimipour et al.'s schemes \cite{KarimipourV652002}
& \makecell{Entanglement swapping between d-level cat states and Bell states} \\

\noalign{\bigskip}

Our schemes
& \makecell{Entanglement swapping of Bell states, multi-particle GHZ states, \\
and SGHZ states in qubit systems}	\\

\noalign{\smallskip}\hline

\end{tabular}
\end{table}

\section{Conclusion}

We have shown the general formalism describing the entanglement swapping
between any number of qubit systems for both Bell states and GHZ states.
We have introduced a class of multi-particle GHZ states, and proposed a series of
entanglement swapping schemes between any number of Bell states and the introduced GHZ states.
We have shown that the proposed schemes are not only useful for the creation of multi-particle GHZ states,
but also have applications in quantum information processing.
In view of the wide applications of entanglement swapping in the fields of 
entanglement purification, quantum repeaters and quantum information science, 
we will further study the usefulness of the proposed entanglement swapping schemes in these fields in the future.
As for the entanglement swapping between more than two qubit systems for the introduced GHZ states,
we only consider the measurements of the first half of the particles in each state,
but do not consider the cases of measuring the second half of the particles in some states. This is an open question.

\section*{Aknowledgements}

The authors thank the anonymous referees for their important suggestions and comments.
This work is supported by the National Natural Science Foundation of China
(Grants No. 61332019, No. 61202386, No. 61402339), and the Major
State Basic Research Development Program of China (973 Program, Grant No. 2014CB340601).

\end{document}